\documentclass[a4paper,11pt,oneside]{report}
\usepackage[DIV=20,BCOR=2mm,headinclude=true,footinclude=false]{typearea}
\usepackage[title,titletoc,toc]{appendix}
\usepackage{amsmath} 
\usepackage{amsthm} 
\usepackage{amssymb}	
\usepackage{graphics} 
\usepackage{multicol} 
\usepackage{float}
\usepackage[section]{placeins}
\usepackage{wrapfig}
\usepackage{subcaption}
\usepackage{pbox}
\usepackage{bbm}
\usepackage[font=small,labelfont=bf]{caption}
\usepackage{framed}
\usepackage{multirow}
\usepackage{tensor}
\usepackage{array}
\usepackage{rotating}
\usepackage[]{siunitx}
\DeclareSIUnit\parsec{pc}

\newcommand\id{\ensuremath{\mathbbm{I}}}
\newcommand\etaperp{\eta^{\bot}}
\newcommand\etapar{\eta^{\|}}
\newcommand\xipar{\xi^{\|}}
\newcommand\xiperp{\xi^{\bot}}
\renewcommand\L{\mathcal{L}}
\renewcommand\H{\mathcal{H}}
\renewcommand\P{\mathcal{P}}

\newcommand\R{\mathcal{R}}
\renewcommand\S{\mathcal{S}}
\newcommand\C{\mathcal{C}}
\newcommand\M{\mathcal{M}}

\DeclareMathOperator{\Tr}{Tr}
\newcommand{\sgn}{\mathop{\mathrm{sgn}}}

\renewcommand{\v}[1]{\ensuremath{\mathbf{#1}}} 
\newcommand{\gv}[1]{\ensuremath{\mbox{\boldmath$ #1 $}}}

\newcommand{\ket}[1]{\left| #1 \right>} 
\newcommand{\bra}[1]{\left< #1 \right|} 
\let\baraccent=\= 
\renewcommand{\=}[1]{\stackrel{#1}{=}} 

\newtheorem{asm}{Assumption}

\theoremstyle{definition}

\theoremstyle{remark}

\usepackage{xcolor,colortbl}
\definecolor{Gray}{gray}{0.9}
\newcolumntype{g}{>{\columncolor{Gray}}p}
\makeatletter

\newcommand*{\@rowstyle}{}
\newcommand*{\rowstyle}[1]{
  \gdef\@rowstyle{#1}%
  \@rowstyle\ignorespaces%
  }

\newcolumntype{=}{
  >{\gdef\@rowstyle{}}%
}

\newcolumntype{+}{
  >{\@rowstyle}%
}

\makeatother

\begin{document}

\title{
	{Multiple Field Inflation and Signatures of Heavy Physics in the CMB}\\
	{\large University of Utrecht}\\
}
\author{\begin{tabular}{r@{ }l}
                & Yvette Welling \\[1ex]
Supervisors:    & Ana Ach\'ucarro \\
                & Tomislav Prokopec
\end{tabular}}

\date{\today}

\maketitle

\chapter*{Abstract}
In this thesis we present our research in the context of multiple field inflation. Current precision measurements of the cosmic microwave background radiation (CMB) provide compelling evidence for the paradigm of inflation. Simple single field inflation can fit the data well, but the precise microphysical origin of inflation remains unknown. We are interested in the case that single field inflation is an effective description of a more fundamental theory containing multiple scalar fields. Even if these extra degrees of freedom are all very heavy compared to the Hubble scale of inflation they still might influence the dynamics of the inflaton. This leads to features in the statistical properties of the temperature fluctuations in the CMB. We study what the possible effects are and how future data might be able to detect physics beyond single field inflation. In this thesis we will first introduce inflation and its current observational status. Then we discuss the master's research, which consists of two projects. The first project is to provide an overview of studies of multiple field inflation in the literature. We translate between the different notations and definitions used in various papers and study the different approximation schemes and their regime of validity. The second project is a numerical study of concrete models of multi-field inflation from recent papers in the literature. We study if the current and future experiments might be able to detect the presence of the additional fields in these models.

\tableofcontents

\chapter{Outline Thesis}
Cosmology is a fascinating area of research. The last decades have provided us with a huge amount of experimental data which in combination with theoretical physics rapidly increased our knowledge about the intriguing history of our universe. The redshift-distance mappings of galaxies established the fact the universe is expanding and supported the idea the universe originated from some extremely hot and dense state. This so called \emph{Hot Big Bang model} could in addition explain the abundances of light elements in the universe and gained important evidence from the observation of the \emph{Cosmic Microwave Background} (CMB). In combination with the galaxy surveys it turned out the history of the universe is very well described by the $\Lambda$CDM concordance model. On the other hand, these observations gave rise to questions concerning the initial homogeneity of the universe and led to the idea of \emph{inflation}. It was soon realized that inflation could also explain the origin of the primordial density perturbations, which are responsible for the large scale structure in the universe. These density fluctuations result into temperature fluctuations of the CMB, because the photons need to climb out the gravitational wells. It was a great triumph for the paradigm of inflation when a more precise measurement of the temperature anisotropies of the CMB established the generic predicted properties by inflation.\\

Different scenarios for inflation lead to very similar predictions, however, they differ in the more subtle details. Because the current data is extremely precise, we can try to use the CMB as a probe for physics beyond the simplest models for inflation. The CMB provides a unique window into the early universe and therefore in general it is extremely interesting to understand the observational signatures of different inflationary models, such that we can recognize hints of new physics or falsify classes of models in the current and future data. In this thesis we focus on multi-field inflation and its signatures in the CMB data. \\

In chapter \ref{Chapter:inflation} we review the basics of the standard model of cosmology. We start with a description of the Big Bang model and explain how inflation can solve certain problems within this theory. Then we discuss the observational evidence of the paradigm of inflation and outline future tests of inflation which can be used to learn more about the physics of the early universe. From chapter \ref{Chapter:multifieldinflation} on we consider the case that single field inflation is an effective description of a more fundamental theory containing multiple scalar fields. In chapter \ref{Chapter:multifieldinflation} we discuss the first research project which is an overview of studies of multiple field inflation in the literature. This allows us to introduce multiple field inflation in full detail. We translate between the different notations and definitions used in various papers and study the different approximation schemes and their regime of validity. In addition we discuss the predictions of observables within the approximation schemes. In chapter \ref{Chapter:dataanalysis} we discuss the second research project in which we perform a numerical study of concrete models of multi-field inflation from recent papers in the literature. We study if the current and future experiments might be able to detect the presence of the additional fields in these models. Finally, in chapter \ref{Chapter:discussionoutlook} we summarize and discuss our findings.

\chapter{The Standard Model of Cosmology}
\label{Chapter:inflation}
The latest decades of precision observations have led to great advances in the understanding of our universe. Observations of supernovae \cite{Perlmutter:1998np, Riess:1998cb}, maps of large scale structure (LSS) \cite{Abazajian:2008wr} and the precision measurements of the temperature anisotropies in the cosmic microwave background radiation (CMB) \cite{Hinshaw:2012aka, Ade:2013zuv} have resulted in the standard model of cosmology, the $\Lambda$CDM model, which is a combination of the Big Bang model and Gaussian, adiabatic and almost scale invariant initial conditions, as predicted by the simplest models of inflation. In this chapter we review the basics of the standard model of cosmology. We start with a description of the Big Bang model and then focus on the role played by inflation, how to realize inflation and its current observational evidence.

\section{The Big Bang Model}
The Big Bang theory can successfully account for the expansion history of the homogeneous universe, the existence of the CMB, the abundance of light elements and the formation of the structure. However it cannot explain why the universe was almost perfectly homogeneous at early times and where the initial density perturbations come from. In this section\footnote{This section is largely based on \cite{BaumannA, Dodelson:2003ft} where much more can be found.} we give a description of the Big Bang model. We first recap some physics of the homogeneous universe and explain the successes of the Big Bang theory and then we address the problems related to initial conditions.

\subsection{Evolution of the homogeneous universe}
Galaxy surveys \cite{Abazajian:2008wr} indicate that on the largest scales the distribution of structure in our universe is independent of the direction we point our telescopes. Assuming that we are not a special observer and that we would have measured the same distribution of large scale structure at any other position in space this results in the cosmological principle. The cosmological principle states that the universe is homogeneous and isotropic on the largest scales. The Big Bang model relies on the validity of the cosmological principle and on the universality of the physical laws, in particular the gravitational laws of general relativity. From these symmetries of the universe it follows that geometry of the spacetime of the universe is described by the Friedmann-Robertson-Walker (FRW) metric
\begin{equation}
ds^2=-dt^2+a^2(t)\left[\frac{dr^2}{1-\kappa r^2}+r^2(d\theta^2+\sin^2\theta d\phi^2)\right],
\label{Equation:FRWmetric}
\end{equation}
which is completely determined by the scale factor $a(t)$ and the curvature $\kappa$ of the spatial hypersurfaces of constant time. The spatial slices can either be flat ($\kappa=0$), positively curved ($\kappa=1$) or negatively curved ($\kappa=-1$). The scale factor represents the relative size of the universe at a given time. Often the line element is written in terms of comoving coordinates $x^i$
$$
ds^2=-dt^2+a^2(t)\gamma_{ij}dx^i dx^j,
$$
where the spatial metric $\gamma_{ij}$ contains the information about the curvature of the spatial slices. Comoving coordinates can be thought of as the grid points on a coordinate grid which expands with the expansion of the universe. This means the comoving distance between two points stays the same all the time. The physical coordinates $a(t)x^i$ however imply a growth of the physical distance between two points on the grid as the universe expands. It turns out that the comoving frame is a very natural frame. Using the geodesic equation one can derive that for both massive and massless particles the physical three momentum, with respect to the comoving frame, decays with the expansion of the universe as $p\sim\frac{1}{a}$. This means that in an expanding universe all particles will eventually come to rest with respect to the comoving frame. Moreover this means that the wavelength of light is stretched with the expansion of the universe. Photons from distant sources will therefore be redshifted, this is quantified by the redshift parameter $z$
\begin{equation}
z(t_1)\equiv\frac{\lambda(t_0)-\lambda(t_1)}{\lambda(t_1)} \quad \rightarrow \quad z+1=\frac{1}{a(t_1)},
\label{Equation:redshift}
\end{equation}
where we have defined $a(t_0)=1$. Redshift is being used as a time variable to describe the history of the late-time universe.\\

The expansion history of the universe or the evolution of the scale factor $a(t)$ can be found if we know the matter content of the universe. Homogeneity and isotropy requires the stress energy tensor to take the form of a perfect fluid
$$
T_{\mu\nu}=(\rho+P)U_\mu U_\nu-P g_{\mu\nu}.
$$
Covariant conservation of energy-momentum yields the continuity equation
\begin{equation}
\dot{\rho}+3H(\rho+P)=0,
\label{Equation:Continuity equation}
\end{equation}
and the Einstein equations yield the Friedmann equations
\begin{equation}
H^2=\frac{\rho}{3M_P^2}-\frac{\kappa}{a^2},
\label{Equation:Friedmann1}
\end{equation}
\begin{equation}
\frac{\ddot{a}}{a}=-\frac{1}{6M_P^2}(\rho+3P).
\label{Equation:Friedmann2}
\end{equation}
where $M_P\equiv\sqrt{\frac{\hbar c}{8\pi G}}$ is the reduced Planck mass and $H\equiv \frac{\dot{a}}{a}$ the Hubble parameter for which the best estimate today is given by $H_0=(67.3 \pm1.2) \SI{}{\km\per\s\per\mega\parsec}$, using the Planck, WMAP low-$l$ polarization and high resolution CMB data \cite{Ade:2013zuv}. Only two of the three equations are independent, but they can all be very useful. From the continuity equation (\ref{Equation:Continuity equation}) we can derive how the energy density changes in an expanding universe. If a cosmological fluid satisfies the following equation of state $P=\omega \rho$ then the continuity equation gives
$$
\rho\sim a^{-3(1+\omega)}.
$$
In the case of matter such as the visible nuclei and electrons we can neglect the pressure and we find $\rho_{m}\sim a^{-3}$. Besides visible matter which is called baryonic matter it is believed there is another matter component in the universe which is called dark matter. In case of a gas of relativistic particles we have $\omega=\frac{1}{3}$ such that $\rho_{r}\sim a^{-4}$. This agrees with the fact that the energy of these particles is redshifted as the universe expands such that it falls of with one more factor of the scale factor. This type of cosmological fluid is called radiation and consists of the massless particles photons and gravitons and other relativistic particles such as neutrinos (only in the early universe). Interestingly, it has turned out that there is another fluid component of the universe which has a negative equation of state $\omega=-1$ which leads to a constant energy density $\rho_\Lambda\sim 1$. This mysterious cosmological fluid is called dark energy. By use of the Friedmann equations (\ref{Equation:Friedmann1}) and (\ref{Equation:Friedmann2}) we can determine the evolution of the scale factor in presence of these three cosmological fluids. Defining a critical density $\rho_c\equiv 3M_P^2 H_0^2$ corresponding to a flat universe today and the density parameters $\Omega_i\equiv\frac{\rho_i}{\rho_c}$ the Friedmann equation becomes
\begin{equation}
H^2=H_0^2\left[\Omega_{m}\left(\frac{1}{a}\right)^3+\Omega_{r}\left(\frac{1}{a}\right)^4+\Omega_\Lambda-\frac{\kappa}{H_0^2}\left(\frac{1}{a}\right)^2\right],
\label{Equation:Friedmann and matter}
\end{equation}
where the matter density parameter consists of one part belonging to baryonic matter and another part of dark matter $\Omega_{m}=\Omega_{dm}+\Omega_{b}$. If we know the values of the density parameters today then we can read off the expansion history of the homogeneous universe. 
The three cosmological fluids scale differently with the scale factor $a$ and therefore there are different epochs in the history of the universe where its content was dominated by one fluid. The first Friedmann equation (\ref{Equation:Friedmann1}) and the continuity equation (\ref{Equation:Continuity equation}) can be combined to find the time evolution of the scale factor for a single fluid
\begin{equation}
aH = \dot{a} \sim a^{-\frac{1}{2}(1+3\omega)} \quad \rightarrow \quad a\sim t^{2/(3+3\omega)}.
\label{Equation:Hversusa}
\end{equation}
First we had an epoch of radiation domination $a\sim t^{1/2}$, then matter domination $a\sim t^{2/3}$ and recently we entered an epoch of dark energy domination $a\sim e^{Ht}$.

\subsection{The Big Bang theory and observations}
The key idea of the Big Bang model is that the universe is expanding which is in agreement with the observed redshift of galaxies. Hubble was the first (1929) to find that galaxies are receding from us at a speed proportional to the distance, which is known as the Hubble law. To understand the evolution of the universe from an observational point of view one needs to determine both the redshift and the comoving distance of the observed objects. This means additional information about distant objects is needed. It is believed that there are standard candles which are objects of known absolute luminosity (for example Cepheid variables and type IA supernovaes) and standard rulers which are objects of known physical size. The relation between the observed flux and the absolute luminosity of a star depends on both the coordinate distance and redshift. The more the universe has expanded and the further away the source the fainter the observed flux becomes. Similarly the ratio of the physical size of an object and its measured angular size yields the coordinate distance at the moment of emission. Combining the two types of observations one can find the coordinate distance as function of redshift. Any matter content of the universe will predict a certain relation between redshift and comoving distance which can be derived using equations (\ref{Equation:redshift}) and (\ref{Equation:Friedmann and matter}). The measurements today are good enough to distinguish between different cosmological models and they strongly disfavor a flat matter-only universe. The best fit from supernovae only (SDSS, SNLS, HST and low-z) is a flat universe with $30\%$ matter \cite{Betoule:2014frx}. With the current precision measurements of the CMB the cosmological parameters have been determined to even greater accuracy. 
These observations favor a flat universe $\kappa=0$ containing $5\%$ baryonic matter, $26\%$ dark matter, $69\%$ dark energy and a tiny fraction of about $0.01 \%$ radiation today \cite{Ade:2013zuv}. The determination of the cosmological parameters using the measurements of the CMB is explained at the end of this chapter.\\

If we go back in time the universe is shrinking. This means that the universe was hotter and denser in the past and the early universe was a thermal soup of particles. Therefore the Big Bang model can make predictions of non-equilibrium processes in the thermal history of the universe using the known laws of particle physics. Two big successes of the Big Bang theory are the existence of the cosmic microwave background radiation and the prediction of the relative abundances of light elements produced during nucleosynthesis. At temperatures above $T\sim 1 eV$ the universe was a plasma of photons and electrons and nuclei. The electrons coupled the photons tightly to the baryons through Thomson and Coulomb scattering. When the universe cooled down to $T\sim 0.26-0.33 eV$ neutral hydrogen was formed such that there were much fewer electrons for the photons to scatter with. Therefore at $T\sim0.23-0.28 eV$ when the universe was 380 000 years old the photons decoupled from the matter and they could travel freely. This is what we observe as the CMB as first discovered by Penzias and Wilson \cite{Penzias:1965wn}. The observed spectrum of the CMB is an almost perfect blackbody with a temperature $T=2.72548 \pm 0.00057 K$ \cite{Fixsen:2009ug}. Going even further back in time the lightest elements could only form when the universe cooled below the relevant binding energies at about $T\sim 100 keV$ when the universe was three minutes old. Using the initial conditions of the universe and the relevant cross-sections it is possible to calculate the abundances of these elements. It turns out that most of the predictions of the Big Bang Nucleosynthesis (BBN) are in reasonable agreement with the current estimates \cite{Hinshaw:2012aka, Cyburt:2003fe}. The predictions depend on the baryon density (proton, electron and neutron density at that time) and therefore BBN puts constraints on the baryon density. Precision observations of the cosmic microwave background yield an independent measurement of the baryon density.\\

The universe is obviously not perfectly homogeneous otherwise there would have been no structure in the universe like galaxies, galaxy clusters or filaments of galaxies. Even on the largest scales of $100 Mpc$ or bigger, homogeneity is just an approximation and there are actually very tiny inhomogeneities as shown by deep redshift surveys \cite{Abazajian:2008wr} and the temperature fluctuations in the CMB \cite{Ade:2013zuv}. The fact that on smaller scales the structure becomes non-linear fits perfectly well with our picture of gravitational instability, inhomogeneities grow due to the attractive nature of gravity. The predicted distribution of galaxies and their evolution within the Big Bang model agree very well with observations \cite{Bertschinger:2001is} and therefore this is considered as another major success of the Big Bang theory. 

\subsection{The initial value problems of the Big Bang theory}
The perfect blackbody spectrum of the CMB indicates that the early universe was in local thermal equilibrium and moreover it turns out that the CMB temperature is almost perfectly isotropic with anisotropies of order $10^{-5}$. Within the Big Bang model this is surprising since as we will see in a moment one would expect many causally disconnected regions of the CMB sky so why do they appear to be in thermal equilibrium? Moreover, where do the primordial inhomogeneities come from? In this subsection we will further elaborate on the initial homogeneity problem by explaining the so-called horizon problem and the flatness problem. In the next section we will see how inflation can resolve all the problems concerning initial conditions at the same time.\\

In the Big Bang theory the universe has a finite age and therefore a special feature of the spacetime of the universe is the existence of a causal horizon which represents the maximum distance of objects whose signals could have reached us within the lifetime of the universe. Moreover because of the expansion of the universe we can also define an event horizon which represents the maximum distance light can travel if it is emitted today. In order to understand the horizon problem we will therefore introduce the comoving particle horizon and the comoving Hubble radius. To study the propagation of light it is convenient to redefine the radial coordinate $d\chi=dr/\sqrt{1-\kappa r^2}$ such that we can rewrite the line element (\ref{Equation:FRWmetric}) as
$$
ds^2=-dt^2+a^2(t)\left[d\chi^2+f(\chi)d\Omega^2\right],
$$
where the precise form of $f(\chi)$ does not matter for the moment because we will only consider radial null geodesics. The \textit{comoving particle horizon} $\chi_p$ is defined as the comoving distance a freely traveling photon traverses between time 0 and time $t$:
\begin{equation}
\chi_p = \int_0^t\frac{dt'}{a(t')}=\int_0^a\frac{1}{aH}d\ln a.
\label{Equation:particlehorizon}
\end{equation}
In standard Big Bang cosmology the universe has up to now been dominated by radiation and matter, therefore a good estimate of the comoving particle horizon is given by $\chi_p  \sim a$ in the radiation era and $\chi_p  \sim a^{1/2}$ in the matter era. Performing the last integral for a fluid with $\omega > -1/3$ one can derive
$$\chi_p=\frac{2}{1+3\omega}\frac{1}{aH}.$$
This means that in all epochs of the standard Big Bang theory the particle horizon coincides more or less with the \textit{comoving Hubble radius} $\frac{1}{aH}$. Therefore one would expect that scales beyond, say, twice the Hubble radius are not in direct causal contact. The particle horizon at the time of the CMB is much smaller than the full CMB sky we observe today. One can compute the number of causally disconnected regions of the CMB sky \cite{Prokopec} which turns out to be about $3\times 10^3$ patches. On the other hand we see the regions in the CMB sky are in almost perfect thermal equilibrium. This is very unnatural and called the \textit{horizon problem}.\\

Moreover we know from the CMB measurements that the universe is very close to flat today $\left|\kappa/a^2H_0^2\right|\leq 0.01$. Defining the critical density corresponding to a flat universe at any time $\rho_c(t)\equiv3M_P^2H^2$ we can parameterize the flatness of the universe by $\Omega_\kappa=\frac{\rho_{tot}-\rho_c}{\rho_c}=-\frac{\kappa}{a^2H^2}$. This means the universe must have been very flat in the far past. If there had been a tiny deviation from flatness initially the universe would have either collapsed or become diluted too quickly. This second part of the very unnatural initial homogeneous conditions constitutes the \textit{flatness problem}.

\section{Inflation}
We have seen in the previous section that the Big Bang theory can successfully account for many observations in our universe. However, within the Big Bang theory it cannot be explained why the universe was so homogeneous at the time of recombination and there is no explanation of where the primordial inhomogeneities come from. In this section we explain how the paradigm of inflation can solve both problems at the same time\footnote{This section is largely based on \cite{Baumann:2009ds, Mukhanov:2005sc, Baumann:2014nda}.}. Furthermore, we explain the mechanism of inflation in its simplest realization and see how its generic predictions are established by current observations. As a historical note, an early stage of inflation was proposed by Starobinsky \cite{Starobinsky:1979ty} and Guth \cite{Guth:1980zm}. Guth explained that it could solve the horizon and flatness problem, however the inflationary models of Guth were problematic in the sense that no radiation was produced after inflation. Shortly afterwards slow-roll inflation was proposed by Linde \cite{Linde:1981mu} and independently by Albrecht and Steinhardt \cite{Albrecht:1982wi}.\\

\subsection{Inflation as a solution to the horizon and flatness problems}
Instead of assuming extremely fine-tuned initial conditions we actually want a mechanism which makes sure that all the regions in the CMB are causally connected such that causal physics can homogenize and flatten the universe. From equation (\ref{Equation:particlehorizon}) we can deduce we need an epoch of decreasing comoving Hubble radius in the very early universe
$$
\frac{d}{dt}\left(\frac{1}{aH}\right)<0.
$$
This means the particle horizon can become arbitrarily large depending on how long this epoch lasts. Therefore the Hubble radius and the particle horizon are not equivalent anymore. The particle horizon depends on the full history of the universe and regions separated by a distance larger than the particle horizon could never have communicated. The Hubble radius does not know about the history of the universe which means that regions separated by a distance larger than the Hubble radius could have communicated in the past. However regions beyond the Hubble radius cannot easily communicate now, because the Hubble radius is approximately the distance a photon can travel within one expansion time. This can be seen if you compute this distance for a fluid with equation of state $\omega$ using equation (\ref{Equation:Hversusa})
$$
\int_a^{2a}\frac{1}{aH}d\ln a = \frac{2}{1+3\omega}\left[2^{\frac{1}{2}(1+3\omega)}-1\right]\frac{1}{aH},
$$
which is indeed equal to the Hubble radius up to a factor close to 1 for the fluids matter, radiation and dark energy. In the case of an decreasing Hubble radius the regions cannot communicate at all because the Hubble radius provides a bound on the distance a photon can travel
$$
\int_a^{\infty}\frac{1}{aH}d\ln a = \frac{2}{-1-3\omega}\frac{1}{aH}.
$$
The Hubble radius is in this case proportional to the \textit{event horizon}.\\

An early epoch of decreasing Hubble sphere is called \textit{inflation}. Alternatively one can derive equivalent definitions for inflation from the Friedmann equations (\ref{Equation:Friedmann1}, \ref{Equation:Friedmann2}) and equation (\ref{Equation:Hversusa}). One could define inflation as a period of accelerated expansion,
$$
\ddot{a}>0,
$$
or a period where a fluid with negative pressure dominates
$$
\rho + 3P < 0 \quad \rightarrow \quad \omega<-\frac{1}{3},
$$
or finally a period of slowly varying Hubble parameter
$$
-\frac{\dot{H}}{H^2}=-\frac{d\ln H}{dN} < 1.
$$
This condition says that the relative change of the Hubble parameter during one expansion time is small. We defined $N$ as the number of expansion times of the universe (e-folds) such that $dN\equiv d\ln a=H dt$. The limit of $H$ constant is called de Sitter inflation, which results in exponential growth of the scale factor $a\sim e^{Ht}$. Since inflation has to end we need to deviate from perfect de Sitter, instead we would like to have \textit{quasi-de Sitter} inflation, and we can introduce the parameter $\epsilon$ to parameterize this deviation
$$
\epsilon \equiv -\frac{\dot{H}}{H^2},
$$
such that for successful inflation we need $0<\epsilon <1$. Since inflation is supposed to solve the horizon and flatness problems, inflation should last long enough. Therefore we can introduce another kinematical parameter $\eta_H$ which represents the relative change of the $\epsilon$ parameter during one e-fold
$$
\eta_H\equiv-\frac{d\ln \epsilon}{2dN}=-\frac{\dot{\epsilon}}{2H \epsilon}.
$$
If this parameter is small then it says that $\epsilon$ does not vary much during one expansion time and it ensures that the period of inflation is elongated. We can compute how much inflation we need in order to solve the horizon and flatness problem. The Hubble radius at the beginning of inflation should be at least as big as the Hubble radius today. We assume the Hubble parameter is more or less constant during inflation $H_I \sim 10^{13} GeV$ (see the next section) and after inflation there is only an epoch of radiation domination. Using equation (\ref{Equation:Hversusa}) this yields the following ratio of the Hubble radius now and at the end of inflation
$$
\frac{a_0 H_0}{a_e H_e} \sim \frac{a_e}{a_0} \sim \frac{\sqrt{H_0}}{\sqrt{H_I}} \sim 10^{-27},
$$
where we used the value for the Hubble radius today $H_0 \sim 10^{-33} eV$. Therefore we can deduce the minimal ratio of the scale factor at the end of inflation to the scale factor at the beginning of inflation
$$
\frac{a_e}{a_i} = \frac{a_e}{a_0}\frac{a_0}{a_i}=10^{-27.5}\frac{a_0}{a_i}\geq 10^{-27}\frac{H_I}{H_0} \sim 10^{27}.
$$
We need therefore $N\sim 27\ln(10) \sim 62$ to solve the horizon and flatness problems.

\subsection{How to realize inflation}
Although a universe filled with a fluid with negative pressure is hard to imagine, it's not hard to realize inflation theoretically. The simplest models consist of a scalar field $\phi$, called the \textit{inflaton}, minimally coupled to gravity:
\begin{equation}
S=\int d^4 x\sqrt{-g}\left[\frac{1}{2}R+\frac{1}{2}g^{\mu\nu}\partial_{\mu}\phi\partial_{\nu}\phi - V(\phi) \right].
\label{Equation:Actionsinglefield}
\end{equation}
We assume we will have sufficient inflation such that we can safely take the FRW metric for $g^{\mu\nu}$ such that the inflaton field is homogeneous $\phi(t,\textbf{x})\equiv \phi(t)$. The Friedmann equations yield the following equations
\begin{align*}
&H^2=\frac{1}{3M_P^2}\left(\frac{1}{2}\dot{\phi}^2+V(\phi)\right),\\
&\dot{H}=-\frac{\dot{\phi}^2}{2M_P^2},
\end{align*}
which are consistent with the equation of motion for the scalar field
$$
\ddot{\phi}+3H\dot{\phi}+\frac{dV}{d\phi}=0.
$$
Let us study these equations in more detail. If we would like to have a period of slowly varying Hubble parameter we need the potential energy $V(\phi)$ to dominate over the kinetic energy $\frac{1}{2}\dot{\phi}^2$. From the equation of motion we see that a Hubble parameter acts as a friction term and the field derivative of the potential as a force. This means that in order to keep the kinetic energy small the acceleration of the field has to be small or equivalently the potential should be flat enough. Because the inflaton rolls slowly down the potential, this type of inflation is therefore called \textit{slow-roll inflation}. We can define slow-roll parameters which keep track of whether this situation is satisfied or not
$$
\epsilon\equiv -\frac{\dot{H}}{H^2}, \quad \quad \eta\equiv-\frac{\ddot{\phi}}{H\dot{\phi}}.
$$
The slow-roll approximation consists now of taking $\epsilon, |\eta| \ll 1$. The second condition implies that the friction term dominates in the equation of motion of the scalar field
$$
3H\dot{\phi}+\frac{dV}{d\phi}\approx 0.
$$
If this approximation is valid then this implies similar smallness conditions on the gradient and Hessian of the potential and one could equivalently define potential slow-roll parameters. If the gradient and Hessian of the potential are small compared to the potential energy density itself, then this implies slow-roll. This situation is however not generalizable to any inflationary model, in particular multiple field inflation which we will study in the main part of this thesis. Therefore we do not discuss them in order to avoid confusion. Comparing the slow-roll parameters with the kinematical parameters $\epsilon$ and $\eta_H$ introduced in the previous subsection we see that if the slow-roll approximation is valid, then we are in a quasi-de Sitter period, as required for successful inflation. During inflation most of the energy density is in the potential energy density. When the inflaton reaches the minimum of the potential most of the energy is transferred to the kinetic energy of the inflaton such that inflation ends. After this the inflaton should decay in all the particles of the Standard Model. This process is called reheating. After thermalization of the particles the standard Hot Big Bang era begins.

\subsection{An example: quadratic potential}
Because there are many possible models for inflation one might be worried that it cannot be falsified. However all the different scenarios for inflation lead to very similar predictions which are confirmed by the data as we will explain later on. On the other hand each particular model may differ in the more subtle details of the predictions which allows one to learn more about the underlying physics of inflation by the current and future cosmological precision measurements. For now we do not worry about what the inflaton represents, but first we try to get more understanding of the predictions of inflation by use of a particular example of slow-roll inflation: chaotic inflation. We will first introduce chaotic inflation in this subsection and in the next section we will use the same model to study more advanced predictions of inflation. Chaotic inflation is a subset of single field models of the form (\ref{Equation:Actionsinglefield}) with a power-law potential which allows for slow-roll inflation. We will take the quadratic potential as concrete example, i.e.
$$
V(\phi)=\frac{1}{2}m^2\phi^2.
$$
Assuming we already entered the stage of slow-roll, the field equation and Friedmann equation become
$$
\dot{\phi}\approx -\frac{m^2\phi}{3H} \quad \text{and} \quad H^2\approx\frac{m^2\phi^2}{6M_p^2}
$$
which leads to the following expression for the slow-roll parameters
\begin{equation}
\epsilon \approx 2\frac{M_P^2}{\phi^2} \quad \text{and} \quad \eta_H \approx -2\frac{M_P^2}{\phi^2}.
\label{Equation:slowrollparameters_chaoticinflation}
\end{equation}
Here we derived $\eta_H$ using the approximate expression for $\epsilon$. We can estimate how many e-folds of inflation awaits us if we start at a particular value for the field $\phi_i$ making the assumption we have slow-roll until the end of inflation. Then the end of inflation is given by the field value
$$
\epsilon = 1 \rightarrow \phi_e = \sqrt{2} M_P.
$$
Now the number of e-folds until the end of inflation can be computed as follows
\begin{equation}
N_e-N_i = \int_{a_i}^{a_e} \ d\ln(a) = \int_{t_i}^{t_e}  \ H dt = \int_{\phi_i}^{\phi_e}  \ \frac{H}{\dot{\phi}}d\phi \approx -\int_{\phi_i}^{\phi_e} \ \frac{\phi}{2 M_P^2}d\phi = \frac{\phi_i^2}{4M_P^2}-\frac{1}{2}.
\label{Equation:numberofefolds_chaoticinflation}
\end{equation}
This expression will be useful when we study the generation of initial density perturbations in the next chapter.

\subsection{Inflation as the origin of structure}
\label{Section:inflation as the origin of structure}
The Big Bang theory can explain the distribution of galaxies and their evolution assuming a certain distribution of initial inhomogeneities which grow due to the attractive nature of gravity. However, the origin of the primordial inhomogeneities remains elusive and this introduced another problem concerning initial conditions. We have seen so far that inflation can make the universe very homogeneous and isotropic. On the other hand tiny quantum fluctuations in the inflaton field are unavoidable and it was soon realized that these fluctuations can account for all the structure in the universe. The physical picture is that quantum fluctuations in the inflaton field are quickly stretched to large scales by the extreme expansion of the universe. These quantum fluctuations are frozen as soon as they reach scales beyond the Hubble radius\footnote{This will be called `Hubble radius crossing' from now on.} where they become the initial density fluctuations of the universe to seed all structure in the universe. Inflation therefore not only solves the horizon and flatness problem, but it also provides a mechanism to produce the initial density perturbations, which explains the structure we see around us, which is extremely exciting. Starobinsky already noted in his early paper \cite{Starobinsky:1979ty} that quantum effects during this inflationary epoch would be important and calculated a spectrum of gravitational waves. Mukhanov and Chibisov calculated the spectrum of density fluctuations for the first time \cite{Mukhanov:1981xt} but it was done independently by several groups at a workshop in Cambridge \cite{Hawking:1982cz, Starobinsky:1982ee, Guth:1982ec, Bardeen:1983qw}. Observations of the CMB confirm these predictions to high accuracy \cite{Smoot:1992td, Hinshaw:2012aka, Ade:2013uln} which provides compelling evidence in favor of inflation. In this section we will outline the calculation of the lowest order statistics of both the density perturbations and the gravitational waves at the end of inflation.

\subsubsection{Cosmological perturbation theory}
We study the behavior of arbitrary quantum fluctuations with respect to the classical spacetime background and fields. Perturbations in the matter fields are related to perturbations in the metric by the perturbed Einstein equations which can be very complicated to keep track of, due to the non-linear nature of the equations. Moreover the physical interpretation of these perturbations is a bit obscured by the freedom in the choice of coordinates, there is no preferred coordinate system. Perturbations can pop up if you pick different coordinates, which means that these are actually not physical. The way to resolve this is to introduce \textit{gauge-invariant} variables, i.e. physical variables which do not depend on the chosen coordinate system. The perturbations are assumed to be small and for our purposes it is sufficient to expand all equations to linear order in the perturbations and therefore we can use standard cosmological perturbation theory\footnote{This subsubsection is based on \cite{BaumannB}}. The treatment of the perturbations is a lot easier at linear order for two reasons. First of all the Fourier modes of the perturbations evolve independently which can be derived using the the spatial translation symmetry of the background. Second because of the spatial rotation symmetry of the background one can derive that the modes with different helicity evolve independently at linear order as well. We will explain this last statement by applying it directly to single field inflation. The most general way to parameterize linear perturbations in this theory is given by
\begin{align*}
ds^2&=(1+2\Phi)dt^2-2a(t)B_idx^idt-a^2(t)[(1-2\Psi)\delta_{ij}+2E_{ij}]dx^i dx^j,\\
\phi(t,\v{x})&=\phi(t)+\delta\phi(t,\v{x}),
\end{align*}
where $E_{ij}$ is symmetric and traceless. The helicity of a perturbation mode is defined by the way the amplitude changes under rotations around its Fourier vector $\v{k}$. The amplitude of a mode of helicity $m$ gets multiplied by $e^{im\theta}$ where $\theta$ is the rotation angle. The helicity of different 3-tensor components can be most easily derived in the helicity basis $\v{e}_3 \sim \v{k}$ and $\v{e}_\pm\equiv (\v{e}_1\pm i\v{e}_2)/\sqrt{2}$, which are defined to transform as $e^{\pm i\theta}\v{e}_\pm$ under rotation around $\v{k}$. Now given an arbitrary component of a contravariant 3-tensor $T$ we have
$$
\tensor{T}{_{j_1 \ldots j_m}^\prime}=e^{i\theta(n_+-n_-)}\tensor{T}{_{j_1 \ldots j_m}},
$$
where $n_+$ and $n_-$ are the number of plus and minus indices in $j_1, \ldots j_m$. This means the perturbations can be divided into scalar, vector and tensor perturbations, which correspond respectively to modes with helicity $0$, $\pm1$ and $\pm2$. The scalars $\Phi$, $\Psi$ and $\delta\phi$ are zero-rank tensors and therefore scalar perturbations. The 3-vector $B_i$ can be subdivided in two vector perturbations and one scalar perturbation. The scalar perturbation is parallel to $\v{k}$ and the vector perturbations are orthogonal to $\v{k}$ which means that in real space we can decompose $B_i$ as follows
$$
B_i = \partial_i B + B_i^V, \quad \text{with} \quad \partial^iB_i^V=0.
$$
Using the same reasoning we can decompose the symmetric and traceless 3-tensor $E_{ij}$ into scalar, vector and tensor perturbations. For the scalar perturbation we find for example
$$
E_{ij}^S = c_1 k^2\delta_{ij} + c_2 k_i k_j = \left(\frac{1}{3}k^2\delta_{ij}-k_i k_j\right)c \quad \rightarrow \quad E_{ij}^S= \left(\frac{1}{3}\delta_{ij}\partial^2-\partial_i \partial_j\right)E,
$$
where we used the traceless property to eliminate one of the constants. Furthermore we have
\begin{align*}
E_{ij}^V &= \partial_i E_j+\partial_j E_i \quad \text{with} \quad \partial^iE_i=0,\\
\partial^i E_{ij}^T &= 0.
\end{align*}
The decomposition of the perturbations in scalar, vector and tensor types reduces the complexity of deriving the perturbation equations. The scalar perturbations are captured by
\begin{align*}
&ds^2=-(1+2\Phi)dt^2-2a(t)\partial_i Bdx^idt+a^2(t)[(1-2\Psi)\delta_{ij}+2\partial_i\partial_jE]dx^i dx^j,\\
&\phi(t,\v{x})=\phi(t)+\delta\phi(t,\v{x}).
\end{align*}
Here we absorbed a part of $E_{ij}^S$ in $\Psi \delta_{ij}$. Note that there are less true degrees of freedom because we still need to incorporate the gauge freedom and the Einstein equations. The scalar gauge transformations
$$
t\rightarrow t+\xi^0, \quad x^i\rightarrow x^i+\partial^i \xi
$$
remove two degrees of freedom. The Einstein energy and momentum constraint equations remove in addition two more degrees of freedom. This means we are left with one scalar degree of freedom. For the actual computation we would like to fix a particular gauge, therefore we should take care that we are only considering physical quantities. It turns out to be useful to work with the following gauge invariant variable
$$
q\equiv a\delta\phi+\frac{a\dot{\phi}}{H}\Psi,
$$
the so-called Mukhanov-Sasaki variable which will be the canonically normalized quantization variable. Another very useful variable is the curvature perturbation on uniform density hypersurfaces
$$\R\equiv\Psi+\frac{H}{\dot{\phi}}\delta\phi,$$
which has the important property to freeze out on super Hubble scales. It becomes constant at a certain time during inflation and it only starts evolving again in an epoch of known physics much later than the epoch of inflation. This means that the relevant modes of the curvature perturbation are insensitive to the unknown physics after inflation and and therefore this variable is used to compute observables. In order to simplify the perturbed Einstein equations one can fix a gauge and rewrite everything in terms of the gauge-invariant variables. Some popular gauges are the Newtonian gauge ($B=E=0$), the uniform density gauge ($\partial T^0_0=E=0$) such that $\R=\Psi$ which explains its name and the comoving gauge ($\partial T^0_i=E=0$) such that $\R=\Psi$ which explains the fact that $\R$ is also called the comoving curvature perturbation. \\

Solving the Einstein equations and the equations of motion of the perturbed field leads to the linearized equations of motion for the gauge-invariant perturbations. This computation is a bit tedious and since we just gave all the ingredients to execute it we just state the result.  After performing a Fourier transform we get
\begin{equation}
q^{\prime\prime} +\left(k^2-\frac{z^{\prime\prime}}{z} \right )q = 0,
\label{Equation:perturbation_q single field}
\end{equation}
with $z\equiv \sqrt{2a^2\epsilon}$. Here the prime denotes a derivative with respect to conformal time $\tau$ defined by $a d\tau=dt$. Before studying this equation in more detail we should not forget about the vector and tensor perturbations. These are fortunately much easier to deal with. First of all vector perturbations decay with the expansion of the universe, therefore even if they were present we still can neglect them. The tensor perturbations are described by
$$
ds^2=-dt^2+a^2(t)[\delta_{ij}+2h_{ij}]dx^idx^j,
$$
with $\partial_i h_{ij}=h^i_i=0$. It is possible to decompose $h_{ij}$ in the eigenmodes of the spatial Laplacian $\nabla^2$ as $h_{ij}=\sum_{+,\times} h(t)e_{ij}^{+,\times}(x)$ such that after performing a Fourier transformation the only equation defining the tensor perturbations is
$$
\ddot{h}+3H\dot{h}+\frac{k^2}{a^2}h=0.
$$
Note we are studying Fourier modes of the perturbations with comoving wavevector $k$. Defining the variable $v\equiv ah$ we can rewrite this equation as
\begin{equation}
v^{\prime\prime}+\left(k^2-\frac{a^{\prime\prime}}{a}\right)v=0.
\label{Equation:perturbation_h single field}
\end{equation}
This equation is almost equivalent to (\ref{Equation:perturbation_q single field}) and if we can solve the equations for the scalar perturbations then we immediately have the solution of the tensor perturbations. Therefore we will only focus on the scalar perturbation equations and come back to the gravitational waves in the very end. Expressing the coefficients of equation (\ref{Equation:perturbation_q single field}) in terms of the slow-roll parameters we get
$$
q^{\prime\prime} +\left(k^2-a^2H^2\left(2+2\epsilon-3\eta+\eta\xi-4\epsilon\eta+2\epsilon^2\right) \right )q = 0.
$$
From here we can see that we can divide the evolution of the perturbations roughly in two regimes. During inflation the Hubble radius $1/aH$ will decrease quasi-exponentially in time. This means that a perturbation of sub-Hubble scale $k\gg aH$ will eventually grow to a perturbation of super-Hubble scale $k\ll aH$ if there is enough inflation. At sub-Hubble scales the coefficient $k^2$ dominates over $a^2H^2$ and by neglecting the $a^2H^2$-term, we can solve the equation exactly. Within the slow-roll approximation it is possible to find an analytical solution of the full perturbation equation. First of all we can derive
$$
\H^\prime=(1-\epsilon)\H \quad \rightarrow \quad \H\approx -\frac{1+\epsilon}{\tau},
$$
such that we get the following equation
$$
q^{\prime\prime} +\left(k^2-\frac{\beta^2-\frac{1}{4}}{\tau^2} \right )q = 0,
$$
where
$$
\beta^2 \approx \frac{9}{4}+3\epsilon.
$$
This equation is equivalent to the Bessel equation and the solution is given by a linear combination of the Hankel functions of the first and second kind
\begin{equation}
q(\tau)= \sqrt{-k \tau}\left(A H^{(1)}_\beta(-k\tau)+B H^{(2)}_\beta(-k\tau)\right).
\label{Equation:slowrollsolutionq}
\end{equation}
The initial conditions of the perturbations are given by the underlying physical idea that they arise as quantum fluctuations.

\subsubsection{Quantizing the perturbations}
To quantize the theory we first need to know the action to quadratic order in perturbations. The simplest way to derive this is to go back to the original action, perturb the inflaton field and ignore the spacetime fluctuations for the moment. We take $\phi(t,\v{x})=\phi_0(t)+\delta\phi(t,\v{x})$ such that the relevant part of the action (\ref{Equation:Actionsinglefield}) becomes
$$
S=\int\ d\tau d^3 x \left[\frac{1}{2}a^2\left((\phi_0^\prime+\delta\phi^\prime)^2-(\partial_i\phi_0+\partial_i\delta\phi)^2\right)-a^4\left(V(\phi_0)+ V_\phi(\phi_0)\delta\phi+\frac{1}{2}V_{\phi\phi}(\phi_0)\delta\phi^2+\ldots\right) \right].
$$
The action to first order in the perturbations vanishes because of the background equations of motion. The action to second order in the perturbations becomes
$$
S^{(2)}=\int\ d\tau d^3 x\left[\frac{1}{2}a^2\left((\delta\phi^\prime)^2-(\partial_i\delta\phi)^2\right)-a^4\frac{1}{2}V_{\phi\phi}(\phi_0)\delta\phi^2 \right].
$$
Changing variables to $\tilde{q}\equiv a\delta\phi$ this becomes after some manipulations
$$
S^{(2)}=\frac{1}{2} \int\ d\tau d^3 x\left[(\tilde{q}^\prime)^2-(\partial_i \tilde{q})^2+\left(\H^\prime+\H^2-a^2V_{\phi\phi}(\phi_0)\right)\tilde{q}^2 \right].
$$
Now from this action follow the same equations of motion as for the Mukhanov-Sasaki variable $q$ as given in equation (\ref{Equation:perturbation_q single field}) up to some slow-roll corrections. If you would have picked the spatially flat gauge then $q=\tilde{q}$ which means this is very close to the quadratic action in perturbations. The correct normalization follows from here and therefore the full action to quadratic order in perturbations becomes
$$
S^{(2)}=\frac{1}{2} \int\ d\tau d^3 x\left[(q^\prime)^2-(\partial_i q)^2+\frac{z^{\prime\prime}}{z}q^2 \right].
$$
Note that we have canonical kinetic terms, which means that the Mukhanov-Sasaki variable is the one we should use for canonical quantization of the perturbations. \\

We know how to quantize the canonical coordinate and momentum fields $q$ and $\pi=q^\prime$. We promote them to operators
$$
q \rightarrow \hat{q}, \quad \text{and} \quad \pi \rightarrow \hat{\pi},
$$
and impose the standard equal-time commutation relations.
\begin{equation}
[\hat{{q}}(\tau, \v{x}), \hat{{\pi}}(\tau, \v{y})] =i\delta(\v{x}-\v{y})
 \label{Equation:commutationrelations_q_singlefield}
\end{equation}
Expanding $q$ in terms of its spatial Fourier modes $q_\v{k}$;
$$
q(\tau,\v{x})= \int\frac{d^3\v{k}}{(2 \pi)^{3/2}}q_\v{k}(\tau)e^{i\v{k}\cdot\v{x}},
$$
the equations of motion are given by equation (\ref{Equation:perturbation_q single field}) which we repeat here while keeping the index $\v{k}$ for clarity
\begin{equation*}
q^{\prime\prime}_{\v{k}}+\left(k^2-\frac{z^{\prime\prime}}{z}\right)q_{\v{k}}=0
\end{equation*}
The solutions of the equations of motion for $q_\v{k}$ are given by complex mode functions $q_k$ (and $\pi_k=q_k^\prime$) such that the general solution of $q_\v{k}$ is determined by a constant complex vector $a_\v{k}$:
$$
q_\v{k}=q_k a^\ast_\v{k}+q^{\ast}_k a_{-\v{k}}
$$
and the solution for $\pi_\v{k}$ becomes
$$
\pi_\v{k}=\pi_k a^\ast_\v{k}+\pi^{ \ast}_k a_{-\v{k}}.
$$
Note that the reality condition $q^{\ast}_\v{k}(\tau)=q_\v{-k}(\tau)$ is manifestly satisfied. After promoting the vector $a_\v{k}$ and its complex conjugate to the operators $a_\v{k}$ and $\hat{a}^\dag_{\v{k}}$, we would like to interpret them as annihilation and creation operators respectively, which satisfy the usual commutation relations
\begin{equation}
\begin{aligned}
&[\hat{a}_{\v{k}},\hat{a}^\dag_{\v{q}}]=\delta(\v{k}-\v{q}).
\end{aligned}
 \label{Equation:commutationrelations_a_singlefield}
\end{equation}
These are consistent with the commutation relations (\ref{Equation:commutationrelations_q_singlefield}) if and only if the mode functions are normalized as follows:
\begin{equation}
q^{ \ast}_k \pi_k- \pi^{\ast}_k q_k = i ,\\
\label{Equation:wronskian_singlefield}
\end{equation}
Another constraint on the mode functions comes from imposing initial conditions. We assume that at the beginning of inflation the initial state of the universe is the vacuum state $\ket{0}$, defined by $\hat{a}_k\ket{0}=0$ for all scales $k$ of interest, such that there is no initial particle production, i.e. the Bunch-Davies vacuum. This statement implies that at the beginning of inflation when $k$ dominates all other scales in the equations of motion the universe is in the state of minimal energy and the Hamiltonian does not contain any terms proportional to $\hat{a}\hat{a}$ and  $\hat{a}^\dag\hat{a}^\dag$. After some algebra the energy of the ground state is given by
$$
\bra{0}\hat{H}\ket{0}=\frac{1}{2}\int d^3 \v{k}\delta^3(0)\left[\pi_k \pi^{\ast}_k+k^2q_k q^{\ast}_k - \frac{z^{\prime\prime}}{z}q_k q^{\ast}_k\right].
$$
Therefore at the beginning of inflation $\tau_0$, which is assumed to be deep inside the sub-Hubble regime where we can assume $k\ggg \frac{z^{\prime\prime}}{z}$, we have in addition to (\ref{Equation:wronskian_singlefield}) the following constraint:
\begin{equation}
|\pi_k|^2+k^2|q_k|^2 \quad \text{is minimized}
 \label{Equation:bunchdavies_singlefield}
\end{equation}
Writing the sub-Hubble solution to the equations of motion for $q_k$ as
$$
q_k=a e^{ik\tau} + b e^{-ik\tau}.
$$
we can derive from the constraints (\ref{Equation:wronskian_singlefield}, \ref{Equation:bunchdavies_singlefield}) that the initial solution is given by\footnote{Strictly speaking, the Bunch-Davies initial conditions can only be imposed when inflation is eternal, such that one can take the limit $\tau\rightarrow -\infty$.}
\begin{equation}
q_k (\tau)= \sqrt{\frac{1}{2k}} e^{ik\tau +i\lambda},
\end{equation}
for some random phase $\lambda$. This completely determines the evolution of $q_\v{k}$. Completing now the analysis in case of slow-roll inflation as given by (\ref{Equation:slowrollsolutionq}) we find the following early time limit
$$
\lim_{-k\tau\rightarrow \infty} q(\tau)=\sqrt{\frac{2}{\pi}}\left(A e^{-ik\tau-i\frac{\pi}{2}(\beta+\frac{1}{2})}+B e^{ik\tau+i\frac{\pi}{2}(\beta+\frac{1}{2})} \right).
$$
Therefore we have $A=0$ and $B=\sqrt{\frac{\pi}{4 k}} e^{i\lambda-i\frac{\pi}{2}(\beta+\frac{1}{2})}$. After the modes have crossed the Hubble radius we find therefore the following late time solution
$$
\lim_{-k\tau\rightarrow 0} q(\tau)=B\frac{\sqrt{2}\Gamma(\beta)}{\pi}\left(\frac{2}{-k\tau}\right)^{\beta-1/2}=\frac{\Gamma(\beta)}{\sqrt{2k\pi}}\left(\frac{2}{-k\tau}\right)^{\beta-1/2}.
$$
Transforming to the curvature perturbation $\R$ we find at late times
\begin{equation}
\R=\frac{1}{\sqrt{2\epsilon}M_p a}q = \frac{H}{2\sqrt{\epsilon}M_p k^{3/2}},
\label{Equation:SolutionRsinglefield}
\end{equation}
here we used the slow-roll approximation for which we have $\beta\approx \frac{3}{2}$ and $-\frac{1}{\tau}\approx aH$. It has been proven by Weinberg \cite{Weinberg:2003sw} that the curvature perturbation $\R$ freezes out at super-Hubble scales. This expression for $\R$ should therefore be evaluated for each mode at Hubble radius crossing $k=aH$.

\subsection{How to characterize the perturbations}
\label{Section:how to characterize the perturbations}
In order to relate the observed temperature anisotropies in the CMB to the predictions of a particular theoretical model of inflation we compute the power spectrum of the density fluctuations. The power spectrum is the Fourier transform of the real-space two-point correlations and contains all the information about the distribution of density perturbations in case they are Gaussian\footnote{More precisely, all information about a Gaussian distribution of density perturbations is contained in the Gaussian density matrix, which is determined by the $\hat{q}\hat{q}$, $\hat{q}\hat{\pi}$ and $\hat{\pi}\hat{\pi}$ two-point correlations.}. For every mode $\v{k}$ the value of the power spectrum represents a variance $\sigma_k$. From the CMB we have learned that the distribution is very close to Gaussian, so it is a good first characterization of the perturbations. To learn more about the distribution of perturbations however, one should compute higher order correlations in addition. The dimensionless powerspectra $\P_{\R}$ for $\R_\v{k}$ are defined as
$$
\bra{0}\hat{\R}_{\v{k}}(\tau) \hat{\R}_{\v{k}^\prime}(\tau)\ket{0}\equiv\delta(\v{k}-\v{k}^\prime)\frac{2\pi^2}{k^3}\P_{\R}(k,\tau),
$$
such that
$$
\bra{0}\hat{\R}(\tau,\v{x}) \hat{\R}(\tau,\v{y})\ket{0} =\int\frac{d^3\v{k}}{4\pi k^3}\P_{\R}(k,\tau)e^{-i\v{k}\cdot(\v{x}-\v{y})}.
$$
In the case of slow-roll inflation we plug in equation (\ref{Equation:SolutionRsinglefield}) into this definition and we find
$$
\P_\R=\frac{H^2}{ 8\pi^2 \epsilon M_P^2}|_{k=aH}.
$$
The power spectrum is evaluated at Hubble radius crossing because of the freeze out of the curvature perturbation. Since we are in the slow-roll regime the functions $H$ and $\epsilon$ are not precisely constant but rather slowly varying in time. This means that the power spectrum has a slight scale dependence, since different modes cross the Hubble radius at different times and therefore the amplitude of the power spectrum changes slightly as well. Before we compute the scale dependence we should not forget about the spectrum of gravitational waves. The computation goes in a similar fashion as for density perturbations and we won't repeat the full analysis. The important steps are to solve equation (\ref{Equation:perturbation_h single field}), to find the canonical quantization variable which is $M_p\frac{v}{2}$. Finally, after transforming back to $h$, the power spectrum for gravitational waves is given by $\P_t=2\P_h$ because there are two polarizations. This results in
$$
\P_t=\frac{2 H^2}{\pi^2 M_P^2}|_{k=aH}.
$$
Note the difference with the power spectrum of scalar perturbations, there is no factor of $\epsilon$. This means that a measurement of the amplitude of the power spectrum of gravitational waves would be a direct probe of the Hubble scale of inflation. Furthermore the power spectrum of tensor perturbations is also scale dependent because $H$ changes in time. For slow-roll inflation the power spectra are determined by the amplitude and its tilt to a very good approximation
\begin{align*}
\P_\R &= A_s(k_\star )\left( \frac{k}{{k_\star }} \right)^{n_s - 1},\\
\P_t &= A_t(k_\star)\left( \frac{k}{{k_ \star }} \right)^{n_t}.
\end{align*}
Here $A_s$ and $A_t$ are the amplitudes of the scalar and the tensor power spectrum and $k_\star$ denotes a pivot scale at which they are evaluated. The parameters $n_s$ and $n_t$ represent the tilt of the power spectra and are called the \textit{scalar spectral index} and the \textit{tensor spectral index} respectively. The scalar spectral index can be computed as follows
\begin{align*}
n_s-1= \frac{d\ln\P_\R}{d\ln k}=\frac{d\ln\P_\R}{dN}\frac{dN}{d\ln k}\approx -2\epsilon+2\eta_H,
\end{align*}
and similarly we find the tensor spectral index
$$
n_t \approx -2\epsilon.
$$
Furthermore we can define the \textit{tensor-to-scalar ratio} as
$$
r=\frac{A_t}{A_s}=16\epsilon,
$$
which indicates the relative amount of tensor modes compared to the scalar modes. The observables $A_s$, $r$, $n_s$ and $n_t$ can be related to experiment. The observable scales cross the Hubble radius $60-53$ e-folds before the end of inflation such that the slow-roll approximation is valid. Slow-roll inflation predicts therefore two spectra with a red tilt, i.e. more power on large scales. The tensor power spectrum can only get a blue tilt if the weak energy condition is violated, $\rho+P >0$. Moreover, slow-roll inflation predicts a background of gravitational waves, although the relative amplitude depends on the precise model. In the next section we will discuss the splendid agreement of the inflationary paradigm and the current observations, but first we
will get some feeling for the values of these observables.

\subsubsection{An example: quadratic potential}
Let us compute the values of the observables in case of chaotic inflation with a quadratic potential. Since the interesting modes cross the Hubble radius between 60 and 53 e-folds before the end of inflation the value of the field can be estimated using equation (\ref{Equation:numberofefolds_chaoticinflation}). This yields
$$
\phi_{60} \approx 15.5 M_P \quad \text{and} \quad \phi_{53} \approx 14.6 M_P,
$$
plugging this into equation (\ref{Equation:slowrollparameters_chaoticinflation}) the slow-roll parameters take the approximate value
$$
\epsilon=-\eta_H \approx \frac{2}{15^2}.
$$
This yields the following estimate for the value of the observables for chaotic inflation
$$
n_s\approx 0.96, \quad \quad r\approx 0.14, \quad \quad n_t\approx 0.02 \quad \text{and} \quad A_s\approx 53\frac{m^2}{M_p^2}.
$$
The measured amplitude of the power spectrum determines the value $m$, and therefore inflation predicts the shape of the spectra and their relative amplitudes.\\

We can use the measured value of the amplitude of the power spectrum of scalar perturbations to determine the value of the Hubble parameter if we know the details of the inflationary model. The best fit for the amplitude $A_s$ as measured by Planck is $A_s=(2.196 \pm 0.06)\cdot 10^{-9}$ and this fixes the value of $m$. We can express $H$ in terms of the tensor-to-scalar ratio $r$ as follows
$$
H^2=\frac{\pi^2 r M_p^2 A_s}{2} \rightarrow H \approx 2\cdot10^{-5} \sqrt{r} M_{Pl},
$$
note that we have changed from the reduced Planck mass $M_p$ to the Planck mass $M_{Pl}$. In case of chaotic inflation therefore we get a Hubble scale of inflation of
$$
H \sim 10^{13} GeV,
$$
which is about 6 orders of magnitude below the Planck scale.

\subsubsection{The Lyth Bound}
From the concrete example of chaotic inflation with a quadratic potential we see that the inflaton has to traverse a large distance in field space $\Delta\phi \sim 15 M_p$ in order to create a fair amount of gravitational waves. More quantitatively, we can deduce a relation between the tensor-to-scalar ratio and the evolution of the inflaton field by inserting the definition of $\epsilon$ into $r=16\epsilon$
$$
\left|\frac{d\phi}{dN}\right|=M_p\sqrt{\frac{r}{8}}.
$$
Assuming we have at least 30 e-folds of slow-roll inflation from the time $N=60$ where the observable modes cross the Hubble radius we find the
Lyth bound \cite{Lyth:1996im}:
$$
\frac{\Delta\phi}{M_P}\geq\int_{30}^{60}\sqrt{\frac{r_\star}{8}}\approx\left(\frac{r_\star}{0.01}\right)^{1/2},
$$
where $r_\star$ is the measured value of the tensor-to-scalar ratio. The Lyth bound provides a lower bound of the distance the inflaton moves in field space for a given tensor-to-scalar ratio. This means that if the tensor-to-scalar ratio turns out to be large enough to be detectable $r\gtrsim 0.01$, then the inflaton will traverse a large distance in field space and therefore the inflaton will therefore explore a larger part of the potential. In particular if inflation is embedded in a multi-field inflationary model it means the inflaton is expected to be more sensitive to the other degrees of freedom. Moreover, we saw in the example of a quadratic potential that the Hubble scale will be about 6 orders of magnitude below the Planck scale if $r\gtrsim 0.01$. If these extra degrees of freedom are much heavier than the Hubble scale, then the dynamics of the inflaton might be sensitive to physics only a couple of magnitudes below the Planck scale.

\section{Inflation and observations}
In the previous section we have seen that inflation can resolve the initial homogeneity problem of the Big Bang theory. At the same time, inflation also creates tiny initial inhomogeneities which seed all structure in the universe. In this section we explain how precision observations of the CMB provide compelling evidence for inflation and moreover how we can use these observations to learn more about the early universe.

\subsection{Evidence in favor of inflation}
The standard model of cosmology, the $\Lambda$CDM model, is the simplest model which combines the Big Bang theory and inflation. It is described by only six free parameters: the density parameters of baryonic matter, dark matter and dark energy, the optical depth (which we do not discuss here), the scalar amplitude $A_s$ and the scalar spectral index $n_s$. Everything else can be derived from these parameters. This simple model agrees well with the cosmological data. The Planck collaboration \cite{Ade:2013zuv} has determined all six parameters to high accuracy by measuring the temperature anisotropies. In particular for the inflationary parameters Planck has found
$$
A_s=(2.196 \pm 0.06)\cdot 10^{-9}, \quad \text{and} \quad n_s=0.9603\pm0.0073,
$$
which are evaluated at the pivot scale $k_\star = 0.05 \SI{}{\per\mega\parsec}$. A scale-invariant power spectrum has been ruled out with almost $6\sigma$. However, the tensor-to-scalar ratio is assumed to be negligible. Gravitional waves contribute to the temperature power spectrum in a specific way and therefore including them in the analysis is expected to weaken the constraints on $n_s$. The Planck collaboration analyzed also an extension of the standard model with $r$ as an additional parameter and found $n_s=0.962\pm0.011$ with $2\sigma$ certainty, which makes the result very solid. Together with the observation of super-Hubble correlations with coherent phases this provides strong evidence in favor of inflation. We will explain these latter statements in words in the section on CMB anisotropies below.\\

Concerning primordial gravitational waves, Planck has set an upper limit on the tensor-to-scalar ratio assuming the $\Lambda$CDM model extended with the tensor-to-scalar ratio and found $r<0.12$ with $2\sigma$ certainty. Probing the value of $r$ requires a measurement of the so-called $B$-mode polarization of the CMB as we explain in the section on CMB anisotropies. Recently a detection of primordial $B$-modes has been reported by the BICEP2 collaboration \cite{Ade:2014xna} with the best-fit value of
$$
r=0.2\pm0.07,
$$
where $r=0$ is disfavored at $7\sigma$. If this is a true detection of primordial $B$-modes, this will be a revolution for the paradigm of inflation. There is some tension with the upper bound provided by Planck, but the measurement of Planck comes from the temperature power spectrum and is therefore model dependent. Moreover, it is noted by the BICEP2 team that accounting for the contribution of foreground dust might shift this value down.

\subsubsection{CMB Anisotropies}
The anisotropies of the CMB temperature and polarization can provide very accurate information about the matter content of the universe and the initial conditions laid down by inflation\footnote{This subsection is an excert of \cite{Hu:2001bc} where everything is explained in much more detail. See also \cite{Dodelson:2003ft}.}. We will explain what information on the cosmological parameters we can extract from the CMB anisotropies and make more clear why the observations are in outstanding agreement with inflation. \\

The temperature fluctuations of the CMB are described by the power spectrum which contains all information about the statistical properties if the fluctuations are Gaussian. The fluctuations are expanded in spherical harmonics. The power spectrum for a given multipole $l$ is proportional to the absolute value squared of the multipole moment of the temperature fluctuation field $C_l$. For low values of $l$ there is an absolute limitation to the accuracy of the power spectrum because there are only few samples to take the mean of. This is called cosmic variance. Furthermore when the measurement is averaged over bands $\Delta l\approx l$ this leads to another error. For $l\sim100$ we already have a very precise measurement if the other sources of noise can be suppressed. \\
Let's see what information we can extract from the temperature power spectrum. Before decoupling the universe consisted of a tightly coupled photon-baryon plasma. By analyzing the behavior of this cosmological fluid it is possible to understand the existence of the peaks and troughs and their location. These properties depend not only on the matter contents of the universe but also on the initial conditions for which a particular form is assumed, namely a nearly scale-invariant power spectrum as predicted by simplest models for inflation. The initial distribution of density perturbations sets the initial conditions for the oscillator equation of the photon-baryon fluid. The pressure gradients in the photon density will act as a restoring force and any initial perturbation will therefore oscillate with the speed of sound $c_s$. The temperature perturbation $\Theta$ at the end of recombination in the most idealized case is given by
$$
\Theta(\eta_\ast)=\Theta(0)\cos\left(ks_\ast\right), \quad \text{with} \quad s_\ast=\int_{\eta_0}^{\eta_\ast} c_sd\eta.
$$
On the largest scales $kc_s\ll1$ the perturbations are still frozen and therefore this provides a direct measure of the initial conditions. On the smaller scales the fluctuations evolve until the end of recombination, such that some of the modes are catched at their maximum and some at their minimum.
These spatial inhomogeneities appear as an angular anisotropy today and we see a series of acoustic peaks. The peak locations depend very sensitively on the curvature of the universe. The spatial curvature of the universe acts like a lens, a positive curvature makes objects look closer than they actually are. The converse holds for a negative curvature. The position of the first peak therefore strongly suggests a flat universe. This is another breakthrough for inflation since it predicts flatness of the universe. At the same time in order to get those peaks there should have been initial fluctuations with super-Hubble correlations with a nearly scale invariant spectrum and furthermore perturbations which start to evolve as soon as they enter the Hubble radius which uniquely determines the relationship between the phases of the oscillations. Inflation explains all these non-trivial observations at the same time. If you would consider an alternative theory where curvature perturbations are generated only inside the Hubble radius where the phases of the perturbations are arbitrary, the peaks would be washed out.\\

The gravitational potential also provides a gravitational force to the oscillator. Gravity compresses the fluid and pressure resists. This means we will have both pressure gradients $k\Theta$ and potential gradients $k\Psi$. In addition one should take into account the redshift of the photons by the gravitational potential. In the absence of baryons the oscillator equation remains the same as without gravity, but then for the `effective temperature perturbation' $\Theta+\Psi$. This is also the measured temperature perturbation since after recombination the photons need to climb out of the potential wells and therefore redshift in temperature. The initial temperature perturbation can be deduced from the initial gravitational potential perturbation. The initial Newtonian potential can be interpreted as a local time shift $\delta t/t=\Psi$. The temperature of the photons goes like the inverse of the scale factor. Therefore, using equation (\ref{Equation:Hversusa}), the initial temperature fluctuation is given by $\Theta=-\frac{2}{3}(1+\omega)^{-1}\Psi$ which equals $\Theta=-\Psi/2$ in the radiation era and $\Theta=-2\Psi/3$ in the matter era. This means for a potential well $\Psi < 0$ the effective temperature is negative which means the plasma begins rarefied in gravitational wells. Due to the gravitational potential the fluid will get compressed and by the resistance of the pressure it gets rarefied again. This means the first peak corresponds to the mode where the fluid exactly has compressed once at recombination. \\

When baryons are included they provide extra inertia to the oscillator. The resulting effect can be understood by understanding the oscillation equations as a spring in a gravitational field where the inclusion of baryons increases the mass. This leads to enhanced oscillations and a shift of the mean value to $\Theta=-(1+R)\Psi$. Here $R$ is the photon-baryon momentum density ratio. Since we still observe the effective temperature $\Theta+\Psi$ only the peaks are enhanced so this has obvious observational consequences. The value of $R$ and hence the amount of shifting depends on the baryon density. \\
Taking into account the correct matter-to-radiation ratio changes the history of the universe and hence the size of the sound horizon at recombination, which is ambiguous with other effects. Fortunately there is another distinguishable effect of radiation, namely it makes the gravitational force evolve with time. The Poisson equation governs the evolution of the gravitational potential. Because of the radiation pressure the background density decreases with time which makes the gravitational potential decay. This induces a driving effect, because when the fluid is compressed the gravitational potential has gone and in addition the redshifting goes away. As soon as the universe is matter dominated the gravitational potential becomes dominated by the behavior of the dark matter. This means the amplitudes of the peaks are increased with a decreasing value of the dark matter-to-radiation ratio. In addition, when the gravitational potential is destroyed, the baryon loading effect should also be eliminated. Therefore the third peak indicates that dark matter exists and that it dominates at recombination.\\

Then finally there is another effect which damps the oscillations. It can be thought as a random walk of the photons that makes hot regions colder and cold regions hotter. The photons exchange momentum with electrons by Thomson scattering and they drag along the protons because of the tight Coulomb coupling. Anisotropies of the order of the damping scale or less therefore decrease and the oscillations beyond the third peak are suppressed. The damping length depends on the amount of baryonic matter.\\

Besides temperature fluctuations in the CMB it is expected that a part of the photons are polarized due to Thomson scattering before the surface of last scattering. Polarization is a vector field therefore and it turns out to be convenient to decompose it in the so-called $E$-modes and $B$-modes \cite{Zaldarriaga:1996xe, Kamionkowski:1996ks}, which are defined as the parity even and odd components of the photon polarization. Scalar perturbations only produce $E$-modes and tensor perturbations create both. This means that a detection of $B$-modes would be an unique signature of primordial gravitational waves. Moreover, the peaks in the $E$-mode spectrum are related to the peaks in the temperature spectrum because they originate from the same physics and therefore a combined probe provides an important cross check of the assumptions of the underlying model. The cross-correlation of the temperature fluctuations and the $E$-modes have been measured and this has provided the strongest check of phase coherence of the initial conditions which is in striking agreement with canonical single field inflation. \\

It should be noted that the observed temperature fluctuations and polarization of the CMB photons also carry along information about the physics after recombination. These are called secondary temperature and polarization anisotropies. This means we can in addition learn many things about the evolution and structure of the universe. On the other hand this limits the precision of the predictions because the primordial physics can be degenerate with foreground effects.

\subsection{Observing physics beyond the simplest models of inflation}
We have seen that precision observations of the CMB provide compelling evidence in favor of inflation. The generic predictions given by single field slow-roll inflation are confirmed by the data. However, if we would like to learn more about the precise microphysical origin of inflation it is necessary to deviate from the standard assumptions. We summarize the main observables to look for signatures beyond canonical single field inflation. Canonical single field inflation is defined as single field slow-roll inflation with canonical kinetic terms, minimally coupled to gravity and with Bunch-Davies initial conditions.
\subsubsection{Deviation from a power law spectrum}
Standard inflation assumes a power law scaling of the power spectrum. A deviation from slow-roll could result in a detectable correction to the power spectrum which allows for a change in the tilt
$$
\P_\R = A_s(k_\star )\left( \frac{k}{{k_\star }} \right)^{n_s - 1+\frac{1}{2}\alpha_s\ln(k/k_\star)}.
$$
The parameter $\alpha_s$ is called the \textit{spectral running} and for slow-roll inflation it is expected to be very small because it is second order in the slow-roll parameters. Using the Planck data alone there are no signs of a deviation from power law. The BICEP2 result however could point out there is a running to compensate for the high value of $r$, which requires new physics to explain. Another possibility to deviate from a power law spectrum is to have features on top of it. This is something we are interested in for this thesis.

\subsubsection{Non-Gaussianity}
The distribution of perturbations is fully characterized by the power spectrum if it is Gaussian. Single field inflation predicts a negligible amount of non-Gaussianity\footnote{In canonical single field inflation, interactions are suppressed by slow-roll parameters and therefore the bispectrum is suppressed by a combination of the slow-roll parameters times the square of the power spectrum, which is too small to be detectable.} and therefore much can be learned from a detection of non-Gaussianity, in particular on interactions of the inflaton with other fields. The main observable related to non-Gaussianities is the $3$-point correlator or its Fourier transform, the bispectrum
$$
\bra{0}\hat{\R}_{\v{k_1}}\hat{\R}_{\v{k_2}}\hat{\R}_{\v{k_3}}\ket{0}\equiv (2\pi)^3 B_\R(k_1,k_2,k_3)\delta(\v{k_1}+\v{k_2}+\v{k_3}),
$$
where the delta-function comes from translational invariance of the statistics. The three momenta therefore form a triangle and one can consider different shapes if also scale invariance of the bispectrum is assumed. The Planck collaboration has tested for different shapes. Local non-Gaussianity corresponds to the limit of squeezed triangles and such a detection would rule out all single field models. Planck \cite{Ade:2013ydc} has put very tight constraints on this type of non-Gaussianity and this excludes for example many multi-field models with more than one light field. Equilateral non-Gaussianity is constrained by Planck but there is room for a future detection. This type of non-Gaussianity can arise in single-field models, so we need a detection or more precise constraints to say more. The Planck collaboration has not published the full bispectrum, which means that there might be more information in the observational data in case the bispectrum is scale dependent. Furthermore, future experiments of large scale structure such as the Euclid mission \cite{Amendola:2012ys} are designed to probe non-Gaussianity to higher accuracy.

\subsubsection{Non-adiabatic perturbations}
Single-field inflation predict only fluctuations in the total energy density of the matter, or equivalently only fluctuations in the net curvature perturbation. This is because there is only one field involved. If one would allow for multiple light fields it is possible to have fluctuations in the relative densities of the different matter components. These type of fluctuations are called non-adiabatic perturbations $\S$. The effect of a surviving spectrum of non-adiabatic perturbations $\P_\S$ on the spectrum of temperature fluctuations is that they will result in peaks which are out of phase with the acoustic peaks coming from $\P_\R$. By means of a detection of the cross-correlation of the $E$-modes spectrum and the temperature spectrum the relative ratio of the power spectrum of non-adiabatic perturbations compared to the power spectrum of adiabatic perturbations is strongly constrained by Planck \cite{Ade:2013uln}
$$
\frac{\P_\S}{\P_\R}<0.036.
$$
How this exactly translates into a constraint on multi-field inflation is model-dependent. Isocurvature perturbations present at the end of inflation can get partially converted to curvature perturbations. This means that inflationary models which produce isocurvature perturbations are not necessarily ruled out. For example, in curvaton models of inflation a significant fraction of the observed adiabatic perturbations could originate from isocurvature perturbations. This is discussed in section 10.3.2 of \cite{Ade:2013uln}. Moreover, the evolution of the non-adiabatic perturbations depends presumably on the physics after inflation.

\subsubsection{Tensor tilt}
If future polarization experiments can confirm the detection of a tensor-to-scalar ratio of order $r\sim O(0.1)$ then there is hope to detect the tensor spectral index as well from the temperature power spectrum. In single field slow-roll inflation we have the following relation between the tensor tilt and the tensor-to-scalar ratio
$$
n_t = -\frac{r}{8},
$$
such that a disagreement with this relation would also be a hint of new physics.

\chapter{Multiple Field Inflation: a Translation of Different Studies in Literature}
\label{Chapter:multifieldinflation}
\section{Introduction}
In the previous chapter we have seen that inflation can explain why the universe is so homogeneous and flat and that it can account for tiny primordial density fluctuations which are the seeds of all structure in the universe. Moreover, the current observations of the temperature and polarization anisotropies of the CMB provide compelling evidence in favor of inflation \cite{Ade:2013uln, Ade:2014xna}.
Now we have very precise measurements of the CMB and hints that inflation took place only a few orders of magnitude below the Planck scale, it is interesting to investigate if we can learn more about fundamental physics beyond the standard model by using the observational data. Since inflation is formulated in a field theoretical framework it is in particular possible to study models motivated by supergravity and string theory. These models typically consist of multiple scalar fields \cite{Baumann:2014nda}. This means that the dynamics of the inflaton could have been influenced by the other degrees of freedom and therefore it is highly interesting to explore observational distinguishable effects compared to canonical single field inflation\footnote{Canonical single field inflation is defined as single field slow-roll inflation with canonical kinetic terms, minimally coupled to gravity and with Bunch-Davies initial conditions.}. For this we need to have a good theoretical understanding of multiple field inflation. In this chapter we provide an overview of a part of the literature on multiple field inflation. The studies use different notations and definitions which can give rise to much confusion when one would like to understand which approximations are precisely made. In the first part we recap the general framework of inflation and discuss the slow roll and turn parameters while we provide dictionaries which translate between the different notations and definitions used in the literature. In the second part the different approximations and their regimes of validity are discussed. An overview of the literature studied can be found in table \ref{Table:literature}. The shortcuts listed in this table are used in the dictionaries to denote the corresponding part of the literature.

\begin{table}[h]
    \begin{tabular}{ | p{11cm} | p{2cm} | p{2cm} |}
    \hline

    Authors          & Papers          & Shortcut          \\
    \hline
    Gordon, Wands, Bassett and Maartens & \cite{Gordon:2000hv} & GW \\
    & &\\
    Groot Nibbelink and van Tent & \cite{GrootNibbelink:2001qt} & TG \\
        & &\\
    Cremonini, Lalak and Turzynski & \cite{Cremonini:2010ua, Lalak:2007vi} & CT \\
         & &\\
    Ach\'ucarro, Atal, Cespedes, Gong, Hardeman, Palma and Patil & \cite{Achucarro:2010da, Achucarro:2010jv, Achucarro:2012yr} & AP\\
         & &\\
    Peterson and Tegmark & \cite{Peterson:2010np, Peterson:2011yt} & PT \\
    \hline
    \end{tabular}
\caption{An overview of the main literature studied in this chapter. The shortcuts are used in the dictionaries to denote the corresponding part of the literature.}
    \label{Table:literature}
\end{table}

\section{Framework of multiple field inflation}
\label{Section:setup}
In this section we recap the general formalism of multiple field inflation with Einstein gravity. We fix our notation and give the translation to notation used in the selected papers in table \ref{Table:dictionarysetup}.
\subsection{Set-up}
We study multiple field inflation in which the matter content of the universe is described by an arbitrary number of fields $\phi^a$ with possible non-canonical kinetic terms which are minimally coupled to gravity. The full theory is given by the following action
\begin{equation}
S=\int d^4 x\sqrt{-g}\left[\frac{M_P^2}{2}\tilde{R}-\frac{1}{2}g^{\mu\nu}G_{ab}(\phi)\partial_\mu\phi^a\partial_\nu\phi^b-V(\phi^a) \right]
\label{Equation:action}
\end{equation}
where $g$ is the determinant of $g_{\mu\nu}$, $M_P$ is the reduced Planck mass and $\tilde{R}$ the Ricci scalar constructed from spacetime quantities. Latin indices are used to denote the scalar fields which are at the same time treated as coordinates on the field manifold they span. Greek indices are used for the usual spacetime coordinates. We assume the universe is homogeneous and isotropic on large scales and spatially flat, which means that the background is described by the Friedmann-Robertson-Walker metric
\begin{equation}
ds^2=-dt^2+a^2(t)\delta_{ij}dx^idx^j=g_{\mu\nu}dx^\mu dx^\nu.
\label{Equation:spacetimemetric}
\end{equation}
The matrix $G_{ab}$ describing the non standard kinetic terms is symmetric, positive-definite and smooth and therefore can be interpreted as a field metric belonging to a smooth field manifold. The potential $V$ is continuous and bounded from below, but besides that completely arbitrary.
\subsection{Covariant formalism}
\label{Section:covariantformalism}
In order to derive compact and field coordinate invariant equations describing the evolution of the classical background and the quantum fluctuations we use the covariant formalism with respect to the field manifold developed in \cite{GrootNibbelink:2001qt}. Moreover, to understand the properties of the background field trajectory it is convenient to change basis in field space by defining directions along and perpendicular to the background trajectory first introduced in \cite{Gordon:2000hv}. \\

We rewrite the equations manifestly in tensor form to make all the equations field coordinate invariant. We use $G_{ab}$ and its inverse $G^{ab}$ for the shorthand notation to lower or raise indices when we take dot products. Furthermore since $\partial_a \tensor{A}{^{b\ldots}_{c\ldots}}$ for some arbitrary tensor $A$ does in general not transform like a tensor, we need to define a covariant derivative $\nabla_a$ such that when applied on tensors it will return another tensor
$$
\nabla_a \tensor{A}{^{b\ldots}_{c\ldots}}=\partial_a \tensor{A}{^{b\ldots}_{c\ldots}}+\Gamma^b_{ax} \tensor{A}{^{x\ldots}_{c\ldots}} - \Gamma^x_{ac} \tensor{A}{^{b\ldots}_{x\ldots}} + \ldots.
$$
The connection coefficients $\Gamma^a_{bc}$ are chosen as
$$
\Gamma^a_{bc}=\frac{1}{2}G^{ad}(\partial_b G_{cd}+\partial_c G_{db}-\partial_d G_{bc}),
$$
such that the covariant derivative is metric compatible.
Furthermore we can also define a covariant derivative on the field manifold with respect to any of the spacetime coordinates
$$
D_{\mu}=(\partial_\mu\phi^a)\nabla_a,
$$
which still creates tensorial quantities because $\phi^a$ transforms as a scalar under spacetime coordinate transformations.
Finally all information of the curvature of the field manifold is contained in the Riemann tensor
$$
\tensor{R}{^a_{bcd}}=\partial_c\Gamma^a_{bd}-\partial_d\Gamma^a_{bc}+\Gamma^a_{cx}\Gamma^x_{bd}-\Gamma^a_{dx}\Gamma^x_{bc},
$$
which can be contracted to the Ricci tensor $R_{ab}\equiv \tensor{R}{^\lambda_{a\lambda b}}$ and the Ricci scalar $R\equiv \tensor{R}{^a_a}$. \\

Let us now study the basis induced by an inflationary trajectory. Consider an arbitrary curve in field space parameterized as $\phi^a(t)$. This allows us to define a proper field distance along the curve given by
$$
\sigma=\int\sqrt{G_{ab}\dot{\phi}^a\dot{\phi}^b}dt.
$$
which is strictly increasing in time for a quasi de Sitter inflationary trajectory. This defines a field speed along the curve
$$
\dot{\sigma}\equiv \sqrt{G_{ab}\dot{\phi}^a\dot{\phi}^b}.
$$
If we use $\sigma$ as parameter along the curve we get the following tangent vector along the curve
\begin{equation}
T^a \equiv \frac{d\phi^a}{d\sigma} = \frac{\dot{\phi}^a}{\dot{\sigma}},
\label{Equation:tangentvector}
\end{equation}
which is automatically normalized. We can define the directional covariant derivative
$$
D_\sigma\equiv\frac{d\phi^a}{d\sigma}\nabla_a,
$$
which can be used to define parallel transport of tensors along this curve as having a vanishing directional covariant derivative. Geodesics in field space can be defined as paths which parallel transport their own tangent vector
$$
D_\sigma T^a=0 \quad \rightarrow \quad D_t T^a=0.
$$
We know the evolution of the fields will be along geodesics unless a force pushes them off. This role is fulfilled by the potential term in the action. If minus the gradient does not point in the same direction as tangent vector of the field trajectory, it will deviate from a geodesic. Therefore it is useful to define a normal unit vector to the trajectory as follows
\begin{equation}
N^a\equiv s_N(t)\left(G_{bc}D_t T^b D_t T^c\right)^{-1/2}D_t T^a,
\label{Equation:normalvector}
\end{equation}
which is only well defined when $D_t T^a$ is nonzero. This is the reason the function $s_N(t)=\pm1$ is introduced\footnote{As pointed out in \cite{Achucarro:2010da}.}, which indicates the orientation of $N^a$ with respect to $D_t T^a$. We let $s_N(t)$ just take some initial value at a given time where $D_t T^a$ is nonzero such that we know the initial orientation of $N^a$. During the intervals where $D_t T^a$ vanishes $N^a$ will be defined such that it is a continuous function of $t$ , furthermore $s_N(t)$ may flip sign such that when $D_t T^a$ becomes nonzero again, $N^a$ stays a continuous function of $t$. This corresponds to a right-handed or left-handed orientation in the two dimensional case. The rate of change of the tangent unit vector $T^a$ can therefore be in both the $+N^a$ direction and in the $-N^a$ direction and will be proportional to the gradient of the potential in the normal direction. \\

In this chapter we try to keep our expressions as concise as possible, which means we will introduce some more simplifying notation, likewise as done in the papers we study in this overview. In particular, a potential source of confusion is the notation used for the gradient and higher order derivatives of the potential and its projection along the tangent and normal directions. For clarity, we fix our notation here. The gradient and Hessian of the potential are denoted by
$$
V_a \equiv \nabla_a V, \quad \quad V_{ab}\equiv \nabla_a \nabla_b V.
$$
In case of two fields, the projections of the gradient and the Hessian of the potential along the tangent vector $T^a$ and the normal vector $N^a$ are written as
$$
V_T \equiv T^aV_a, \quad V_N\equiv N^a V_a, \quad \text{and} \quad V_{TT}\equiv T^a T^b V_{ab}, \quad  V_{TN}=V_{NT}\equiv T^a N^b V_{ab}, \quad  V_{NN}\equiv N^a N^b V_{ab}.
$$
We will use similar notation if we project other quantities, which have not been defined yet, along the tangent and normal directions by using the indices $T$ and $N$.\\

As a final note, the equations are given as function of coordinate time $t$, but one could equivalently use another time variable. Two particularly useful time variables in the context of inflation are conformal time $\tau$ as defined by the relation $dt=ad\tau$ and number of e-folds of inflation $N$ as given by $Hdt=dN$. We don't work with a special time coordinate but interchange between $t$, $\tau$ and $N$ depending on which variable is most convenient. Derivatives with respect to these time variables are denoted as
$$
\frac{d}{dt}(\ldots)\equiv\dot{(\ldots)}, \quad \quad \frac{d}{d\tau}(\ldots)\equiv(\ldots)^\prime, \quad \quad \frac{d}{dN}(\ldots)\equiv(\ldots)^; .
$$
A dot, a prime and a semicolon denote derivatives with respect to time $t$, conformal time $\tau$ and number of e-folds $N$ respectively. \\

In the rest of this section we first explain how to derive the equations describing the evolution of the classical background and introduce the slow-roll parameters and the notion of turns. Then we continue with the perturbations equations and their quantization for any number of fields. In the end we simplify to the case of two fields and we project the perturbations equations along the tangent and normal directions to study the influence of the background dynamics on the evolution of the perturbations. We provide dictionaries to translate between the different notations and definitions used in literature along the way.

\subsection{Background equations and slow-roll and turn parameters}
\label{Section:slowroll_turns}
Assuming the cosmological principle and that the universe is spatially flat, we can use the spacetime metric defined in equation (\ref{Equation:spacetimemetric}). Minimizing the action yields the homogeneous equations of motion for the scalar fields and the Friedmann equations,
\begin{equation}
\begin{aligned}
D_t\dot{\phi}^a&+3H\dot{\phi}^a+V^a=0, \\
H^2&=\frac{1}{3M_P^2}\left(\frac{1}{2}\dot{\sigma}^2+V\right), \\
\dot{H}&= - \frac{\dot{\sigma}^2}{2M_P^2}, \\
\end{aligned}
\label{Equation:background}
\end{equation}
where $H=\dot{a}/a$ is the Hubble parameter $V^a$ a short-hand notation for the gradient of the potential $G^{ab}\nabla_b V$. In order to solve the equations of motion for the perturbations later on, we need to understand how the perturbations are influenced the background dynamics. Based on our intuition from single field inflation we can use a slow-roll approximation to simplify the equations. Since the inflaton is in general not moving in a single field direction it is convenient to change basis by using the unit vectors pointing along and perpendicular to the background trajectory as defined in section \ref{Section:covariantformalism}. This allows us to define slow-roll parameters in multiple field inflation. Furthermore the trajectory can be forced to deviate from a geodesic in field space by the shape of the potential and therefore the tangent and perpendicular basis vectors may rotate into each other. This can be parameterized by so called turn parameters. We now make the notion of slow-roll and turns more precise and compare our definitions with the definitions used in the literature.

\subsubsection{Geometric slow-roll parameters and turn parameters}
We project the background equations of motion (\ref{Equation:background}) along the tangent $T^a$ and normal direction $N^a$ and find
\begin{equation}
\begin{aligned}
\end{aligned}
\label{Equation:backgroundprojected}
\end{equation}
with $V^a=\nabla^a V=V_T T^a+ V_N N^a$. The first equation coincides with the equations of motion for single field inflation. The second equation shows that the inflationary trajectory deviates from a geodesic when the normal component of the gradient of the potential is nonzero. This motivates the following definitions
\begin{equation}
\epsilon=-\frac{\dot{H}}{H^2},
\label{Equation:epsilon}
\end{equation}
$$
\eta^a=-\frac{1}{H\dot{\sigma}}D_t \dot{\phi}^a,
$$
which can be divided into the slow roll parameters $\epsilon$ and $\etapar$
\begin{equation}
\etapar\equiv\eta^a T_a=-\frac{\ddot{\sigma}}{H\dot{\sigma}},
\label{Equation:etapar}
\end{equation}
and the turn parameter $\etaperp$
\begin{equation}
\etaperp\equiv\eta^a N_a=\frac{V_N}{\dot{\sigma}H}.
\label{Equation:etaperp}
\end{equation}
Only if the background trajectory is a geodesic of field space then $D_t T^a =0$ and therefore $\etaperp$ parameterizes the deviation of the inflationary trajectory from a geodesic. For later use we also define a third slow roll parameter
\begin{equation}
\xipar=-\frac{\dddot{\sigma}}{H\ddot{\sigma}},
\label{Equation:xipar}
\end{equation}
and a second turn parameter
\begin{equation}
\xiperp=-\frac{\dot{\etaperp}}{H\etaperp},
\label{Equation:xiperp}
\end{equation}
which are only well defined if both $\ddot{\sigma}$ and $\etaperp$ are nonzero, but in all the equations they will appear as $\xipar\etapar$ and $\xiperp\etaperp$ anyway so we do not need to worry about this. Note that the definition of the normal vector $N^a$ implies the following signatures
$$\sgn(\etaperp(t))=-s_N(t) \quad \rightarrow \quad \sgn(\etaperp(t)N^a)=-1,$$
which means that for any equation the number of $\etaperp$'s and $N^a$'s in front of each term should be all odd or even.\\

The slow roll approximation can now be defined as $\epsilon, \etapar, \xipar\ll1$. The first two approximations $\epsilon, \etapar\ll1$ represent that the potential is flat in the tangent direction, such that the field rolls slowly down the potential. To see this rewrite the first order slow roll approximation as
\begin{equation}
\begin{aligned}
&\epsilon\ll1 \quad \rightarrow \quad \dot{\sigma}^2\ll V \quad \rightarrow \quad H^2 \approx \frac{V}{3M_P^2},\\
&\etapar\ll1 \quad \rightarrow \quad \frac{V_T^2}{9H^2}\ll V \quad \rightarrow \quad \frac{1}{3}M_P^2\left(\frac{V_T}{V}\right)^2\ll 1,\\
&\xipar\ll1 \quad \rightarrow \quad 3(\etapar+\epsilon)\approx \frac{\dot{V}_T}{H^2 \dot{\sigma}} \quad \rightarrow \quad 3M_P^2\frac{T^a\nabla_aV_{T}}{V}\ll 1,
\end{aligned}
\label{Equation:geometricSRinV}
\end{equation}
with $V_T=T^a\nabla_a V$, which is the projection of the gradient of the potential along the tangent direction. The approximation $\xipar\ll1$ assures the first two approximations will be prolonged because it says that the field acceleration $\ddot{\sigma}$ does not change much during an expansion time. Similarly we can define the slow-turn approximation $\etaperp, \xiperp \ll 1$, but note that this approximation has nothing to do with facilitating quasi de Sitter inflation as the slow-roll approximation does. Moreover, if the slow-turn approximation is valid it will suppress the multi-field effects, because they are parameterized by $\etaperp$.

\subsubsection{Kinematical slow-roll parameters}
Equivalently one can use a kinematical definition of the slow roll parameters in order to define various orders of the slow roll approximation. This is in particular useful if $\epsilon$ changes adiabatically during the times where the observable modes cross the Hubble radius. In this case one can construct a derivative expansion of $\epsilon$ where the corrections will be suppressed. The first slow roll parameter $\epsilon$ parameterizes the deviation from the de Sitter inflation such that the zeroth order slow roll approximation $\epsilon=0$ is equivalent to the de Sitter inflation. Since inflation should end this parameter is in general nonzero and at the end of inflation it becomes of order 1. Many e-folds before the end of inflation we expect it is possible to assume the first order slow-roll approximation $\epsilon \ll1$ and $\epsilon\approx const.$, which means we assume quasi-exponential inflation, i.e. the Hubble parameter does not change much during one expansion time. The second order slow roll approximation allows $\epsilon$ to vary in time but assures the quasi-de Sitter regime will be prolonged. We extend the same reasoning to the third order slow-roll approximation. This motivates the alternative definition of the slow roll parameters
\begin{equation}
\begin{aligned}
\epsilon_H & \equiv -\frac{\dot{H}}{H^2}= \epsilon, \\
\eta_H & \equiv -\frac{\dot{\epsilon}_H}{2H\epsilon_H} = \etapar-\epsilon,\\
\xi_H & \equiv -\frac{\dot{\eta}_H}{H\eta_H} = \frac{\etapar(\xipar-\etapar-3\epsilon)+2\epsilon^2}{\etapar-\epsilon},
\end{aligned}
\label{Equation:nongeometricSR}
\end{equation}
where the index $H$ belongs to the kinematical slow-roll parameters. We expressed the kinematical slow roll parameters in terms of the geometric ones to be able to easily interchange them. From the expressions in terms of the field speed and its accelerations it seems that, when the background and perturbation equations are expressed as function of number e-folds, these slow roll parameters will appear naturally, therefore we will choose to work with these parameters later on. Similarly to the various orders of the slow roll approximation we can define a zeroth, first and second order slow turn approximation. We list all order slow roll approximations in table \ref{Table:SRSTapproximation}.
\begin{table}[h]
    \begin{tabular}{ | p{3cm} |  p{5.5cm} p{3.5cm} | p{4.5cm} | }
    \hline
                        & Slow roll approximation              & Neglect                        & Slow turn approximation                   \\
    \hline
    General meaning     &   $\epsilon, \etapar, \xipar\ll1$  or $\epsilon_H, \eta_H, \xi_H\ll1$   &  No prescription          & $\etaperp, \xiperp \ll 1$ \\
    During transition   & && \\
    \quad Zeroth order        & $\epsilon_H=\eta_H=\xi_H=0$   & $O(\epsilon_H,\eta_H,\xi_H)$& $\etaperp = \xiperp = 0$                   \\
    \quad First order         & $\epsilon_H\lll 1$ and  $\eta_H=\xi_H=0$ &$O(\epsilon_H^2,\eta_H,\xi_H)$&  $|\etaperp| \lll1$ and $\xiperp=0$         \\
    \quad Second order        & $|\eta_H|<\epsilon_H\ll1$ and  $\xi_H=0$ &$O(\epsilon_H^3,\epsilon_H\eta_H,\eta_H^2,\xi_H)$&  $|\xiperp|<|\etaperp| \ll1$  \\

    \hline

    \end{tabular}
\caption{Summary of the slow roll and slow turn approximations introduced in section \ref{Section:slowroll_turns}. The first row gives the slow-roll and the slow-turn approximation in the general sense. The second row states the approximations at various orders, which can be used when one is considering a period consisting of only a couple of expansion times, such as the transition from the sub-Hubble to the super-Hubble regime.}
\label{Table:SRSTapproximation}
\end{table}
\subsubsection{Slow-Roll-Slow-Turn (SRST) approximation}
\label{Section:SRSTapproximation}
The SRST approximation is defined as taking both the slow roll and the slow turn approximation \cite{Peterson:2010np}. We define the first (or second order) SRST approximation as a combination of the first (second) order slow-roll and the first (second) order slow-turn approximation. Within this approximation we can express the non-geometric slow roll parameters and slow turn parameters in terms of the gradient and the Hessian of the potential.
Assuming $\etapar_H, \etaperp \ll 1$ we get the following SRST approximation of the background equations of motion for the fields
$$
3\phi^{;a}+\frac{V^a}{H^2}\approx 0,
$$
where the semicolon denotes a derivative with respect to the number of e-folds $N$ and where we denoted $V^a\equiv\nabla^a V$. This yields the following expressions
\begin{equation}
\begin{aligned}
\epsilon^{(1)}&=\frac{M_P^2}{2}\left(\frac{V_T}{V}\right)^2,\\
\eta_H^{(1)}&=M_P^2\left(\frac{V_{TT}}{V}-\left(\frac{V_T}{V}\right)^2\right),\\
{\etaperp}^{(1)}&=M_P^2\left(\frac{V_{NT}}{V}-\frac{V_N V_T}{V^2}\right).
\end{aligned}
\label{Equation:firstorderSRST}
\end{equation}
Here we used the notation $V_{TT}=T^a T^b \nabla_a\nabla_b V$, $V_{TN}=V_{NT}=T^a N^b \nabla_a\nabla_b V$ and $V_{NN}=N^a N^b \nabla_a\nabla_b V$. Plugging these expressions into the background equations we get the following improved expressions
\begin{equation}
\begin{aligned}
\epsilon^{(2)}&=\epsilon^{(2)}\left(1+\eta_H^{(1)}\right)^2,\\
\eta_H^{(2)}&=\eta_H^{(1)}-M_P^2\left(2\sigma^;\left(\frac{V_T^3}{V^3}-\frac{V_T V_{TT}}{V^2}\right)+\frac{\etaperp V_{TN}+V_{TT}^;}{V}\right),\\
{\etaperp}^{(2)}&={\etaperp}^{(1)}-M_P^2\left(2\sigma^;\left(\frac{V_T^2 V_N}{V^3}-\frac{V_T V_{NT}}{V^2}\right)+\frac{\etaperp V_{TN}+ V_{TT}^;}{V}\right).
\end{aligned}
\label{Equation:secondorderSRST}
\end{equation}
This allows us to express the observables in terms of the original model defined by the potential and field metric. The tangent and normal vector are however not determined without solving the background equations.\\

We notice that the SRST approximation implies that the gradient squared $M_P^2V_aV_b/V^2$ and the Hessian of the potential $M_P^2\nabla_bV_{a}/V$, except for $M_P^2V_{NN}/V$, are small along the inflationary trajectory. We expect that a converse relation also holds. If one assumes the slow roll approximation and if the full Hessian of the potential is small, we expect that the turn parameters are of the same order as the slow roll parameters. We did not manage to find a solid proof, but presumably the argument goes as follows. The slow roll approximation implies that $\left(\frac{M_PV^T}{V}\right)^2$ is small along the trajectory. Assuming there is one point along the trajectory where the turn rate vanishes such that $V_N=0$ we expect that the smallness of the Hessian implies that $V_N$ can only increase by $V_{NT}$ times the distance along the trajectory by the estimate
$$
V_N \approx \int d\phi^b\ N_a\nabla_b V^a = \int d\sigma\ V_{NT}.
$$
A more physical argument is that we expect that a very flat potential cannot force the inflationary trajectory to deviate from a geodesic too much. 

\subsubsection{Useful expressions}
\label{Section:usefulexpressions}
We derive a couple of useful expressions which we often use for manipulating the equations. First of all the derivatives of $\epsilon$ and $\etapar$ can be used to convert from the geometric to the kinematical slow roll parameters
\begin{align*}
\epsilon^;&=2 \epsilon(\epsilon-\etapar),\\
(\etapar)^;&=-\etapar (\xipar-\epsilon-\etapar),
\end{align*}
which gives in particular
$$
\etapar\xipar=\eta_H\xi_H+\eta_H(\eta_H+5\epsilon)+2\epsilon^2.
$$
Next we can express the slow-roll parameters in terms of the field speed and its accelerations
$$
\epsilon = \frac{(\sigma^;)^2}{2M_P^2}, \quad \quad \eta_H  =-\frac{\sigma^{;;}}{\sigma^;}, \quad \quad \xi_H  =-\frac{\sigma^{;;;}}{\sigma^{;;}}+\frac{\sigma^{;;}}{\sigma^;}
$$
With the following expression we can easily convert the Hubble parameter to the potential and vice versa.
$$
V=(3-\epsilon)M_P^2H^2.
$$
Finally during the transition regime we can express $\H$ and $\frac{a^{\prime\prime}}{a}$ explicitly in conformal time. We assume the second order slow-roll approximation and a Taylor expansion of $\epsilon$ around Hubble radius crossing yields
$$
\epsilon=\epsilon_\ast-k\epsilon_\ast\eta_{H\ast}(\tau-\tau_\ast),
$$
where $\tau_\ast$ is the time of Hubble radius crossing for a given mode $k$ and $\epsilon_\ast$ the value of $\epsilon$ at that time.
We can neglect $\epsilon_\ast\eta_{H\ast}$ in the second order slow-roll approximation (see section \ref{Section:slowroll_turns} and Table \ref{Table:SRSTapproximation}) and therefore we can derive
\begin{equation}
\begin{aligned}
\H^\prime=(1-\epsilon)\H^2 \quad \longrightarrow \quad &\H\approx-\frac{1}{\tau}(1+\epsilon_\ast+\epsilon^2_\ast) \quad \text{and} \quad  \frac{a^{\prime\prime}}{a} \approx \frac{2+3\epsilon_\ast+4\epsilon^2_\ast}{\tau^2}, \\
\text{with} \quad &k\tau_\ast\approx -(1+\epsilon_\ast+\epsilon^2_\ast),
\end{aligned}
\label{Equation:expressionH}
\end{equation}
where we fixed the value of the conformal time at Hubble radius crossing for each mode $k$ in order to keep the expressions as simple as possible. The first order slow-roll approximation renders the same expressions but then up to linear order in $\epsilon$.

\subsubsection{Dictionary slow-roll and turn parameters}
Finally we compare our slow-roll parameters with the ones used in the other papers in Table \ref{Dictionary:slowroll}. Note that the slow-roll parameters used in the papers of TG and PT are `contaminated' with turn parameters. This means their slow-roll approximation implicitly makes use of the slow-turn approximation. Their parameters can therefore only be used in the SRST approximation. In the more general case where one would like to study the influence of turns, one could use one of the sets of slow-roll parameters and the turn parameters defined section \ref{Section:slowroll_turns}. These parameters allow one to study the slow-roll contributions and the contributions from turns in the trajectory separately. This is important in order to understand the observational signatures of multi-field inflation.
\begin{table}[h]
    \begin{tabular}{ | p{3cm} | g{1.5cm}  | p{0.75cm} | p{4cm} | p{0.75cm} | p{0.75cm} | p{5.5cm}|}
    \hline
    Quantity                & Here                  &  AP                 &  TG                                               & GW                       &CT                         & PT                 \\
    \hline
                            &                      &                     &                                                   &                          &                           &                    \\
    Slow-roll parameters    &     $\epsilon$ \ \  or  $\epsilon_H$      &   $\epsilon$        &    $\tilde{\epsilon}$                             &-                         &$\epsilon$                 &$\epsilon$          \\
                            &     $\etapar$  or $\eta_H$      &   $\eta_{\|}$       &$\tilde{\eta}^{\|}=$
                                                                             \underline{\color{blue}$-\etapar$}                         &-                         &-                           &$\eta_1=$ \underline{\color{blue}$-\sigma^;\eta_H$}           \\
                            &     $\xipar$  or   $\xi_H$     &   $\xi_{\|}$        &$\tilde{\xi}^{\|}=$
                                                                            \underline{\color{blue}$-{\etaperp}^2+\xipar\etapar$}       &-                         &-                           &$\xi_1=$  \underline{\color{blue}$-\sigma^;((\etaperp)^2+\eta_H\xi_H+\eta_H^2)$}           \\
    Turn parameters         &     $\etaperp$       &   $\eta_{\bot}$     &$\tilde{\eta}^{\bot}=$
                                                                            \underline{\color{blue}$-\etaperp$}                         &$-\frac{1}{H}\dot{\theta}$&-                           &$\eta_2=$ \underline{\color{blue}$-\sigma^;\etaperp$}           \\
                            &     $\xiperp$         &  $\xi_{\bot}$       &$\tilde{\xi }^{\bot}=$
                                                                           \underline{\color{blue}$\etaperp(\epsilon+2\etapar+\xiperp)$}&-                         &-                           &$\xi_2=$ \underline{\color{blue}$\sigma^;\etaperp(2\eta_H+\xiperp)$}            \\
                            &                       &                     &                                                   &                          &                            &                    \\
    \hline

    \end{tabular}
\caption{Comparison of the slow-roll parameters defined here and the ones used in the other papers. The abbreviations used in the first row are explained in Table \ref{Table:literature}. When there is a difference in definition we indicate this in blue and with an underline. }
    \label{Dictionary:slowroll}
\end{table}

\subsection{Dictionary background equations}
A translation of all notation introduced so far to the notation used in the other papers can be found in Table \ref{Table:dictionarysetup}. In this dictionary we exclude the slow-roll and turn parameters, since these are already translated in Table \ref{Dictionary:slowroll}. Besides the slow-roll and turn parameters, the definitions of the quantities introduced so far to describe the set-up of multiple field theory are equivalent in all papers. This means the dictionary is a pure translation of symbols.

\begin{table}[h]
    \begin{tabular}{ | p{5cm} || g{1.8cm} | p{1.8cm} | p{1.8cm} | p{1.8cm} | p{1.8cm} | p{1.8cm} |}
    \hline

    Quantity                                        & Notation here          & Notation AP           & Notation TG                  &Notation GW              &Notation CT                 &Notation PT                         \\
    \hline

                                                    &                        &                       &                              &                          &                            &                                    \\
    Spacetime metric                                & $g_{\mu\nu}$           & $g_{\mu\nu}$          & $g_{\mu\nu}$                 & $g_{\mu\nu}$             &$g_{\mu\nu}$                & $g_{\mu\nu}$                       \\
    Spacetime derivative                            &  $\partial_\mu$        &  $\partial_\mu$       & $\partial_\mu$               & $\partial_{\mu}$         &$\partial_\mu$              &$\frac{\partial}{\partial x^{\mu}}$ \\
    Cov. derivative on spacetime                    & $\nabla_\mu$           & $\nabla_\mu$          &   $\nabla_\mu$               & -                        &-                           &-                                  \\
    Conformal time                                  & $\eta$                 & $\tau$                & $\eta$                       & -                        &$\tau$                      &-                                   \\
    Comoving time                                   & $t$                    & $t$                   & $t$                          & $t$                      &$t$                         &$t$                                 \\
    Number of e-folds                               & $N$                    & $N$                   & $N$                          & -                        &$N$                         &$N$                                 \\
    Derivative w.r.t. $\eta$                        & $(..)^{\prime}$        & $(..)^{\prime}$       & $(..)^{\prime}$              & -                        &$(..)^{\prime}$             &-                                   \\
    Derivative w.r.t. $t$                           & $\dot{(..)}$           & $\dot{(..)}$          & $\dot{(..)}$                 & $\dot{(..)}$             &$\dot{(..)}$                & $\dot{(..)}$                       \\
    Derivative w.r.t. $N$                           & $(..)^{;}$             & -                     & -                            & -                        &-                           &$(..)^{\prime}$                     \\

                                                    &                        &                       &                              &                          &                            &                                    \\
    Field metric                                    &  $G_{ab}$              & $\gamma_{ab}$         & $G_{ab}$                     & -                        &-                           &$G_{ij}$                            \\
    Field derivative                                &  $\partial_a$          &   $\partial_a$        & $\partial_a$                 & -                        &$(..)_{\phi}$               &-                                   \\
    Metric compatible connection                    & $\Gamma^a_{bc}$        &  $\Gamma^a_{bc}$      &  $\Gamma^a_{bc}$             & -                        &-                           &$\Gamma^i_{jk}$                     \\
    Riemann tensor                                  &   $R^a_{bcd}$          &  $\mathbb{R}^a_{bcd}$ & $ R^a_{bcd}$                 & -                        &-                           &-                                   \\
    Cov. derivative on $\mathcal{M}$                & $\nabla_a$             &  $\nabla_a$           &   $\nabla_a$                 & -                        &-                           &$\nabla_i$                          \\
    Cov. der. on $\mathcal{M}$ w.r.t. $x^\mu$       &  $D_{\mu}$             & $\frac{D}{dx^{\mu}}$  &  $\mathcal{D}_{\mu}$         & -                        &-                           &$\frac{D}{dx^{\mu}}$                \\

                                                    &                        &                       &                              &                          &                            &                                    \\
    Field coordinates                               &  $\phi^a$              & $\phi_0^a$            &  $\gv{\phi}^a$               & $\varphi_I$              &$\chi, \phi$                & $\phi^i$                           \\
    Field velocity vector                           & $\dot{\phi}^a$         & $\dot{\phi}_0^a$      &  $\dot{\gv{\phi}}^a$         &$\dot{\varphi}_I$         &$\dot{\chi},\dot{\phi}$     &$\dot{\phi}^i$                      \\
    Field speed                                     &  $\dot{\sigma}$        & $\dot{\phi}_0$        &  $|\dot{\gv{\phi}}|$         & $\dot{\sigma}$           &$\dot{\sigma}$              &$Hv$                                \\
    Field acceleration                              &  $\ddot{\sigma}$       &   $\ddot{\phi}_0$     &  $|\dot{\gv{\phi}}|\dot{}$   & $\ddot{\sigma}$          &$\ddot{\sigma}$             &$H(Hv)^{\prime}$                    \\
    Field-gradient potential                        & $V_a$                  &  $V_a$                &  $\nabla_a V$                &$V_{\varphi_I}$           &$V_{\chi}, V_{\phi}$        & $\nabla_i V$                       \\
    Double gradient potential                       & $\nabla_b V_a$         &  $\nabla_b V_a$       &  $\nabla_b \nabla_a V$       &$V_{\varphi_I \varphi_J}$ &$V_{\chi\chi}$ etc. & $\nabla_j \nabla_i V$              \\
    Tangent direction                               &  $T^a$                 &  $T^a$                &   $\v{e}_1^a$                & -                        &$E^I_{\sigma}$              & $\v{e}_1^i$                        \\
    Normal direction                                &  $N^a$                 & $N^a$                 & $\v{e}_2^a$                  & -                        &$E^I_{s}$                   & $\v{e}_2^i$                        \\
    Projection of $V_a$ along $T^a$                 & $V_{T}$                &  $V_{\phi}$           & -                            &$V_{\sigma}$              &$V_{\sigma}$                & -                                  \\
    Projection of $V_a$ along $N^a$                 & $V_{N}$                & $V_{N}$               & -                            &$V_s$                    &$V_s$                       & -                                  \\
    \hline
    \end{tabular}
\caption{This dictionary provides a translation of all notation introduced up to section \ref{subsection:perturbationeqns} to the notation used in the other papers. This concerns notation to describe the set-up of the multiple field theory and the background equations. The abbreviations used in the first row are explained in table \ref{Table:literature}.}
    \label{Table:dictionarysetup}
\end{table}

\subsection{Perturbation theory}
\label{Section:perturbationequations}
Ultimately we would like to compare the predictions from multi-field inflation with the small anisotropies in the cosmic microwave background radiation. The background fields are responsible for a spatially homogeneous and isotropic universe. Therefore we have to study the behavior of arbitrary quantum fluctuations with respect to the classical spacetime background and fields. Assuming they are small we can expand all equations to linear order in the perturbations. The observed temperature fluctuations can be related to these quantum perturbations. By assumption the expectation value of the perturbations vanishes therefore we can characterize the perturbations by the Fourier transform of the two-point correlation function which is a measure of the variance of their distribution. In this subsection we explain how to derive the equations of motion for the perturbations and in section \ref{Section:quantization} they will be quantized and related to observables. There will be overlap with chapter \ref{Chapter:inflation} section \ref{Section:inflation as the origin of structure}, in particular when we discuss the quantization, but we repeat this in order to have a consistent chapter.

\subsubsection{Cosmological perturbation theory}
Because we treat the perturbations to linear order we can therefore use standard cosmological perturbation theory, see for instance \cite{Baumann2013}. Using the spatial translation symmetry of the background it follows that at linear order the Fourier modes of the perturbations evolve independently. Moreover because of the spatial rotation symmetry of the background one can derive that the modes with different helicity evolve independently at linear order as well. This means the real perturbation of the metric can be divided into scalar, vector and tensor perturbations, which correspond respectively to modes with helicity $0$, $\pm1$ and $\pm2$. The most general way to parameterize linear scalar perturbations is then given by
\begin{align*}
&ds^2=-(1+2\Phi)dt^2+2a(t)\partial_i Bdx^idt+a^2(t)[(1-2\Psi)\delta_{ij}+2\partial_i\partial_j E]dx^i dx^j,\\
&\phi^a(t,\v{x})=\phi^a(t)+\delta\phi^a(t,\v{x}).
\end{align*}
Note there are fewer true degrees of freedom because we still need to incorporate the gauge freedom and the Einstein equations. The scalar gauge transformations
$$
t\rightarrow t+\xi^0, \quad x^i\rightarrow x^i+\partial^i \xi
$$
remove two degrees of freedom. The Einstein energy and momentum constraint equations set in addition two more degrees of freedom to zero. This means we are left with $n$ scalar degrees of freedom, where $n$ is the number of scalar fields in our multiple field model. For the actual computation we would like to fix a particular gauge, therefore we should take care that we are only considering physical quantities. It turns out to be useful to work with the following gauge invariant variables
$$
Q^a\equiv\delta\phi^a+\frac{\dot{\phi}}{H}\Psi,
$$
or equivalently the generalized Mukhanov-Sasaki variables $q^a\equiv aQ^a$. By using the basis induced by the inflaton trajectory we can separate $Q^a$ into perturbations $Q^T$ along the trajectory and $n-1$ perturbations orthogonal to the trajectory of which $Q^N$ is one. Adiabatic or curvature perturbations proportional to $Q^T$ correspond to perturbations in the total energy density or to the curvature of equal time hypersurfaces. The entropy or isocurvature perturbations (proportional to $Q^N$ and the other components of $Q^a$) correspond to relative perturbations in the energy densities of the different fields while leaving the total energy density and curvature of the spatial slices unchanged. In the end we will compute the two point correlation function of the curvature perturbation on uniform density hypersurfaces
$$\R\equiv\frac{H}{\dot{\sigma}}Q^T,$$
which has the important property to be conserved on super-Hubble scales $k\ll aH$ when the isocurvature perturbations are negligible. If that is the case it means that the curvature modes can be evaluated after they are well outside the Hubble radius and they will only evolve in time when they enter the Hubble radius again at late times where the physics determining their dynamics is well understood. What we precisely mean by entering the Hubble radius will be explained after we derived the perturbation equations. Next one can fix a gauge to simplify the calculations. Some popular gauges are the Newtonian gauge ($B=E=0$), the uniform density gauge ($\delta \rho=E=0$ such that $\R=\Psi$, which explains its name) and the comoving gauge ($\delta q=E=0$ such that $\R=\Psi$ and therefore it is also called the comoving curvature perturbation).
\subsubsection{Perturbation equations}
\label{subsection:perturbationeqns}
Computing the Einstein equations and the equations of motion of the perturbed fields leads to the following linearized equations of motion for the gauge-invariant perturbations.
\begin{equation}
\begin{aligned}
&D_t^2 Q^a + 3HD_t Q^a - \frac{\nabla^2}{a^2}Q^a+\frac{\tensor{C}{^a_b}}{a^2} Q^b = 0,  \\
&\frac{\tensor{C}{^a_b}}{a^2}=\underbrace{\nabla_bV^a-\dot{\sigma}^2 \tensor{R}{^a_{cdb}}T^cT^d}_{\text{'Mass matrix' } \tensor{(M^2)}{^a_b}}+2\epsilon\frac{H}{\dot{\sigma}}\left(T^aV_b+T_bV^a\right)+2\epsilon(3-\epsilon)H^2T^aT_b.\\
\end{aligned}
\label{Equation:perturbation_Q}
\end{equation}
In conformal time these equations simplify by using the Mukhanov-Sasaki variables:
\begin{equation}
D_{\tau}^2 q^a +\left(-\nabla^2\tensor{\id}{^a_b}+\tensor{\Omega}{^a_b} \right )  q^b = 0,
\label{Equation:perturbation_q}
\end{equation}
with
\begin{equation}
\tensor{\Omega}{^a_b}=\tensor{C}{^a_b} -(2-\epsilon)\H^2\tensor{\id}{^a_b},
\label{Equation:omegamatrix}
\end{equation}
with $\H=aH$. Besides the scalar perturbations we also have tensor perturbations. Since the extra fields do not provide any additional tensor perturbations compared to canonical single field inflation, we can get exactly the same equations as we found in section \ref{Section:inflation as the origin of structure}.
Because we will be working in the slow-roll regime this means it reduces to the single field slow roll expression for the power spectrum of the tensor perturbations. Expressed in terms of the slow-roll parameters, the tensor power spectrum is the same for every multi-field model. The definition of the power spectrum and the solution for tensor perturbations are given in section \ref{Section:quantization}. Going back to the scalar perturbations we can divide the evolution of the perturbations roughly in three regimes. This can be most easily seen when the equations are expressed in terms of e-folds because then the coefficients are all dimensionless
$$
D_{N}^2 Q^a + \left(3-\frac{\sigma^{;2}}{2M_P^2}\right)D_{N} Q^a + \frac{k^2}{a^2 H^2}Q^a+\frac{\tensor{C}{^a_b}}{a^2 H^2} Q^b = 0.
$$
Here $D_N$ is the covariant derivative on the field manifold with respect to the number of e-folds $N$, which is \textit{not equal to} $N^a\nabla_a$.
Note we are studying Fourier modes of the perturbations with comoving wavevector $k$. During inflation $aH$ will increase quasi-exponentially with $N$, where $1/aH$ is the so-called Hubble radius. This means we can identify three scales of interest, sub-Hubble scales $k\gg aH$, transition scales $k\sim aH$ and super-Hubble scales $k\ll aH$. At sub-Hubble scales the coefficient $\frac{k^2}{a^2 H^2}$ dominates over the other coefficients, which means the differential equations decouple and we can solve them exactly. One has to be careful though with additional mass scales which may introduce another regime of interest. At super-Hubble scales we can neglect $\frac{k^2}{a^2 H^2}$ which allows us in certain cases to estimate the behavior of the solutions. During the transition region the curvature modes under consideration cross the Hubble radius and get frozen if the isocurvature modes are small enough. We will go into more detail in solving these equations in section \ref{Section:analyticalapproximations}, but first we will discuss the quantization of the perturbations in section \ref{Section:quantization} and express the coefficients of the perturbations equations in terms of the slow-roll and turn parameters in section \ref{section:Perturbation equations and slow-roll and turns}.

\subsection{Quantization of perturbations}
\label{Section:quantization}
We use canonical quantization to quantize the perturbation fields.
\subsubsection{Canonical coordinates and momenta of $q^a$ in any frame}
The equations of motion for $q^a$ can be expressed in any frame, given by a set of vielbeins $\{\tensor{e}{^I_a}\}$. In this section we stay with a general frame but keep in mind that there is a preferred frame given by the tangent and normal directions defined by the inflationary trajectory $\{T_a, N_a, \ldots\}$. Defining $q^I=\tensor{e}{^I_a} q^a$ the equations of motion become:
\begin{equation}
q^{I\prime\prime}+2\tensor{Z}{^I_J}q^{J\prime}+\left(\tensor{Z}{^I_J^\prime}+Z^I_KZ^K_J\right)q^J+\left(-\nabla^2\id^I_J+\Omega^I_J \right ) q^J=0,
\label{Equation:perturbation_anyframe_q}
\end{equation}
with
\begin{equation}
\tensor{Z}{^I_J}=\tensor{e}{^I_a}\left(D_{\tau}\tensor{e}{^a_J}\right) \quad \text{and} \quad \Omega^I_J=e^I_a\Omega^a_be^b_J.
\label{Equation:zmatrix}
\end{equation}
The equations of motion for $q^I$ correspond to the following Lagrangian density
$$
\L=\frac{1}{2}\left[(q^{I\prime}+\tensor{Z}{^I_J}q^{J})^2 - (\nabla q^I)^2 - q^I \Omega_{IJ} q^J \right ],
$$
where the overall time-independent factor is fixed by the limiting single field case, see for instance \cite{Mukhanov:2005sc}.
The Lagrangian has no canonical kinetic terms, but that can be resolved by a redefinition of the fields
$$
q^I\equiv R^I_J(\tau)\tilde{q}^J \quad \text{where $R$ satisfies} \quad \tensor{R}{^I_J^\prime}=-Z^I_K R^K_J \quad \text{with} \quad R^I_J(\tau_i)=\id^I_J \quad \text{for some initial time $\tau_i$}.
$$
Now the equations of motion for $\tilde{q}^I$ reduce to:
\begin{equation}
\tilde{q}^{I\prime\prime}+\left(-\nabla^2\id^I_J+(R^{-1}\Omega R)^I_J \right ) \tilde{q}^J=0,
\label{Equation:eomqtilde}
\end{equation}
and the Lagrangian density becomes
$$
\L=\frac{1}{2}\left[(\tilde{q}^{I\prime})^2 - (\nabla \tilde{q}^I)^2 - \tilde{q}^I (R^{-1}\Omega R)_{IJ} \tilde{q}^J \right],
$$
which has standard kinetic terms. Therefore we know how to quantize the canonical coordinate and momentum fields $\tilde{q}^I$ and $\tilde{\pi}^I=\tilde{q}^{I\prime}$. We promote them to operators and impose the standard equal-time commutation relations. From this we can derive that the pair $\{q^I, \pi^I=q^{I\prime}+Z^I_J q^J\}$ are the corresponding coordinate and momentum fields in the non-canonical frame.
\begin{equation}
\begin{aligned}
&\tilde{q}^I \rightarrow \hat{\tilde{q}}^I, \quad \tilde{\pi}^{I}\rightarrow \hat{\tilde{\pi}}^{I}, \quad  && q^I \rightarrow \hat{q}^I, \quad  \pi^{I} \rightarrow \hat{\pi}^{I}, \\
&[\hat{\tilde{q}}^I(\tau, \v{x}), \hat{\tilde{\pi}}^J(\tau, \v{y})] =i\delta^{IJ}\delta(\v{x}-\v{y}), \text{ hence } && [\hat{q}^I(\tau, \v{x}), \hat{\pi}^J(\tau, \v{y})]=R^I_K R^J_L[\hat{\tilde{q}}^K(\tau, \v{x}),\hat{\tilde{\pi}}^L(\tau, \v{y})]=i\delta^{IJ}\delta(\v{x}-\v{y}).\\
\end{aligned}
 \label{Equation:commutationrelations_q}
\end{equation}

\subsubsection{Creation and annihilation operators and initial conditions}
Before we quantize the theory, we rewrite the solution to the equations of motion for $q^I$ and $\pi^I$ in a convenient way. Expanding $q^a$ in terms of its spatial Fourier modes $q_\v{k}^I$;
$$
q^I(\tau,\v{x})= \int\frac{d^3\v{k}}{(2 \pi)^{3/2}}q_\v{k}^I(\tau)e^{i\v{k}\cdot\v{x}},
$$
the equations of motion become
\begin{equation}
q^{I\prime\prime}_{\v{k}}+2\tensor{Z}{^I_J}q^{J\prime}_{\v{k}}+\left(\tensor{Z}{^I_J^\prime}+Z^I_KZ^K_J\right)q^J_{\v{k}}+\left(k^2\id^I_J+\Omega^I_J \right ) q^J_{\v{k}}=0.
\label{Equation:eom_qk}
\end{equation}
The solutions of the equations of motion for $q_\v{k}^I$ are given by complex mode functions $\tensor{(q_k^I)}{_\alpha}$ (and $\tensor{(\pi_k^I)}{_\alpha}=\tensor{(q_k^I)}{_\alpha^\prime}+Z^I_J \tensor{(q_k^J)}{_\alpha}$) , where $\alpha$ runs from $1$ to $n$. Each $\alpha$ represents a set of solutions $\{\tensor{(q_k^I)}{_\alpha}, \tensor{(q^{I \ast}_k)}{_\alpha}\}$ (possibly zero) such that the general solution of $q_\v{k}^I$ is determined by a constant complex vector $a_\v{k}$:
$$
q_\v{k}^I=\sum_\alpha\left[\tensor{(q_k^I)}{_\alpha}(a^\ast_\v{k})_\alpha+\tensor{(q^{I \ast}_k)}{_\alpha}(a_{-\v{k}})_\alpha \right]
$$
and the solution for $\pi_\v{k}^I$ becomes
$$
\pi_\v{k}^I=\sum_\alpha\left[\tensor{(\pi_k^I)}{_\alpha}(a^\ast_\v{k})_\alpha+\tensor{(\pi^{I \ast}_k)}{_\alpha}(a_{-\v{k}})_\alpha \right].
$$
Note that the reality condition $q^{I \ast}_\v{k}(\tau)=q_\v{-k}^I(\tau)$ is manifestly satisfied. After promoting the vectors $(a_{\v{k}})_\alpha$ and their complex conjugates to the operators $(\hat{a}_{\v{k}})_\alpha$ and $(\hat{a}^\dag_{\v{k}})_\alpha$, we would like to interpret them as annihilation and creation operators respectively, which satisfy the usual commutation relations
\begin{equation}
\begin{aligned}
&[(\hat{a}_{\v{k}})_\alpha,(\hat{a}^\dag_{\v{q}})_\beta]=\delta_{\alpha\beta}\delta(\v{k}-\v{q}).
\end{aligned}
 \label{Equation:commutationrelations_a}
\end{equation}
These are consistent with the commutation relations (\ref{Equation:commutationrelations_q}) if and only if the mode functions are normalized as follows:
\begin{equation}
\begin{aligned}
&\sum_\alpha\left[(q^{I \ast}_k)_\alpha(\pi^{J }_k)_\alpha- (\pi^{J\ast}_k)_\alpha (q^{I }_k)_\alpha\right ] = i \delta^{IJ},\\
&\sum_\alpha\left[(q^I_k)_\alpha (q^{J \ast}_k)_\alpha- (q^J_k)_\alpha (q^{I \ast}_k)_\alpha\right ] = 0,\\
&\sum_\alpha\left[(\pi^I_k)_\alpha (\pi^{J \ast}_k)_\alpha- (\pi^J_k)_\alpha (\pi^{I \ast}_k)_\alpha\right ] = 0.\\
\end{aligned}
 \label{Equation:wronskian}
\end{equation}
Another constraint on the mode functions comes from imposing initial conditions.
\begin{asm}
At the beginning of inflation the initial state of the universe is the vacuum state $\ket{0}$, defined by $\hat{a}_k^I\ket{0}=0$ for all scales $k$ of interest, such that there is no initial particle production, i.e. the Bunch-Davies vacuum.
\label{asm:BunchDavies}
\end{asm}
Since we are only interested in the observable modes which are sub-Hubble at the beginning of inflation (UV modes) and we are working in the linear approximation where different modes do not interact, this statement implies that we assume an initial state of minimal energy for these UV modes\footnote{This means we do not specify the initial conditions for the infrared modes $k\ll aH$. It is however possible to define a global Bunch-Davies vacuum following the lines of \cite{Glavan:2013mra}.}. Therefore the Hamiltonian does not contain any terms proportional to $\hat{a}\hat{a}$ and  $\hat{a}^\dag\hat{a}^\dag$. After some algebra the Hamiltonian can be written as:
\begin{equation*}
\begin{aligned}
\hat{H} &= \int d^3\v{x}\left[\sum_I(\hat{\pi}^I\hat{q}^{I \prime}) -\L  \right ]\\
&= \frac{1}{2}\sum_{\alpha\beta}\int d^3\v{k}\left[ (F_k)_{\alpha\beta}(\hat{a}^\dag_{\v{k}} )_\alpha (\hat{a}^\dag_{-\v{k}} )_\beta + (G_k)_{\alpha\beta}(\hat{a}^\dag_{\v{k}} )_\alpha (\hat{a}_{\v{k}} )_\beta + h.c. \right ],
\end{aligned}
\end{equation*}
with
\begin{equation*}
\begin{aligned}
&(F_k)_{\alpha\beta}= \sum_I \left[(\pi^I_k)_\alpha (\pi^I_k)_\beta - 2 (\pi^I_k)_\alpha Z^I_J (q^J_k)_\beta + (q^I_k)_\alpha k^2 (q^I_k)_\beta +(q^I_k)_\alpha  \Omega^I_J (q^J_k)_\beta
\right ],\\
&(G_k)_{\alpha\beta}= \sum_I \left[(\pi^I_k)_\alpha (\pi^{I\ast}_k)_\beta - 2 (\pi^I_k)_\alpha Z^I_J (q^{J\ast}_k)_\beta + (q^I_k)_\alpha k^2 (q^{I\ast}_k)_\beta +(q^I_k)_\alpha  \Omega^I_J (q^{J\ast}_k)_\beta
\right ].
\end{aligned}
\end{equation*}
The energy of the ground state is given by
$$
\bra{0}\hat{H}\ket{0}=\frac{1}{2}\sum_I\int d^3 \v{k}\delta^3(0)\sum_\alpha\left[(\pi^I_k)_\alpha (\pi^{I\ast}_k)_\alpha+k^2(q^I_k)_\alpha (q^{I\ast}_k)_\alpha - 2(q^I_k)_\alpha Z^I_J(\pi^{J\ast}_k)_\alpha+(q^I_k)_\alpha\Omega^I_J(q^{J\ast}_k)_\alpha\right].
$$
Here we interpret $\delta(\v{0})$ as the infinite total volume of space\footnote{See \cite{mukhanov2007introduction} p.72.} $V$ and we factor it out by demanding that the energy \textit{density} $\bra{0}\hat{H}\ket{0}/V$ is minimized. Therefore at the beginning of inflation $\tau_0$, which is assumed to be deep inside the sub-Hubble regime where we can assume $k\ggg Z^I_J$ and $k^2\ggg \Omega^I_J$, we have in addition to (\ref{Equation:wronskian}) the following constraints:
\begin{equation}
\begin{aligned}
&\sum_I\sum_\alpha\left[|(\pi^I_k)_\alpha|^2+k^2|(q^I_k)_\alpha|^2\right] \quad \text{is minimized}\\
&\sum_I \left[(\pi^I_k)_\alpha (\pi^I_k)_\beta + k^2(q^I_k)_\alpha (q^I_k)_\beta\right]=0
\end{aligned}
 \label{Equation:bunchdavies}
\end{equation}
Writing the sub-Hubble solution to the equations of motion for $q^I_\alpha$ as
$$
q^I_\alpha=a^I_\alpha e^{ik\tau} + b^I_\alpha e^{-ik\tau}.
$$
and denoting $a^I_\alpha$ and $b^I_\alpha$ as the matrices $A$ and $B$, we can rewrite the constraints (\ref{Equation:wronskian}, \ref{Equation:bunchdavies}) as the following matrix equations
\begin{align*}
&A^\ast A^T+A^{\ast T}A-(B^\ast B^T+B^{\ast T}B)=\frac{1}{k}\id, \\
&AB^{\ast T}=\left(AB^{\ast T}\right)^T, \quad BA^{\ast T}= \left(BA^{\ast T}\right)^T\\
&\Tr\left(AA^{\ast T}+BB^{\ast T}\right) \quad \text{is minimized,}\\
&B^TA+A^TB=0.
\end{align*}
Combining the first and third equation and using the cyclic property of the trace we get
$$
\Tr\left(B_1B_1^{T}+B_2B_2^{T}\right) \quad \text{is minimized, with} \quad B=B_1+iB_2,
$$
and since $B_1$ and $B_2$ are real matrices this trace is always nonnegative and therefore the solution is given by $B=0$. We are free to choose $A^I_\alpha \sim \delta^I_\alpha$ and we can deduce that at early times $\tau$ the full solution is given by
\begin{equation}
(q^I_k)_\alpha (\tau)= \sqrt{\frac{1}{2k}} e^{ik\tau +i\lambda^I}  \delta^I_\alpha,  \\
\end{equation}
for some random phase $\lambda^I$. This completely determines the evolution of $q^I_\v{k}$, however the equations of motion are not exact solvable therefore we provide an overview of analytical approximations in the next chapters.
\subsubsection{Power spectrum}
In the end we would like to compare theory to observations, therefore the main quantities of interest are the $n$-point correlation functions characterizing the distribution of the fluctuations. The lowest order statistics are given by two-point correlators. The power spectrum is the dimensionless Fourier transform of the two-point correlation function and corresponds to the variance of the distribution of fluctuations of a given mode $k$.
The powerspectra $\P_{q}^{IJ}$ for $q^I_\v{k}$ are defined as
$$
\bra{0}\hat{q}^I_{\v{k}}(\tau) \hat{q}^J_{\v{k}^\prime}(\tau)\ket{0}\equiv\delta(\v{k}-\v{k}^\prime)\frac{2\pi^2}{k^3}\P_{q}^{IJ}(k,\tau),
$$
such that
$$
\bra{0}\hat{q}^I(\tau,\v{x}) \hat{q}^J(\tau,\v{y})\ket{0} =\int\frac{d^3\v{k}}{4\pi k^3}\P_{q}^{IJ}(k,\tau)e^{-i\v{k}\cdot(\v{x}-\v{y})}.
$$
This will in particular yield the following powerspectrum for the comoving curvature perturbation
\begin{equation}
\P_\R=\frac{H^2}{a^2 \dot{\sigma}^2}\P_{q}^{TT}
\label{Equation:powerspectrumR}
\end{equation}
In case of two fields we define the isocurvature perturbation as $\S=\frac{H}{a\dot{\sigma}}q^N$ and this yields in addition the following power spectra
\begin{equation}
\P_\S=\frac{H^2}{a^2 \dot{\sigma}^2}\P_{q}^{NN},
\label{Equation:powerspectrumS}
\end{equation}
\begin{equation}
\P_{\R\S}=\frac{H^2}{a^2 \dot{\sigma}^2}\P_{q}^{TN}.
\label{Equation:powerspectrumRS}
\end{equation}
The power spectra for the curvature and isocurvature modes are evaluated at the end of inflation which is equal to their value shortly after Hubble radius crossing if the isocurvature modes decay quickly enough. The gravitational waves are characterized by the power spectrum $\P_T$ which is expressed in terms of the slow-roll parameters the same for every multi-field model, namely
\begin{equation}
\P_T\equiv 2\P_h=2\left(\frac{H_\ast}{\pi M_P}\right)^2\left[1+2(2-\ln2-\gamma)\epsilon_\ast\right],
\label{Equation:powerspectrumtensor}
\end{equation}
with $\gamma$ Euler's constant and where the asterisk denotes that the quantities are evaluated at Hubble radius crossing. In this expression first order slow-roll corrections\footnote{More precisely, the multiplicative factor reflects corrections to $\beta\approx \frac{3}{2}$ and $-\frac{1}{\tau}\approx aH$ as stated at the very the end of section \ref{Section:inflation as the origin of structure}.} are taken into account compared to our computations in section \ref{Section:how to characterize the perturbations} and therefore we get the multiplicative factor $1+2(2-\ln2-\gamma)\epsilon_\ast$, see for example \cite{Stewart:1993bc}. The power spectra evaluated at Hubble radius crossing depend on the wavenumber $k$. Assuming the powerspectrum can be approximated by a power law, it is convenient to write
\begin{align*}
&\P_\R=A_s(k_\star)\left(\frac{k}{k_\star}\right)^{n_s - 1 + 1/2\alpha_s\ln k/k_\star+\ldots},\\
&\P_T=A_t(k_\star)\left(\frac{k}{k_\star}\right)^{n_t + 1/2\alpha_t\ln k/k_\star+\ldots},
\end{align*}
where $A_s$ and $A_t$ are the scalar and tensor amplitudes, $n_s$ and $n_t$ the scalar and tensor spectral index and $\alpha_s\equiv dn_s/d\ln k$ and $\alpha_t\equiv dn_t/d\ln k$ the running of the scalar and the tensor spectral indices. The Planck collaboration \cite{Ade:2013uln} has found
\begin{align*}
& A_s = (2.196\pm0.060)\times 10^{-9}  \quad \text{for a pivot scale $k_\star=0.05 \SI{}{\per\mega\parsec}$}, \\
& n_s = 0.9603 \pm 0.0073 \quad \text{assuming $A_t=0$},
\end{align*}
but the BICEP collaboration \cite{Ade:2014xna} contradicts this latter assumption by the following claimed measurement of the tensor-to-scalar power ratio
$$
r\sim 0.2 \quad\text{with}\quad r\equiv \frac{A_t}{A_s}.
$$
This means one should take into account the degeneracy between $A_t$ and $n_s$. The scales which can be related to the CMB measurements and to future large scale structure surveys \cite{Carrasco:2012cv} run from $k=0.002 \SI{}{\per\mega\parsec}$ to $k=0.100 \SI{}{\per\mega\parsec} $ corresponding to scales which leave the Hubble radius 60 to 53 e-folds before the end of inflation.

\subsubsection{Dictionary perturbation equations}
Let us take stock of the variables introduced to describe the perturbation equations and the quantization procedure discussed in sections \ref{subsection:perturbationeqns} and \ref{Section:quantization}. A translation of all notation introduced in these section to the notation used in the literature is given in Table \ref{Dictionary:perturbations}. Most of the definitions are equivalent, so for these the dictionary is a pure translation of symbols. When there is a difference in definition however, this is indicated in blue and with an underline.

\begin{table}[h]
    \begin{tabular}{ | p{4cm} | g{1.8cm} | p{2.5cm} | p{2.5cm} | p{1.8cm} | p{3.2cm}|}
    \hline
    Quantity                        & Notation here         & Notation AP             & Notation TG                       &Notation GW             &Notation PT         \\
    \hline
                                    &                       &                         &                                   &                        &                  \\
    Field perturbations             &$\delta\phi^a$         &$\delta\phi^a$           &$\v{\delta\phi}$                   &$\partial\varphi_I$     &$\v{\delta\phi}$  \\
    Metric perturbation             &$\Phi$                 &$\psi$                   &$\Phi$                             &$\psi$                  &$\Psi$            \\
    Gauge inv. pert.                &$Q^a$                  &$Q^a$                    &-                                  &$Q_I$                   &$\v{\delta\phi_f}$\\
    Mukhanov-Sasaki var.            &$q^a$                  &$v^a$                    &$\v{q}$                            &                        &$\v{q}$           \\
    Com. curv. pert.                &$\R$                    &$\R$                    &-                                  &$\R$                    &$\R$              \\
    Entropy pert.                   &$\S$                    &$\S$                    &-                                  &$\delta s=$
                                                                                        \underline{\color{blue}$\frac{\dot{\sigma}}{H}\S$}                   &$\S$              \\
                                    &                       &                         &                                   &                        &                  \\
    Auxiliary functions             &$\tensor{C}{^a_b}$     &$\tensor{C}{^a_b}=$
                                        \underline{\color{blue}$\frac{1}{a^2}\tensor{C}{^a_b}$} &-                                  &-                       &-                 \\
                                    &$\tensor{(M^2)}{^a_b}$ &-                        &$\tilde{\v{M}}^2     $             &-                       &$\tensor{\tilde{\v{M}}}{^i_j}=\underline{\color{blue}\frac{\tensor{M}{^a_b}}{V}-\frac{V^aV_b}{V^2}}$ \\
                                    &$\tensor{\Omega}{^a_b}$&$\tensor{\Omega}{^a_b}$  &$\v{\Omega}=$
                                                                             \underline{\color{blue}$\frac{1}{a^2H^2}\Omega^a_b$}   &-                       &-                 \\
                                    &                       &                         &                                   &                        &                  \\
    Vielbeins for any frame         &$e^I_a$                &$e^I_a$                  &-                                  &-                       &-                 \\
    Preferred frame                 &$T^a, N^a, \ldots$     &$T^a, N^a, \ldots$       &$\v{e}_n$                          &-                       &$\v{e}_n$         \\
    Coordinates of $q$              &$q^I$                  &$v^I$                    &$q_n$                              &-                       &$q_n$             \\
    Coordinates of $Q$              &$Q^I$                  &$Q^I$                    &-                                  &$Q_I$                   &$\delta\phi_n$\\
    Coordinates of $\Omega$         &$\Omega^I_J$           &$\Omega^I_J$             &$\Omega_{mn}$                      &-                       &-                 \\
    Rotation matrix                 &$R^I_J$                &$R^I_J$                  &$R_{nm}$                           &-                       &$U_{nm}$          \\
    Z-matrix                        &$\tensor{Z}{^I_J}$     &$\tensor{Z}{^I_J}$       &$Z_{mn}=$
                                                                                  \underline{\color{blue}$\frac{1}{aH}Z_{IJ}$}      &-                       &$Z_{mn}=\underline{\color{blue}\frac{1}{aH}Z_{IJ}}$ \\
    Rotated perturbation            &$\tilde{q}^I$          &$u^I$                    &$\tilde{q}_n$                      &-                       &$\tilde{q}_\pm$  \\
    Fourier modes $q^I$             & $q^I_\v{k}$           &$v^I(\v{k},\tau)$        &$q_n$                              &-                       &$q_n$             \\
    Solutions to EOM $q^I_\v{k}$    &$(q^I_k)_{\alpha}$     &$v^I_\alpha(k,\tau)$     &$Q_{nm}$                           &-                       &$q_n$             \\
                                    &                       &                         &                                   &                        &                  \\
    Power spectra                   &$\P_q^{IJ}$            &$\P_v^{IJ}$              & -                                 &-                       &-                 \\
                                    &$\P_Q^{IJ}$            &$\P_Q^{IJ}$              &  -                                &-                       &-                  \\
                                    &$\P_\R$                &$\P_\R$                  & -                                 &-                       &-                 \\

                                    &                       &                         &                                   &                        &                 \\
    \hline
     \end{tabular}
\caption{This dictionary provides a translation of the notation and definition introduced in section \ref{subsection:perturbationeqns} and section \ref{Section:quantization} to the notation used in the other papers. This concerns notation used to describe the perturbations equations and the quantization procedure. When there is a difference in definition we indicate this in blue and with an underline. The abbreviations used in the first row are explained in table \ref{Table:literature}. We leave out CT because in their set of papers they only consider two fields. This is discussed in section \ref{Section:slowroll_turns}.}
    \label{Dictionary:perturbations}
\end{table}

\subsection{Perturbation equations in terms of the slow-roll and turn parameters}
\label{section:Perturbation equations and slow-roll and turns}
Finally we project the perturbations equations along the tangent and normal directions to get more insight in the behavior of these equations by rewriting the coefficients in terms of the slow roll and turn parameters. From now on we will only consider the case of two fields and leave the case of more fields to future work. We will see that $\etaperp$ couples the curvature mode to the isocurvature mode whenever there is a mismatch between the geodesics of field space and the valley of the potential. Therefore $\etaperp$ parameterizes the multi-field effects as expected. In order to keep things clear we choose to work with the kinematical slow roll parameters. We will however switch between the time variables, depending on what is most convenient for the next section, in which we discuss analytical approximations of the solutions.

\subsubsection{Projection of the perturbation equations with two fields}
We simplify our analysis to two fields and project the perturbation equations (\ref{Equation:perturbation_Q}) and (\ref{Equation:perturbation_q}) along the tangent $T^a$ and normal direction $N^a$, which are defined in section \ref{Section:covariantformalism}. Therefore we first express the entries of the $\Omega$ matrix defined in (\ref{Equation:omegamatrix}) in terms of the slow-roll and turn parameters. We give the four components $\Omega_{TT},\ \Omega_{TN},\ \Omega_{NT}$ and $\Omega_{NN}$, separately where the index $T$ or $N$ stands for a projection along the tangent or normal direction, e.g. $\Omega_{TN}=T^aN^b\Omega_{ab}$. Because $\Omega$ is symmetric we have $\Omega_{TN}=\Omega_{NT}$ and the four entries are therefore fully specified by
\begin{align*}
&\Omega_{TT}=-a^2H^2\left(2-\epsilon+\eta_H(-3+\epsilon+\eta_H+\xi_H)-(\etaperp)^2\right),\\
&\Omega_{NT}=a^2H^2\etaperp\left(3-\epsilon-2\eta_H-\xiperp\right),\\
&\Omega_{NN}=-a^2H^2\left(2-\epsilon-m^2\right),
\end{align*}
with $m^2\equiv M^2_{NN}/{H^2}$. Moreover, if we wish to express the perturbation equations in the preferred basis $\{T_a, N_a\}$, we can see from equation (\ref{Equation:perturbation_anyframe_q}) that we also should express the $Z$ matrix in terms of the slow-roll and turn parameters. The $Z$ matrix, which is defined in equation (\ref{Equation:zmatrix}), is anti-symmetric and therefore in case of two fields it is given by $Z_{TN}=-Z_{NT}=aH \etaperp$,
if we pick the vielbeins $\{e^I_a\}$ to be $\{T_a, N_a\}$.\\

Now projecting the equations for the generalized Mukhanov-Sasaki variables (\ref{Equation:perturbation_q}) as function of conformal time along the tangent and normal direction, we get the following system of coupled differential equations
\begin{equation}
\begin{aligned}
q^{T\prime\prime}&+2\etaperp aH q^{N\prime}+\left(k^2+a^2H^2\left(-2+\epsilon+(3-\epsilon)\eta_H+\eta_H^;-\eta_H^2\right) \right )q^T+\etaperp a^2H^2\left(4-2\epsilon-2\eta_H-2\xiperp \right )q^N=0,\\
q^{N\prime\prime}&-2\etaperp aH q^{T\prime}+\left(k^2+a^2H^2\left(m^2-2+\epsilon-(\etaperp)^2\right)\right )q^N+\etaperp a^2H^2\left(2-2\eta_H \right )q^T=0.
\end{aligned}
\label{Equation:perturbation_q_twofields}
\end{equation}
These are the equations describing the evolution of the perturbations which are canonically quantized and therefore useful to study during the sub-Hubble regime in section \ref{Section:subhorizon} where the initial conditions are imposed. We see that the differential equations couple to each other as soon as $\etaperp\neq 0$, i.e. if the inflationary trajectory is curved with respect to the geodesics in field space. If there are no turns, then the equations decouple and the equation for $q^T$ becomes equal to the canonical single field version. As expected, the turn parameter $\etaperp$ carries the most important information about the multi-field effects and therefore should be considered separately from the slow-roll parameters. The variables we are ultimately interested in to compare with experiment are the curvature and isocurvature perturbation $\R$ and $\S$. We can rewrite the full equations of motion for the perturbations (\ref{Equation:perturbation_Q_twofields}) in terms of $\R\equiv\frac{Q^T}{\sigma^;}$ and $\S\equiv\frac{Q^N}{\sigma^;}$ and get
\begin{equation}
\begin{aligned}
\left(\frac{d}{dN}+3-\epsilon_H-2\eta_H \right)\left(\R^;+2\etaperp\S\right)+ \frac{k^2}{a^2H^2}\R&=0,\\
\S^{;;}+(3-\epsilon_H-2\eta_H)\S^;+\left(\frac{k^2}{a^2H^2}+\frac{M^2}{H^2}-(\etaperp)^2+\eta_H(\eta_H-3+\epsilon_H)-\eta_H^;\right)\S&=2\etaperp \R^;.
\end{aligned}
\label{Equation:perturbation_R_twofields}
\end{equation}
Remember that the curvature perturbation $\R$ corresponds to fluctuations in the fields along the inflationary trajectory and represents the spatial curvature perturbation in the uniform density gauge. The isocurvature perturbation $\S$ keeps the spatial curvature perturbation and total energy density of the fields constant, but mixes up the relative ratios of the energy densities of the fields. If the isocurvature perturbations decay on super-Hubble scales, then the curvature perturbation will freeze out \cite{Weinberg:2003sw}. This means the value of the mode functions of $\R$ evaluated shortly after Hubble radius crossing will stay constant until the mode enters the Hubble radius again in an epoch of known physics and therefore we can relate the predicted distribution of the curvature modes to the observed temperature fluctuations in the CMB. The same remarks on the perturbation equations (\ref{Equation:perturbation_q_twofields}) for $q^a$ also apply here. Concerning the signs of coefficients, whether the turn is to the right or to the left it should not change the way the curvature and isocurvature modes are affected, because it is just a matter of convention. Note that we do not need to take absolute values of $\etaperp$ for the source terms of the differential equations, because each projection on $N^a$ also contains the same left/right information given by $s_N(t)$. All the terms in the curvature perturbation equation therefore contain an even number of $\etaperp$'s and $N^a$'s whereas all the terms in the isocurvature perturbation equation contain an odd number. The equations (\ref{Equation:perturbation_R_twofields}) will be extensively used to discuss the analytical approximations of the solution of the perturbation equations in the super-Hubble regime in section \ref{Section:superhorizon} and for one of the approximation schemes in the transition regime in \ref{section:transition}. Besides equations (\ref{Equation:perturbation_R_twofields}), sometimes the equations for the variables $Q^a$ (\ref{Equation:perturbation_Q}) as function of number of e-folds are used to describe the super-Hubble evolution
\begin{equation}
\begin{aligned}
&Q^{T;;}+(3-\epsilon)Q^{T;}+\left(\frac{k^2}{a^2H^2}+(3-\epsilon)\eta_H+\eta_H^;-\eta_H^2\right)Q^T = -2\etaperp\left[ Q^{N;}+(3-\eta_H-\epsilon-\xiperp)Q^N\right],\\
&Q^{N;;}+(3-\epsilon)Q^{N;}+\left(\frac{k^2}{a^2H^2}+m^2-(\etaperp)^2\right)Q^N = 2\etaperp\left[{Q}^{T;}+\eta_HQ^T\right].
\end{aligned}
\label{Equation:perturbation_Q_twofields}
\end{equation}
We will only use these equations to study the `slow-roll approximation on perturbations' in section \ref{Section:superhorizon}. Finally, during transition the equations can be simplified when we assume for example the first order SRST approximation. Using use the expressions (\ref{Equation:expressionH}) we get
\begin{equation}
\begin{aligned}
q^{T\prime\prime}&+\left(k^2-\frac{2+3\epsilon}{\tau^2} \right )q^T = 2\etaperp\left(\frac{1+\epsilon}{\tau} q^{N\prime} - \frac{1+2\epsilon}{\tau^2}\left(2-\epsilon \right )q^N\right),\\
q^{N\prime\prime}&+\left(k^2+\frac{1+2\epsilon}{\tau^2}\left(m^2-2+\epsilon\right)\right )q^N = -2\etaperp\left(\frac{1+\epsilon}{\tau}q^{T\prime}+\frac{1+2\epsilon}{\tau^2}q^T\right),
\end{aligned}
\label{Equation:perturbation_q_slowroll_twofields}
\end{equation}
where we kept everything to linear order in $\epsilon$ and $\etaperp$. This is useful for one of the approximation schemes to solve the perturbation equations in the transition regime studied in section \ref{Section:analyticalapproximations}.

\subsubsection{Dictionary perturbation equations with two-fields}
Some of the papers we include in this overview consider the particular case of two fields. In Table \ref{Dictionary:two-field} we provide a dictionary to translate the variables introduced in section \ref{section:Perturbation equations and slow-roll and turns} to the notation used in these particular papers. When there is a difference in definition of the variables we indicate this in blue and with an underline.
\begin{table}[h]
    \begin{tabular}{ | p{4cm} | g{1.8cm} | p{1.8cm} | p{3.5cm} | p{4.4cm}|}
    \hline
    Quantity                        & Notation here         & AP                           & GW                    & CT                            \\
    \hline
                                    &                       &                              &                       &                               \\
    Two fields                      &$\phi, \chi$           &$\phi^1, \phi^2$              &$\phi, \chi$           &$\chi, \phi$                   \\
    Gauge inv. pert.                &$Q^T, Q^N$             &$Q^T, Q^N$                    &$Q_{\sigma}, \delta s$ &$Q_{\sigma}, \delta s$         \\
    Mukhanov-Sasaki var.            &$q^T, q^N$             &$v^T, v^N$                    &-                      &$u_{\sigma}, u_s$              \\
    Com. curv. pert.                &$\R$                   &$\R$                          &$\R$                   &$\R$                           \\
    Isocurvature pert.              &$\S$                   &$\S$                          &$\delta s=$
                                                                           \underline{\color{blue}$\frac{\dot{\sigma}}{H}\S$}&$\S$                           \\
    Auxiliary functions             &$M^2$                  &$M^2$                         &$\mu_s^2=$
                                                                               \underline{\color{blue}$M^2+3H^2{\etaperp}^2$}&-                              \\
                                    &$\Omega_{TT}$          &$\Omega_{TT}$                 &-                      &$\M_{11}=$
                                                                                                  \underline{\color{blue}$\frac{\Omega_{TT}}{a^2 H^2}+2$}   \\
                                    &$\Omega_{TN}$          &$\Omega_{TN}$                 &-                      &$\M_{12}=$
                                                                                                   \underline{\color{blue}$\frac{\Omega_{TN}}{a^2 H^2}$}    \\
                                    &$\Omega_{NN}$          &$\Omega_{NN}$                 &-                      &$\M_{22}=$
                                                                                                  \underline{\color{blue}$\frac{\Omega_{NN}}{a^2 H^2}+2$}   \\
                                    &$C_{TT}$               &-                             & -                     &$C_{\sigma\sigma}=$
                                                                                            \underline{\color{blue}$\frac{C_{TT}-Z_{TN}^2}{a^2}$}            \\
                                    &$C_{TN}$               &-                             &-                      &$C_{\sigma s}=$
                                                                              \underline{\color{blue}$\frac{C_{TN}+a \dot{Z}_{TN}+2aHZ_{TN}}{a^2}$}          \\
                                    &                       &                              &                       &$C_{s \sigma}=$
                                                                              \underline{\color{blue}$\frac{C_{TN}+a \dot{Z}_{TN}-2aHZ_{TN}}{a^2}$}           \\
                                    &$C_{NN}$               &-                             &-                      &$C_{ss}=$                  \underline{\color{blue}$\frac{C_{NN}-Z_{TN}^2}{a^2}$}           \\
                                    &$V_{TT}$               &$V_{\phi\phi}$                &$V_{\sigma\sigma} $    &$\eta_{\sigma\sigma}=$
                                                                                \underline{\color{blue}$\frac{V_{TT}+T^aT^b\Gamma^c_{ab}V_c}{3H^2}$}         \\
                                    &$V_{TN}$               &$V_{\phi N}$                  &-                      &$\eta_{\sigma s}=$
                                                                                  \underline{\color{blue}$\frac{V_{TN}+T^aN^b\Gamma^c_{ab}V_c}{3H^2}$}       \\
                                    &$V_{NN}$               &$V_{NN}$                      &$V_{ss} $              &$\eta_{ss}=$
                                                                                 \underline{\color{blue}$\frac{V_{NN}+N^aN^b\Gamma^c_{ab}V_c}{3H^2}$}        \\
                                    &$Z_{TN}$               &$Z_{TN}$                      &-                      &-                              \\
    Power spectra                   &$\P_\R$                &$\P_\R$                       &$\P_{Q_\sigma}$        &$\P_\R$                        \\
                                    &$\P_\S$                &$\P_\S$                       &$\P_{\delta s}=?$      &$\P_\S$                        \\
                                    &$\P_{\R\S}$            &$\P_{\R\S}$                   &$\C_{Q_\sigma\delta s}=?$ &-                           \\
                                    &                       &                              &                       &                               \\
    \hline
     \end{tabular}
\caption{This dictionary provides a translation of the notation used to describe the perturbations equations in case of two fields. When there is a difference in definition we indicate this in blue and with an underline. The abbreviations used
in the first row are explained in Table \ref{Table:literature}. This dictionary complements the dictionary given in Table \ref{Dictionary:perturbations}.}
    \label{Dictionary:two-field}
\end{table}

\section{Analytical approximations}
\label{Section:analyticalapproximations}
Now we have introduced all the relevant notions, in particular the important turn parameter which parameterizes the multi-field effects, we continue studying the various approximation schemes to solve the perturbation equations in case of two fields. In this section we mainly work with the time variable number of e-folds of inflation $N$ and the dimensionless positive time variable $z\equiv-k\tau$. We denote a derivative with respect to $N$ as semi colon and a derivative with respect to $z$ exactly like we denoted the derivative with respect to $\tau$, with a prime. As explained at the end of section \ref{Section:perturbationequations} the evolution of the modes of the perturbations can be divided into three regimes, sub-Hubble $a\ll\frac{k}{H}$, transition $a\sim\frac{k}{H}$ and super-Hubble $a\gg\frac{k}{H}$. We will denote the beginning and the end of transition as $z_-$ and $z_+$, or $N_-$ and $N_+$, respectively. A quantity $A$ evaluated at Hubble radius crossing is denoted by $A_\ast$. We fix the value of $z$ at Hubble radius crossing at $z_\ast\approx 1$ as explained in section \ref{Section:usefulexpressions}. Finally we will be sloppy with jargon and entitle all variables $q^T, \ Q^T, \ \R$ as the curvature or adiabatic perturbation and all variables $q^N, \ Q^N, \ \S$ as the isocurvature or entropy perturbation. \\

It turns out that only two types of approximations are made in the papers under consideration. The approximation schemes can roughly be summarized as follows
$$
\text{The slow-roll approximation and either a mass hierarchy} \quad m^2\gg 1 \quad \text{or the slow-turn approximation}.
$$
In all papers the slow-roll approximation is assumed to be true from deep inside the sub-Hubble regime until far beyond the transition regime in the super-Hubble regime. One part of the literature (CT, AP, see Table \ref{Table:literature}) assumes a large mass hierarchy and an effective single field description is derived. The other part of the literature (GW, TG and PT) assumes a slow-turn approximation and the perturbation equations are solved by treating the slow-roll and turn parameters on equal footing. We study the regime of validity of the two approximation schemes by rederiving their analytical solutions of the perturbation equations. Many details on the computations can be found in Appendix \ref{Appendix:analytical approximations}. We will first discuss the early and late time evolution of the perturbations and then solve the equations in the transition regime in the two approximation schemes. Finally in section \ref{Section:overview} the approximations and their validity and the predictions are summarized in Table \ref{Table:overviewapproximations}.

\subsection{Sub-Hubble}
\label{Section:subhorizon}
In the sub-Hubble regime we have $k^2\gg a^2 H^2$ or equivalently $z^2\gg1$ and the perturbation equations (\ref{Equation:perturbation_q_twofields}) reduce to\footnote{Remember we are working with the variable $z\equiv -k\tau$ and a prime denotes a derivative with respect to $z$ in this section.}
\begin{align*}
q^{T\prime\prime}_\alpha&+q^T_\alpha = 0,\\
q^{N\prime\prime}_\alpha&+q^N_\alpha = 0,
\end{align*}
where the solution is determined by assuming the initial state of the universe is the Bunch-Davies vacuum
\begin{align*}
q^T_\alpha=&\sqrt{\frac{1}{2k}}e^{-iz}\delta^1_\alpha,\\
q^N_\alpha=&\sqrt{\frac{1}{2k}}e^{-iz}\delta^2_\alpha,
\end{align*}
like we found in section \ref{Section:quantization}. \\

This is not the complete story however, because if there exists a mass hierarchy $m^2\gg 1$, the frequency of the isocurvature modes changes before the beginning of the transition region. This happens at the scale $k^2/a^2H^2\sim m^2$. In this case the sub-Hubble regime should be divided into two regimes, one where $k$ dominates all other scales in the perturbation equations $k^2/a^2H^2\gg m^2 \gg 1$ and where we can impose the initial conditions, the other where the isocurvature mass scale dominates $ m^2 \gg k^2/a^2H^2\gg 1$. In the second part of the sub-Hubble regime we expect the following behavior of the solution of the perturbations equations
\begin{align*}
&q^T_\alpha = (A_-)_\alpha e^{i\int dt\omega_- }+(A_+)_\alpha e^{i\int dt\omega_+ },\\
&q^N_\alpha = (B_-)_\alpha e^{i\int dt\omega_- }+(B_+)_\alpha e^{i\int dt\omega_+ },
\end{align*}
where $\omega_-^2\approx k^2/a^2$, $\omega_+^2 \approx H^2m^2+k^2/a^2$ and $(A_+)_\alpha=(B_-)_\alpha=0$ when there are no turns. In case of a constant turn rate $\etaperp$ we find that the high frequency dissociates even more from the low frequency such that in any case $\omega_+^2\gg \omega_-^2 \gg H^2$. The solution becomes approximately
\begin{equation}
\begin{aligned}
&\omega_\pm^2\approx \frac{k^2}{a^2}+\frac{1}{2}H^2\left(m^2+3(\etaperp)^2\right)\left(1\pm 1\right) \pm 4\frac{k^2(\etaperp)^2}{a^2\left(m^2+3(\etaperp)^2\right)},\\
&(A_+)_\alpha\approx\frac{-2i\etaperp H\omega_+}{\omega_+^2-k^2/a^2}(B_+)_\alpha, \\
&(B_-)_\alpha\approx\frac{-2i\etaperp H\omega_-}{H^2\left(m^2-(\etaperp)^2\right)+k^2/a^2-\omega_-^2}(A_-)_\alpha.
\end{aligned}
\label{Equation:frequenciessubhorizon}
\end{equation}
Using the quantization conditions (\ref{Equation:wronskian}) we can estimate $(A_-)_\alpha\sim \frac{1}{\sqrt{2a\omega_-}}\delta^T_\alpha$ and $(B_+)_\alpha\sim \frac{1}{\sqrt{2a\omega_+}}\delta^N_\alpha$, which means that the high frequency modes are relatively suppressed. Since the difference between the frequencies becomes larger in case of a constant turn rate we expect that these conclusions also apply for the more general case of slow roll inflation and any turn rate which does not vary too quickly in time. In the next section we will show in addition that the high frequency modes will decay on super-Hubble scales. This means we can neglect the high frequency solutions as soon as the mode enters the part of the sub-Hubble regime where $m^2\gg k^2/a^2H^2$ which allows us to compute an effective theory in section \ref{section:transition}. More calculational details can be found in Appendix \ref{Appendix:analytical approximations}.

\subsection{Super-Hubble}
\label{Section:superhorizon}
We now study the behavior of the perturbation equations in the super-Hubble regime where we can neglect $k^2/a^2H^2$ or equivalently take $z\ll 1$. We start with a simplified case where we assume that the zeroth order slow roll approximation is valid and that both $m^2$ and $\etaperp$ are constant. Therefore we can use the perturbation equations (\ref{Equation:perturbation_q_slowroll_twofields}) where we put $\epsilon=k=0$ and change variables to $z$
\begin{align*}
q^{T\prime\prime}_\alpha&-\frac{2}{z^2}q^T_\alpha = 2\etaperp\left(\frac{1}{z} q^{N\prime}_\alpha - \frac{2}{z^2}q^N_\alpha\right),\\
q^{N\prime\prime}_\alpha&+\frac{1}{z^2}\left(m^2-2-(\etaperp)^2\right)q^N_\alpha = -2\etaperp\left(\frac{1}{z}q^{T\prime}_\alpha+\frac{1}{z^2}q^T_\alpha\right).
\end{align*}
Using the ansatz $q^I_\alpha=A^I_\alpha z^P$, where $A^I_\alpha$ are constants and $P$ some power, we find four solutions to the perturbation equations. The full solution is given by
\begin{align*}
q^T_\alpha&=\delta^T_\alpha \left(Az^{-1}+ B z^{2}\right) +  \delta^N_\alpha C\frac{4\etaperp z^{\frac{1}{2} \left(1+\sqrt{9-4 \left(3 (\etaperp)^2+m^2\right)}\right)}}{3+\sqrt{9-4(m^2+3(\etaperp)^2)}} +
\delta^N_\alpha D\frac{4\etaperp z^{\frac{1}{2} \left(1-\sqrt{9-4 \left(3 (\etaperp)^2+m^2\right)}\right)}}{3-\sqrt{9-4(m^2+3(\etaperp)^2)}} , \\
q^N_\alpha&=\delta^T_\alpha\left( B \frac{6\etaperp}{(\etaperp)^2-m^2}z^{2}\right)+ \delta^N_\alpha\left(C z^{\frac{1}{2} \left(1+\sqrt{9-4 \left(3 (\etaperp)^2+m^2\right)}\right)} + D z^{\frac{1}{2} \left(1-\sqrt{9-4 \left(3 (\etaperp)^2+m^2\right)}\right)}\right),
\end{align*}
to zeroth order in slow roll. Note that the constants $C$ and $D$ are determined by the initial conditions for $q^N$ and therefore the combinations $\etaperp C$ and $\etaperp D$ have a sign independent of the convention for $s_N$ at the beginning of inflation. From this example we read off the general behavior of the perturbations at super-Hubble scales. We have a growing mode $q^T\sim z^{-1}$ and a decaying mode $q^I \sim z^{2}$ whose frequencies goes to zero. If there is no mass hierarchy $m^2\ll 1$ and a very small turn rate $\etaperp \ll 1$, then we have another pair of growing and decaying modes $q^N\sim z^{-1}$ and $q^N \sim z^2$, which means the isocurvature modes will not decay and for non vanishing $\etaperp$ they will keep sourcing the curvature perturbations. If there is a large mass hierarchy $m^2 \gg 1$ then the other modes are decaying $q^I \sim z^{1/2}$ with a high frequency determined by the effective mass squared $M^2_{\text{eff}}=(m^2+3(\etaperp)^2) H^2$. In this case we find that after Hubble radius crossing the curvature mode freezes out,  $\R\sim z^0$, while the isocurvature modes oscillate rapidly and decay, $\S\sim z^{3/2}$ . In Appendix \ref{Appendix:analytical approximations} we derive the full solution to first order in slow roll and we find the same behavior.  \\

Next we study the more general super-Hubble solution. We use the full equations of motion (\ref{Equation:perturbation_R_twofields}) for the curvature and isocurvature perturbations $\R$ and $\S$ as function of number of e-folds. These equations yield the following exact solution in the super-Hubble regime
$$
\R^;=c e^{-\int_{N_+}^N(3-\epsilon-2\eta_H)d\tilde{N}}-2\etaperp\S,
$$
where the exponential corresponds to the decaying solution $\R\sim a^{-1}q^I \sim a^{-3}$ we found above and if there are no turns the isocurvature modes stop sourcing $\R$ and it will become constant. Assuming we pick $N_+$ large enough, such that we can neglect the decaying part, the system reduces to
\begin{equation}
\begin{aligned}
&\R^; = -2\etaperp\S,\\
&\S^{;;}+(3-\epsilon-2\eta_H)\S^;+\left(m^2+3(\etaperp)^2+\eta_H(\eta_H-3+\epsilon)-\eta_H^;\right)\S=0.
\end{aligned}
\label{Equation:superhorizonS}
\end{equation}
We will approximate the solutions in case of a mass hierarchy $m^2\gg 1$ and in case of the SRST approximation.

\subsubsection{Case 1: a mass hierarchy $m^2\gg 1$}
In first order slow roll and when $\S$ is heavy enough we expect $\S\sim a^{-3/2}$ in the super-Hubble regime because $\S\sim q^N/a$ and we found two decaying solutions $q^N\sim a^{-1/2}$ in case of constant turns and constant mass. This means all the isocurvature perturbations decay and we have a freeze out of the curvature perturbation on super-Hubble scales. Assuming $m^2\gg1$ and approximately constant we can derive the conditions on $\etaperp$ such that this expectation is true. Rewriting the equation as
\begin{align*}
&\S^{;;}+f_1\S^;+f_2 \S = 0,\\
&\tilde{S}^{;;}=f_3\tilde{S}, \quad \text{with} \quad \tilde{S}=\exp\left[\int(f_1/2)dN\right]\S \quad \text{and} \quad  f_3=\frac{1}{4}f_1^2+\frac{1}{2}f_1^;-f_2,
\end{align*}
we can use the WKB approximation to find
\begin{align*}
\S=\ & A\exp\left[\int_{N_+}^N\left(\sqrt{f_3}-\frac{f_1}{2}-\frac{1}{4}\frac{f_3^;}{f_3}+\frac{1}{8\sqrt{f_3}}\left(\frac{f_3^{;;}}{f_3}-\frac{5}{4}\left(\frac{f_3^;}{f_3}\right)^2\right)+\ldots\right)d\tilde{N}\right]\\
+\ & B\exp\left[\int_{N_+}^N\left(-\sqrt{f_3}-\frac{f_1}{2}-\frac{1}{4}\frac{f_3^;}{f_3}+\frac{1}{8\sqrt{f_3}}\left(\frac{f_3^{;;}}{f_3}-\frac{5}{4}\left(\frac{f_3^;}{f_3}\right)^2\right)+\ldots\right)d\tilde{N}\right],
\end{align*}
where the functions $f_1$ and $f_3$ expressed in the slow-roll and turn parameters take the form
\begin{align*}
&f_1=3-\epsilon-2\eta_H,\\
&f_3=\frac{1}{4}\left(9-4(m^2+3(\etaperp)^2)-6\epsilon+\epsilon^2+2\epsilon\eta_H\right).
\end{align*}
Only if $f_3$ is varying rapidly this might lead to a non-decaying solution for $\S$, therefore we can derive the following rough constraint on $\xiperp$ to assure that the isocurvature modes will decay on super-Hubble scales
\begin{equation}
\left|\frac{f_3^;}{2f_3}\right|<f1 \quad  \rightarrow \quad \left|\xiperp\right|<\left(3-\epsilon-2\eta_H\right)\left(1+\frac{m^2}{(\etaperp)^2}\right) \quad \text{assuming} \quad m^2>\frac{9}{4}-3(\etaperp)^2>\frac{9}{4}.
\label{Condition:superhorizonxi}
\end{equation}
Note that the decaying solutions are also oscillating as $e^{\pm i\sqrt{m^2+3(\etaperp)^2-9/4}N}$ while the frequency of the curvature mode goes to zero. As remarked in section \ref{Section:subhorizon} the frequency of the isocurvature mode changes after a time $z<m$. 
The curvature and isocurvature modes are described by a system of coupled harmonic oscillators, but do not `synchronize' because they are coupled by a derivative interaction. Instead they will tend to oscillate with a low frequency and a high frequency which dissociate even more if the coupling is turned on. Based on the analysis in this section and in section \ref{Section:subhorizon} we expect the high frequency modes to be highly suppressed with respect to the low frequency modes already after a time $z<m$ and therefore they do not contribute to the final amplitude of $\R$. Therefore one only needs to take into account the low frequency mode for which the time derivatives of the isocurvature perturbations can be neglected, because they correspond to the slowly varying oscillations with a frequency much smaller than $m$. In section \ref{section:transition} we will \textit{integrate out the heavy mode} which basically comes down to neglecting the high frequency solutions and approximating the low frequency solution using the mass hierarchy.

\subsubsection{Case 2: the SRST approximation}
In the studies (TG, PT) the \textit{slow roll approximation on perturbations} is made, which comes down to neglecting $Q^{T;;}$ and $Q^{N;;}$ in the perturbation equations. It is based on the observation that on super-Hubble scales the background fields $\phi^a$ satisfy the same equations of motion as the perturbed fields $\phi^a+\delta\phi^a$. Assuming both the slow-roll and the slow-turn approximation we can neglect $D_N\phi^{a;}$. This means we can use the same approximation on the perturbed fields and therefore the difference, the equation for $Q^a$ given in (\ref{Equation:perturbation_Q}), does not contain a double derivative. This means the perturbation equations (\ref{Equation:perturbation_Q_twofields}) for two fields become
\begin{align*}
&(3-\epsilon)Q^{T;}+\left((3-\epsilon)\eta_H+\eta_H^;-\eta_H^2\right)Q^T = -2\etaperp\left[ Q^{N;}+(3-\epsilon-\eta_H-\xiperp)Q^N\right],\\
&(3-\epsilon)Q^{N;}+\left(m^2-(\etaperp)^2\right)Q^N = 2\etaperp\left[{Q}^{T;}+\eta_HQ^T\right].
\end{align*}
Transforming back to the the variables $\R$ and $\S$ we get
\begin{align*}
&(3-\epsilon)\R^;+\left(\eta_H^;-\eta_H^2\right)\R = -2\etaperp\left[ \S^;+(3-\epsilon-2\eta_H-\xiperp)\S\right],\\
&(3-\epsilon)\S^;+\left(m^2-(\etaperp)^2-(3-\epsilon)\eta_H\right)\S = 2\etaperp\R^;.
\end{align*}
The equation for $\R$ does not simplify so we only use the second equation for $\S$ and stay with $\R^;=-2\etaperp \S$ which yields
$$
\R^;=-2\etaperp\S_+\exp\left[\int^N_{N_+}\frac{1}{3}\left(1+\frac{\epsilon}{3}\right)\left(m^2+3(\etaperp)^2-(3-\epsilon)\eta_H\right)d\tilde{N}\right],
$$
where $\S_+$ times the exponential is the solution for $\S$. Therefore at super-Hubble scales we have the following expression for the curvature perturbations
\begin{equation}
\R=\R_+-\S_+\int_{N_+}^N\left\{2\etaperp \exp\left[\int^{\tilde{N}}_{N_+}\frac{1}{3}\left(1+\frac{\epsilon}{3}\right)\left(m^2+3(\etaperp)^2-(3-\epsilon)\eta_H\right)d\tilde{\tilde{N}}\right] \right\}d\tilde{N}.
\label{Equation:curvaturesuperhorizonSRST}
\end{equation}
Note that the decaying solution $\S\sim e^{-3N}$ is automatically discarded. If there is no large mass hierarchy $m^2\ll 1$, but a nonzero (small) turn rate, the curvature mode does not freeze out completely and its evolution needs to be tracked until the end of inflation. If there is a small mass hierarchy $m^2\sim 1$ then it is expected that the curvature mode will freeze out quickly after Hubble radius crossing.


\subsection{Transition}
\label{section:transition}
Next we need to solve the perturbation equations in the transition regime in order to match the sub-Hubble initial conditions to the super-Hubble solution. In case of a mass hierarchy between the curvature and isocurvature mode we can compute an effective single field description from the moment that $z<m$. If we assume the SRST approximation we can try to solve the full differential equation in this regime.

\subsubsection{Case 1: effective single field description in case of a large mass hierarchy}
As explained in section \ref{Section:superhorizon} we integrate out the heavy modes when $z<m$. The low frequency solution allows us to neglect the time derivatives of $\S$ resulting in an effective equation for $\R$
\begin{equation}
\R^{;;}+\left(3-\epsilon-2\eta_H-2\frac{c_s^;}{c_s}\right)\R^;+\frac{c_s^2k^2}{a^2H^2}\R=0.
\label{Equation:effectiveequationR}
\end{equation}
This equation has a variable speed of sound $c_s$ and therefore differs from the equations which follow from a canonical single field inflationary theory.
The speed of sound $c_s$ in terms of the background quantities is given by
\begin{equation}
\frac{1}{c_s^2}\equiv 1+\frac{4(\etaperp)^2}{\frac{k^2}{a^2H^2}+m^2-(\etaperp)^2+\eta_H(\eta_H-3+\epsilon)-\eta_H^;}.
\label{Equation:speedofsound}
\end{equation}
This allows us to derive the effective action for the perturbations up to quadratic order where we can fix the overall normalization by putting $c_s=1$ and using the canonical single field result discussed in chapter \ref{Chapter:The Standard Model of Cosmology}
\begin{equation}
S_{\text{eff}} = M_P^2\int dNd^3\v{x}\ \frac{a^3\epsilon H}{c_s^2}\left[(\R^;)^2-\frac{c_s^2}{a^2H^2}(\partial_i \R)^2\right].
\label{Equation:effectiveaction}
\end{equation}
The calculational details to find the effective action (\ref{Equation:effectiveaction}) can be found in Appendix \ref{Appendix:analytical approximations}. \\

Next we discuss the validity of the effective single field description. This effective single field description only holds for times $z<m$ which means the number of e-folds before Hubble radius crossing at $z=m$ equals $\Delta N=\ln m$. If you want to impose initial conditions on $\R$ deduced from the sub-Hubble solution you need therefore
\begin{equation}
\Delta N \geq 2.3 \rightarrow  m\geq 10.
\label{Condition:lowerboundm}
\end{equation}
Moreover, we can express the validity of the effective theory in terms of \textit{adiabatic conditions} on the turn rate and the speed of sound following \cite{Achucarro:2012yr}. The derivations can be found in Appendix \ref{Appendix:analytical approximations}. We find the following constraints on the speed of sound
\begin{equation}
\begin{aligned}
&\left|\frac{u^;}{u}\right| \ll \min\left(\frac{|m^2-(\etaperp)^2|}{3}, \frac{\left|m^2-(\etaperp)^2\right|}{\left|2k/aH-2\right|} \right) ,\\
&\left|\frac{u^{;;}}{u}\right| \ll \left| m^2-(\etaperp)^2 \right|,
\end{aligned}
\label{Condition:adiabaticu}
\end{equation}
with $u\equiv 1- c_s^{-2}$ and independent on the fact whether $u$ is small or not. When $u\ll 1$ 
it is also possible to express these adiabatic conditions in terms of the turn parameter and its derivatives and we get approximately
\begin{equation}
\begin{aligned}
&\left|\frac{{\etaperp}^;}{\etaperp}\right| \ll \sqrt{|m^2-(\etaperp)^2|} ,\\
&\left|{\xiperp}^;\right| \ll |m^2-(\etaperp)^2|.
\end{aligned}
\label{Condition:adiabaticturnparameter}
\end{equation}
These conditions ensure that the timescale of the duration of the turn is much bigger than the period of the oscillations about the flat direction of the potential and therefore we can safely neglect the high frequency oscillations. \\

If the adiabatic conditions are satisfied it is possible to solve the effective single field equation (\ref{Equation:effectiveequationR}) for $\R$. We consider two cases, one for which we have a constant turn rate $\etaperp$ for all times and another in which we have a transient turn for which $c_s$ is assumed to go to 1 for the limits $\tau\rightarrow -\infty$ and $\tau\rightarrow 0$. For the case of a constant turn rate, the initial conditions for the low frequency solution $\R$ change in the second part of the sub-Hubble regime $z<m$ as explained in section \ref{Section:subhorizon}. Moreover, when $c_s<1$ the modes will cross the Hubble radius earlier because the sound horizon decreases with $c_s$ which can be seen in equation (\ref{Equation:effectiveequationR}). Taking both effects into account, we find the following power spectrum
\begin{equation}
\P_\R=\frac{H^2_\ast}{8\epsilon_\ast \pi^2 c_s M_P^2}\approx \frac{1}{c_s}\P_{\R 0},
\label{Equation:effectivepowerspectrumRconstantturn}
\end{equation}
which is about $\frac{1}{c_s}$ times the powerspectrum $\P_{\R 0}$ one would achieve when naively computing the truncated single field version.
Actually one should also take into account the change of $H_\ast$ and $\epsilon_\ast$, because if $c_s<1$ the modes will cross the Hubble radius earlier which means that $H_\ast$ becomes bigger and $\epsilon_\ast$ smaller compared to the value one gets in the truncated single field version. This means the amplitude of the power spectrum will be enlarged even more. This effect is negligible when $c_s$ is close to 1 and this approximation is therefore valid for a large mass hierarchy in case of a constant turn rate satisfying $(\etaperp)^2 \ll m^2$. For values of $c_s$ which are smaller we expect to have another small correction on top of this. In section \ref{section:predictions} we will discuss how the observables change. \\

Assuming a transient turn rate and the adiabatic conditions (\ref{Condition:adiabaticu}) we can interpret the contribution of the change in the speed of sound to the effective quadratic action for perturbations (\ref{Equation:effectiveaction}) as a small perturbation
$$
S_{\text{eff}} = S_0+S_{\text{per}}=M_P^2\int dNd^3\v{x}\ a^3\epsilon H\left[(\R^;)^2-\frac{1}{a^2H^2}(\partial_i \R)^2\right]-M_P^2\int dNd^3\v{x}\ a^3\epsilon H u (\R^;)^2,
$$
with $u\equiv 1- c_s^{-2}$. Using the in-in formalism, see for example \cite{LimA}, the change in the power spectrum to first order in $u$ is found  \cite{Achucarro:2012fd} to be
\begin{equation}
\frac{\Delta \P_\R}{\P_\R}(k)=k\int_{-\infty}^0d\tau\ u(\tau) \sin(2k\tau),
\label{Equation:effectivepowerspectrumR}
\end{equation}
This means in case of a transient turn we get a slight scale-dependent power spectrum as in canonical single field inflation but with features on top of that.


\subsubsection{Case 2: full solution to the perturbation equations using the SRST approximation}
We now solve the perturbation equations in the transition regime while assuming the first order slow-roll and slow-turn approximation, see Table \ref{Table:SRSTapproximation}. We can diagonalize the perturbation equations (\ref{Equation:perturbation_q_slowroll_twofields}) which become in this approximation
\begin{equation}
\begin{pmatrix}
 D_\tau^2 + k^2+a^2H^2\left(-2+\epsilon +m^2_-\right)  & 0 \\
 0  & D_\tau^2 + k^2 +a^2H^2\left(-2+\epsilon +m^2_+\right)  \end{pmatrix} \begin{pmatrix} u^T   \\ u^N  \end{pmatrix}  =0,
\label{Equation:transitionqnoturns}
\end{equation}
by rotating the fields to the new fields
$$
u^I(\tau)\equiv (U^{-1})^I_J q^J(\tau),
$$
where the rotation matrix is given by
$$
U=\begin{pmatrix}
 \cos\theta  &  -\sin\theta \\
 \sin\theta  &  \cos\theta \end{pmatrix}, \quad \tan2\theta=\frac{2\etaperp(3-\epsilon)}{(\etaperp)^2-m^2}.
$$
In Appendix \ref{Appendix:analytical approximations} the derivation of equation (\ref{Equation:transitionqnoturns}) can be found. In the studies TG and PT this equation is solved by Hankel functions, which are solutions to the Bessel equation. However, this equation is only equivalent to the Bessel equation if $D_\tau^2=\partial_\tau^2$, which is true if there are no turns during the transition period. Therefore in general we actually should start from equation (\ref{Equation:eomqtilde}) and diagonalize it such that we can solve it with Hankel functions. It turns out that the turn rate must be very small in order to make this identification. In Appendix \ref{Appendix:analytical approximations} we derive the following constraints on the slow-roll and turn parameters which are necessary to solve the perturbation equations in the transition regime by Hankel functions
\begin{align}
& \epsilon, \etaperp\approx const \quad \text{and} \quad |\etaperp|\lesssim 0.04.
\label{Condition:etaperptransition}
\end{align}
We assumed $N_\pm=N_\ast\pm2.3$ as the beginning and the end of the transition regime. If these conditions are fulfilled the solutions of the equations of motion for $u^I$ are given by a linear combination of the Hankel functions of the first and second kind
\begin{equation}
u^I_\alpha=\sqrt{z}H^{(1)}_{\nu^I}(z) A^I_\alpha+\sqrt{z}H^{(2)}_{\nu^I}(z) B^I_\alpha,
\label{Equation:Hankel_u}
\end{equation}
where the constants $\nu^I$ are given by
\begin{align*}
(\nu^T)^2&=\frac{1}{4}+\frac{2-\epsilon_\ast-m^2_-}{(1-\epsilon_\ast)^2}\approx\frac{9}{4}+3\epsilon_\ast+9\frac{(\etaperp)^2}{m^2},\\
(\nu^N)^2&=\frac{1}{4}+\frac{2-\epsilon_\ast-m^2_+}{(1-\epsilon_\ast)^2}\approx\frac{9}{4}+3\epsilon_\ast-(1+2\epsilon_\ast)m^2-9\frac{(\etaperp)^2}{m^2}.
\end{align*}
Here we assumed first order slow roll and $m^2\gg(\etaperp)^2$, where the latter assumption is motivated by the discussion in section \ref{Section:SRSTapproximation}, although we would rather expect from this discussion that $m^2>(\etaperp)^2$. We come back to this later.
The constants $A^I_\alpha$ and $B^I_\alpha$ in equation (\ref{Equation:Hankel_u}) can be found by matching the solution to the initial conditions for $q^I_\alpha$ when $z=z_-\gg1$. It turns out that the coefficients $A^I_\alpha$ in front of the Hankel function of the first kind become zero. To find the solution at the end of transition we can use the asymptotic behavior of the Hankel function of the second kind in the limit $z\ll 1$. We neglect the decaying solution $z^{\nu^T+1/2}$ and transforming back to $q^I_\alpha$ we get at the end of transition
\begin{align*}
q^T_\alpha(z_+) = \ &\sqrt{\frac{\pi}{4k}}\left(\cos^2\theta\delta^T_\alpha+\cos\theta\sin\theta\delta^N_\alpha\right)f(\nu^T)\left(\frac{2}{z_+}\right)^{\nu^T-1/2}\\ \ -&\sqrt{\frac{\pi}{4k}}\left(\cos\theta\sin\theta\delta^N_\alpha-\sin^2\theta\delta^T_\alpha\right)\left(f(\nu^N)\left(\frac{2}{z_+}\right)^{\nu^N-1/2}+ g(\nu^N)\left(\frac{2}{z_+}\right)^{-\nu^N-1/2}\right),\\
q^N_\alpha(z_+)= \ &\sqrt{\frac{\pi}{4k}}\left(\cos^2\theta\delta^N_\alpha-\cos\theta\sin\theta\delta^T_\alpha\right)\left(f(\nu^N)\left(\frac{2}{z_+}\right)^{\nu^N-1/2}+g(\nu^N)\left(\frac{2}{z_+}\right)^{-\nu^N-1/2}\right)\\ \ +&\sqrt{\frac{\pi}{4k}}\left(\cos\theta\sin\theta\delta^T_\alpha+\sin^2\theta\delta^N_\alpha\right)f(\nu^T)\left(\frac{2}{z_+}\right)^{\nu^T-1/2} .
\end{align*}
Here $\theta$ is defined at the beginning of this subsection and the functions $f$ and $g$ are introduced to shorten the expressions and given by
\begin{align*}
&f(\nu^I)=e^{-i\nu^I\pi/2-i\lambda^I+i\pi/4}\frac{i\sqrt{2}\Gamma(\nu^I)}{\pi},\\
&g(\nu^I)=e^{-i\nu^I\pi/2-i\lambda^I+i\pi/4}\frac{\sqrt{2}(1-i\cot(\pi\nu^N))}{\Gamma(1+\nu^N)}
\end{align*}
Now there are multiple possibilities. If $m^2\ll1$ then $\nu^N\sim 3/2$ and we can neglect the decaying solution $z^{\nu^N+1/2}$ as well, but the first term is obviously a growing solution and this might produce a large spectrum for the isocurvature modes. If $m^2\gtrsim \frac{9}{4}$ then the real part of $\nu^N$ becomes zero and both terms $z^{\pm\nu^N+1/2}$ go like $z^{1/2}$ and these modes decay like massive modes. We consider both cases separately.\\

If $m^2\gtrsim \frac{9}{4}\gg \etaperp$ we can approximate $\cos\theta =1$, $\sin\theta=\frac{-\etaperp (3-\epsilon)}{m^2}$ and $\nu^T=\frac{3}{2}+\epsilon_\ast$ we find that the curvature perturbation is completely frozen out at $z_+$ such that we can derive the following power spectra at the end of inflation
\begin{equation}
\begin{aligned}
&\P_\R\approx\frac{H^2_\ast}{8\epsilon_\ast \pi^2  M_P^2}\left(1+2(C-1)\epsilon_\ast\right)\left(1+\frac{9(\etaperp_\ast)^2}{m^4}\right),\\
&\P_\S\approx\frac{H^2_\ast}{8\epsilon_\ast \pi^2  M_P^2}\left(1+2(C-1)\epsilon_\ast\right)\left(\frac{9(\etaperp_\ast)^2}{m^4}\right).
\end{aligned}
\label{Equation:powerspectrumR_SRSTapproximation}
\end{equation}
In this limit, the power spectrum of the isocurvature modes can be neglected and the power spectrum of the curvature modes are indistinguishable from canonical single field inflation because the turn rate is extremely small compared to $m$. \\

If $m^2\ll1$ we can approximate 
$\nu^T=\nu^N=\frac{3}{2}+\epsilon_\ast$ and we find both curvature and isocurvature perturbations which evolve independently. Therefore we both get a spectrum of curvature modes and isocurvature modes at the end of the transition regime
\begin{align*}
&\P_\R\approx\frac{H^2_\ast}{8\epsilon_\ast \pi^2  M_P^2}\left(1+2(C-1)\epsilon_\ast\right),\\
&\P_\S\approx\frac{H^2_\ast}{8\epsilon_\ast \pi^2  M_P^2}\left(1+2(C-1)\epsilon_\ast\right).
\end{align*}
What happens after $z_+$ depends on what value the turn rate will take in the super-Hubble regime and on the physics after inflation. This makes it very hard to make any predictions of observables. The fact that the curvature modes and isocurvature modes evolve independently comes from the assumption $m^2\gg(\etaperp)^2$, which means in this case that we have put $\etaperp$ to zero during transition. This assumption should be weakened to $m^2>(\etaperp)^2$ and we expect that the biggest effects of the turn appear when $m^2\gtrsim(\etaperp)^2$. However, since we still expect to get a sizable spectrum of isocurvature modes which we do not know how to relate to experiment, we do not perform the calculation.

\subsection{Overview}
\label{Section:overview}
We summarize the approximation schemes and their regime of validity in Table \ref{Table:overviewapproximations}. In the papers we have been studying, two different approximation schemes are being used. In the studies TG, PT and GW the perturbations equations are solved by assuming both the slow-roll approximation and the slow-turn approximation (SRST approximation). In the studies CT and AP the slow-roll approximation and a large mass hierarchy is assumed. We found that the computations performed in the corresponding studies are valid if the conditions in the second column of the Table are satisfied. Moreover, we computed the power spectrum of the curvature mode at the end of inflation which are shown in the third column. \\

Let us discuss each study of multi-field inflation separately to see how they fit in this overview. Pioneering studies of curvature perturbations and isocurvature perturbations in multiple field inflation were performed by GW and TG \cite{Gordon:2000hv, GrootNibbelink:2000qt}. In these papers the general formalism is developed to describe the evolution of both types of perturbations. The importance of $\etaperp$ as carrier of multi-field effects is recognized as it turns on the coupling between the curvature and isocurvature perturbations. However, only super-Hubble effects are taken into account and moreover, the turn parameter is treated as a slow-roll parameter. A few years later it was realized that the role of the turns can be more important and lead to observable features which can be tested against the data. In the studies of AP, CT and PT \cite{Cremonini:2010ua, Achucarro:2010da, Peterson:2010np} the effects of turns during Hubble radius crossing are taken in account and analytical expressions for the power spectra at the end of inflation are derived. In the extensive treatment provided by PT \cite{Peterson:2010np} the slow-turn approximation is introduced and the equations are solved to second order in the slow-roll and turn parameters. One of their conclusions is, however, that a large turn rate results in a decay of the isocurvature modes. This can be understood since their approximation scheme is actually only valid in case of negligible turns during the transition regime and during the super-Hubble regime the isocurvature modes will indeed decay if the turn rate is large for a sufficient period of time according to equation (\ref{Equation:curvaturesuperhorizonSRST}). However, it is not so generic that the turn rate is negligible until the end of the transition regime whereafter it becomes large. If the turn rate became nonzero during transition, then we have seen that we get different effects. In the papers \cite{Cremonini:2010ua, Achucarro:2010jv, Achucarro:2012sm, Achucarro:2012yr} the slow-roll approximation and a large mass hierarchy is assumed which results in an effective single field description with a reduced speed of sound. The results of CT apply for constant turn rates and the results of AP are more general and in particular valid for transient turn rates, but constant turn rates can also described within their formalism. \\

Finally, we will elaborate a bit more about the predictions of observables within the two approximation schemes. In case of the SRST approximation we have seen that we either get the same predictions for canonical single field inflation or we don't know how to compute the power spectrum at the end of inflation and after inflation because of the presence of isocurvature modes. Considering that non-Gaussianities of the local form and isocurvature modes are ruled out by observations we expect that the second case is ruled out by the data. But of course one can always argue that the turn rate can stay very small for all times and the isocurvature modes might decay after inflation. In the case of the slow-roll approximation and a large mass hierarchy we have studied the constant turn $\etaperp=const$ and the transient turn for which $\etaperp$ goes to zero at early and late times. For a constant turn we have found a rescaling of the power spectrum of scalar perturbations. Since the power spectrum for tensor modes is unchanged this means we get the following values for the observables
$$
n_s-1\approx-2\epsilon+\eta_H, \quad \quad n_t\approx-2\epsilon, \quad \quad r=16\epsilon_\ast c_s.
$$
This means a constant turn changes the consistency relation between $r$ and $n_t$. Moreover, another observational consequence is that the bispectrum gets renormalized
$$
B_\R = \frac{1}{c_s^2}B_{\R 0},
$$
which means that the relative amplitude of the power spectrum and bispectrum changes with respect to canonical single field inflation\footnote{Canonical single field inflation is defined as single field slow-roll inflation with canonical kinetic terms, minimally coupled to gravity and with Bunch-Davies initial conditions.}. Arguably, the largest observable effects are created by a transient turn rate during or very close to the transition regime. The change in the power spectrum is given by equation (\ref{Equation:effectivepowerspectrumR}) which tells us that a transient turn will result in oscillations in the power spectrum. If the width of the oscillation is small compared to the size of the observable power spectrum, then the power spectrum can be described by a canonical single field power spectrum with oscillations. In this case the amplitude, spectral indices and tensor-to-scalar ratio are unchanged, but new parameters are needed to describe the oscillations. If the width of the oscillation is large compared to the size of the observable power spectrum, then this might change the scalar spectral index and the amplitude. Moreover, it has been shown \cite{Achucarro:2012fd} that the bispectrum will have correlated oscillations and therefore a joint analysis of both spectra could reveal the signatures.



\begin{table}[h]
    \begin{tabular}{ | p{4cm} | p{7.2cm} | p{6cm} |}
    \hline
    Approximation scheme            & Regime of validity         &  Predicted power spectrum    \\
    \hline
                                    &                       &                               \\
    The Slow-Roll-Slow-Turn (SRST) approximation \newline\newline
    Used in: \newline TG, PT and GW
    \newline \cite{Gordon:2000hv, GrootNibbelink:2000qt, Peterson:2010np}                      &  During transition:
                                        $$\epsilon, \etaperp\approx const \quad \text{and}
                                        \quad |\etaperp|\lesssim 0.04$$                     &   If $m^2\gtrsim \frac{9}{4}$ then:
                                                                                                $$
                                                                                                \P_\R\approx\frac{H^2_\ast}{8\epsilon_\ast \pi^2  M_P^2}\left(1+\frac{9(\etaperp_\ast)^2}{m^4}\right).
                                                                                                $$
                                                                                                \newline In other cases it is unknown because of the presence of isocurvature modes. \\
                                    &                       &                               \\
    \hline
                                    &                       &                               \\
    The slow-roll approximation
    and a large mass hierarchy
    $m^2\gg1$        \newline\newline
    Used in: \newline CT and AP
    \cite{Cremonini:2010ua, Achucarro:2010da}
                                    & One needs $m\geq 10$ and the adiabatic conditions:
                                   $$
                                   \left|\frac{{\etaperp}^;}{\etaperp}\right| \ll \sqrt{|m^2-(\etaperp)^2|},
                                   $$
                                   $$
                                   \left|{\xiperp}^;\right| \ll |m^2-(\etaperp)^2|.
                                   $$
                                    Or equivalently:
                                   $$
                                   \left|\frac{u^;}{u}\right| \ll \min\left(\frac{|m^2-(\etaperp)^2|}{3}, \frac{\left|m^2-(\etaperp)^2\right|}{\left|2k/aH-2\right|} \right),
                                   $$
                                   $$
                                   \left|\frac{u^{;;}}{u}\right| \ll \left| m^2-(\etaperp)^2 \right|.
                                   $$                                                               &  In the case of a constant turn rate:
                                                                                                           $$
                                                                                                            \P_\R=\frac{H^2_\ast}{8\epsilon_\ast \pi^2 c_s M_P^2}\approx \frac{1}{c_s}\P_{\R 0}.
                                                                                                            $$
                                                                                                        \newline  In the case of a transient turn which vanishes at early and late times:
                                                                                                        $$
                                                                                                        \frac{\Delta \P_\R}{\P_\R}(k)=k\int_{-\infty}^0d\tau\ u(\tau) \sin(2k\tau).
                                                                                                        $$
                                                                                                          \\
                                    &                       &                               \\
    \hline

     \end{tabular}
\caption{We summarize the approximation schemes and their regime of validity. In the papers we include in our overview two different approximation schemes are being used, these are given in the first row. In the studies TG, PT and GW (see Table \ref{Table:literature}) the perturbations equations are solved by assuming the SRST approximation (see section \ref{Section:SRSTapproximation}). In the studies CT and AP the slow-roll approximation and a large mass hierarchy is assumed. In the second column we give the conditions to render the computations performed in the corresponding papers valid. In the third column we give the analytical predictions of the power spectrum of the curvature mode at the end of inflation. Here $m^2\equiv M^2_{NN}/H^2$ as defined in section \ref{section:Perturbation equations and slow-roll and turns}.}
    \label{Table:overviewapproximations}
\end{table}

\section{Conclusion and Discussion}
\label{Section:conclusiondiscussionproject1}
In this chapter we have studied different papers of multiple field inflation in literature given in Table \ref{Table:literature}. The notation and definitions used in these studies differ and therefore we provide dictionaries to translate between the papers. These dictionaries can be found in Table \ref{Dictionary:slowroll}, \ref{Table:dictionarysetup}, \ref{Dictionary:perturbations}, and \ref{Dictionary:two-field}. In particular the definition of the slow-roll and the turn parameters vary among the papers. We define a turn as a deviation from a geodesic of the inflaton in field space, which will couple the inflaton to the other degrees of freedom. This means the turn parameter is very important to describe the multi-field effects and should be considered separately from the slow-roll parameters. However, some papers mix up the slow-roll parameters and the turn parameters as can be seen in Table \ref{Dictionary:slowroll}. \\

Moreover, we studied the approximation schemes used in the papers which are used to solve the perturbation equations. We find that only two different approximation schemes are used and we study their regime of validity and the analytical predictions of the power spectrum of the curvature mode. The overview of the approximation schemes can be found in Table \ref{Table:overviewapproximations}. In the studies TG, PT and GW the perturbations equations are solved by assuming the Slow-Roll-Slow-Turn (SRST) approximation (see section \ref{Section:SRSTapproximation}). In the studies CT and AP the slow-roll approximation and a large mass hierarchy is assumed. The first approximation scheme turns out to be valid if the turn rate is negligible during the transition regime. This results in a power spectrum for the curvature mode which is either indistinguishable from single field inflation or unknown at the end of inflation due to the presence of the isocurvature modes. The second case is most likely ruled out by the current experimental data. This means that this approximation scheme is very limited in the sense that it can only be used to describe multi-field inflationary models which are effectively canonical single field models. The second approximation scheme is valid when the isocurvature mass is heavy enough and when certain adiabatic conditions on the turn rate are satisfied. For a constant turn rate this results in a renormalization of the power spectrum. In case of a transient turn rate this leads to oscillations in the power spectrum. Both cases influence also other observables. This means that this approximation scheme can be used to study multi-field inflationary models which might have distinct signatures compared to canonical single field models.\\

The two approximation schemes do not cover the full range of possible multi-field models. In particular, it would be interesting to get more understanding of the intermediate case with semi heavy fields $m\gtrsim 9/4$ and the possibility of turns. In this case the isocurvature modes will probably decay on super-Hubble scales which means it is possible to make predictions which can be related to experiment. First of all one could look for a method to solve the perturbation equations analytically. Moreover, it is possible to improve our understanding by consulting numerical methods. \\

In this overview we focused mainly on the case of two fields. It would be interesting to generalize the second approximation scheme to more than two fields where one field is light and the others much heavier, for two heavy fields see \cite{Cespedes:2013rda}. In this case one could try to integrate out all the heavy fields which we expect to induce similar corrections to the final single field version as in the two field case. These corrections will presumably only depend on additional turn rates and masses coming from the additional fields. \\

Finally, we should perform numerical studies of concrete models for multi-field inflation in order to check the analytical results and to improve our understanding of multi-field inflation under less restrictive conditions. We already made a start with this and describe a numerical study in the next chapter.

\chapter{Numerical Study of Concrete Models in Literature}
\label{Chapter:dataanalysis}
\section{Introduction}
\label{Section: data analysis introduction}
We have seen in chapter \ref{Chapter:inflation} that the latest precision measurements of the CMB put inflation on a firm footing. Canonical single field inflation\footnote{Canonical single field inflation is defined as single field slow-roll inflation with canonical kinetic terms, minimally coupled to gravity and with Bunch-Davies initial conditions.} can fit the current CMB data very well, however the precise microphysical origin of inflation remains a mystery. Single-field models could be fundamental theories but from a theoretical point of view they have their limitations, because they have no ultraviolet completion and, with the possible exception of Higgs inflation, no connection with the rest of physics \cite{Baumann:2014nda}. We are interested in the case that single field inflation is an effective description of a more fundamental theory containing multiple scalar fields. The current precision data puts strong constraints on the possible interactions terms in a multi-field model. However if one can find something in the data which is distinct from canonical single field models, this indicates the presence of new physics during inflation and this might a big step towards understanding physics of the very early universe. Therefore we would like to understand the falsifiable observational signatures of multiple field inflation in the current and future data. We have seen in chapter \ref{Chapter:inflation} that if the detection of B-mode polarization by BICEP2 is a result of a tensor-to-scalar ratio of order $r\sim O(0.1)$ then this suggests a high Hubble scale of inflation $H\sim10^{13} GeV$ and a super-planckian excursion of the inflaton in field space. This latter implication means that the dynamics of the inflaton is more sensitive to features in the potential or more generally to any mismatch between the geodesics in field space and the valley of the potential. In chapter \ref{Chapter:multifieldinflation} we have seen that for multi-field models where one field is light and the others much heavier compared to the Hubble scale of inflation it is possible to integrate out the heavy fields to get an effective single field description under the conditions (\ref{Condition:lowerboundm}) and (\ref{Condition:adiabaticturnparameter}). When there are sudden turns in the trajectory this will result in oscillatory features in the power spectrum and correlated features the bispectrum \cite{Achucarro:2012fd}. In \cite{Achucarro:2013cva} a search for these correlated oscillations in the Planck CMB data has been performed which shows that there are hints of oscillatory features in the data. However the full data on the bispectrum has not been released yet. Interestingly, this shows that the current data might already be precise enough to detect signatures of new physics. This motivates us to continue the study of multi-field inflation and much can be learned by studying concrete models for multi-field inflation. It is highly interesting to understand what kind of turns appear in these models and how they will influence the observables. At the same time these models can be used to check if our theoretical understanding is correct. In various models of inflation in supergravity, inflation is embedded in a multiple scalar field theory with non-canonical terms. Because of the presence of non-canonical kinetic terms the trajectory of the inflaton in field space could easily be curved. We study some models from recent papers in the literature \cite{Ferrara:2014ima, Ferrara:2014fqa, Kallosh:2014qta, Ellis:2014gxa}. In general, once the heavy fields are stabilized, the possibility of presence of multi-field effects due to turns is not taken into account in these papers. We would like to study whether their 'na\"ive' single field truncation is valid within the current observational constraints or that we might be able to detect the presence of the additional fields in these models. Furthermore we study another model given in \cite{Dong:2010in} which is a toy model for certain aspects of axion monodromy for which we would like to study the same question.

\section{Characterization of the models}
\label{Section: characterization of the models}
In this section we give a description of the models we have been studying. The specifications of
\cite{Ferrara:2014ima, Ferrara:2014fqa, Kallosh:2014qta, Ellis:2014gxa} are different from \cite{Dong:2010in} and therefore we discuss them separately.
\subsection{No-scale supergravity models}
\label{Subsection: no-scale supergravity models}
The papers \cite{Ferrara:2014ima, Ferrara:2014fqa, Kallosh:2014qta, Ellis:2014gxa} all provide examples of no-scale supergravity models \cite{Cremmer:1983bf} which have a K\"ahler potential of the form
$$
K=-3\ln(T+\bar{T}) + \ldots,
$$
and a superpotential $W$ satisfying
$$
\partial_T W =0,
$$
which imply a vanishing cosmological constant. This can be seen by deriving the scalar potential from the K\"ahler potential and the superpotential (see for example \cite{Achucarro:2007qa}) through
\begin{align*}
V &= {e^K}\left( {{K^{T\bar T}}\left( {{\partial _T}W + W{\partial _T}K} \right)\left( {{\partial _{\bar T}}\bar W + \bar W{\partial _{\bar T}}K} \right) - 3{{\left| W \right|}^2} +  \ldots } \right) \\
&= {e^K}\left( {\frac{{{{\left( {T + \bar T} \right)}^2}}}{3}\left( {\frac{{ - 3W}}{{T + \bar T}}} \right)\left( {\frac{{ - 3\bar W}}{{T + \bar T}}} \right) - 3{{\left| W \right|}^2} +  \ldots } \right)\\
&= {e^K}\left( {0 +  \ldots } \right),
\end{align*}
such that we indeed have a vanishing cosmological constant\footnote{Be aware of the fact that $T$ denotes a matter field and not the tangent vector used in the previous chapter, moreover, $\partial_T$ denotes a partial derivative with respect to the field $T$, i.e. $\partial_T \equiv \partial/\partial T$.}. Terms with additional matter fields are then added to the K\"ahler potential which determine the final scalar potential. The overall exponential factor $e^K$ in front of the potential will result in a tendency to minimize $K$ and therefore the additional fields get probably fixed at zero which means their influence on the inflaton dynamics will be negligible. Of course this depends on the precise form of these extra terms and it should be checked that their influence can be ignored for each model. The kinetic term for the two components of the complex field $T$ is given by
\begin{equation}
{K_{T\bar T}}\partial T\partial \bar T = \frac{3}{{{{\left( {T + \overline T } \right)}^2}}}\partial T\partial \bar T = \frac{1}{2}\left( {\partial {\psi ^2} + {e^{ \mp 2\sqrt {2/3} \psi/M_p }}\partial {\phi ^2}} \right),
\label{Equation: kinetic term Cecotti}
\end{equation}
where we split $T$ in a real and imaginary part because the kinetic term only depends on the real part of $T$
$$
T =  e^{ \pm \sqrt {2/3} \psi/M_p }M_p + i\sqrt {2/3} \phi.
$$
The exponent and the factor $\sqrt {2/3}$ are introduced in order to make the kinetic term belonging to $\psi$ canonical. We see a non-canonical kinetic term $G_{\phi\phi}= e^{ \mp 2\sqrt {2/3} \psi/M_p }$ arises naturally from no-scale supergravity models. \\

Let us now see what matter fields are added to the K\"ahler potential and what kind of superpotentials are used in the papers. In \cite{Ferrara:2014ima, Ferrara:2014fqa, Kallosh:2014qta} the `Cecotti model' \cite{Cecotti:1987sa} and `imaginary Starobinksy model' are studied for which the K\"ahler potential and superpotential are given by
\begin{align*}
&K=-3\ln(T+\bar{T}-X\bar{X}),\\
&W= 3MX(T-f),
\end{align*}
where it is assumed that the field $X$ is stabilized at $X=0$. However, this is not shown explicitly but a reference to \cite{Kallosh:2013lkr} is given in which is explained how this can be realized. We expect that the value of $f$ plays a role here, but after putting $X=0$ its precise value is irrelevant and by a rescaling of $T$ we can put $f$ to one, which we will do later on. Under this assumption the kinetic part of the Lagrangian is given by (\ref{Equation: kinetic term Cecotti}) and the potential energy density becomes
$$
V=3M^2\frac{|T-f|^2}{(T+\bar{T})^2}=\frac{3}{4}M^2 M_p^2\left(1-f e^{\mp\sqrt {2/3} \psi/M_p}\right)^2+\frac{1}{2}M^2e^{\mp2\sqrt {2/3} \psi/M_p}\phi^2.
$$
If the imaginary part of $T$ is zero along the inflationary trajectory then this model coincides with the Starobinsky model. If inflation happens along the imaginary part of $T$ instead, one will get a chaotic inflationary model. This is exactly what is aimed at in these papers. In order to realize this the real part of $T$ should be stabilized and therefore the K\"ahler potential is modified. This is done in several ways and the K\"ahler potential together with the resulting Lagrangians are listed in Table \ref{Table:Overview no-scale models}. \\

In \cite{Ellis:2014gxa} the `no-scale inflationary model' is studied with the following K\"ahler potential and superpotential
\begin{align*}
&K=-3\ln(T+\bar{T})+\frac{X\bar{X}}{(T+\bar{T})^3},\\
&W= 3MX(T-f),
\end{align*}
where by studying some slices of the potential it is concluded that the field $X$ is stabilized at $X=0$. The function $f$ presumably plays a role here, where for small enough values $f\leq1$ the exponential factor in front of the potential constraints the value of $X$, namely $e^K \sim e^{|X|^2/(2f)^3}$. In this paper several stabilization terms are considered as parameterized by $\theta$ which represents a line in the $(\operatorname{Re} T, \operatorname{Im} T))$-plane. The K\"ahler potential is given in Table \ref{Table:Overview no-scale models}. We do not show the resulting Lagrangian density, because we haven't studied this model in more detail yet.
\pagestyle{plain}

\begin{sidewaystable}

    \begin{tabular}{ | p{3cm} |  p{7cm} | p{15cm} | }
    \hline
    Name & K\"ahler potential  & Scalar potential and kinetic term\\
    \hline
    & & \\
    The modified Cecotti model \cite{Kallosh:2014qta} & $K=-3\ln(T+\bar{T}-X\bar{X}+\alpha X\bar{X}(T+\bar{T}))$ & $V=\frac{\frac{3}{4}M^2M_p^2\left(1-f e^{\mp\sqrt {2/3} \psi/M_p}\right)^2+\frac{1}{2}M^2e^{\mp2\sqrt {2/3} \psi/M_p}\phi^2}{1-2\alpha e^{\pm\sqrt{2/3}\psi/M_p}}$ \\
    & & \\
    & & $\L_{kin}= \frac{1}{2}\left( \partial \psi ^2 + e^{ \mp 2\sqrt {2/3} \psi/M_p }\partial \phi ^2\right)$\\
    & & \\
    \hline
    & & \\
    The imaginary supersymmetric Starobinsky model \cite{Ferrara:2014ima} & $K=-3\ln(T+\bar{T}-X\bar{X}+m(T+\bar{T})^n))$ & $V=\frac{\frac{3}{4}M^2 M_p^2\left(1-f e^{\mp\sqrt {2/3} \psi/M_p}\right)^2+\frac{1}{2}M^2e^{\mp2\sqrt {2/3} \psi/M_p}\phi^2}{\left(1+2^{n-1} m e^{(n-1)\sqrt {2/3} \psi/M_p}\right)^2}$ \\
    & & \\
    & & $\L_{kin}= \frac{1+mn(2 e^{\pm\sqrt {2/3} \psi/M_p})^{n-2}\left[2(3-n)e^{\pm\sqrt {2/3} \psi/M_p}+m(2e^{\pm\sqrt {2/3} \psi/M_p})^n\right]}{2(1+m(2e^{\pm\sqrt {2/3} \psi/M_p})^{n-1})^2}\left( \partial \psi ^2 + e^{ \mp 2\sqrt {2/3} \psi/M_p }\partial \phi ^2\right)$\\
    & & \\
    \hline
    & & \\
    \multirow{ 6}{3cm}{The $(R+R^2+R^4)$-extension of the imaginary supersymmetric Starobinsky model \cite{Ferrara:2014fqa}} & $K=-3\ln(T+\bar{T}-X\bar{X}-Q\bar{X}-\bar{Q}X-\frac{z}{12}X\bar{X}B\bar{B}))$ & $V=\frac{3\cosh^2\left[\frac{1}{3}\cosh^{-1}\left(1+9zM^2\left(\frac{2}{3}\phi^2+M_p^2( e^{\pm\sqrt {2/3} \psi/M_p}-f)^2\right)\right)\right]}{4z e^{\pm 2\sqrt {2/3} \psi/M_p}(1+m(2 e^{\pm\sqrt {2/3} \psi/M_p})^{n-1})^2}$ \\
    &  & \\
    &  & $\quad\quad-\frac{3\cosh\left[\frac{1}{3}\cosh^{-1}\left(1+9zM^2\left(\frac{2}{3}\phi^2+M_p^2( e^{\pm\sqrt {2/3} \psi/M_p}-f)^2\right)\right)\right]}{4z e^{\pm 2\sqrt {2/3} \psi/M_p}(1+m(2 e^{\pm\sqrt {2/3} \psi/M_p})^{n-1})^2}$\\
    &  & \\
    &  & \\
    & & \\
    & & $\L_{kin}= \frac{1+mn(2 e^{\pm\sqrt {2/3} \psi/M_p})^{n-2}\left[2(3-n)e^{\pm\sqrt {2/3} \psi/M_p}+m(2e^{\pm\sqrt {2/3} \psi/M_p})^n\right]}{2(1+m(2e^{\pm\sqrt {2/3} \psi/M_p})^{n-1})^2}\left( \partial \psi ^2 + e^{ \mp 2\sqrt {2/3} \psi/M_p }\partial \phi ^2\right)$\\
    & & \\
    \hline
    & & \\
    The no-scale inflationary model \cite{Ellis:2014gxa} & $K=-3\ln(T+\bar{T}-c(\cos\theta(T+\bar{T}-1)-i\sin\theta(T-\bar{T}))^4)+\frac{X\bar{X}}{(T+\bar{T})^3}$ &
    N/A \\
    & & \\
    & & \\
    & & \\
    \hline
    \end{tabular}
\caption{In this table we present an overview of the no-scale supergravity models of inflation coming from the papers \cite{Ferrara:2014ima, Ferrara:2014fqa, Kallosh:2014qta, Ellis:2014gxa}. The names in the first column come from the papers themselves. We give the K\"ahler potential in the second column. The superpotential is the same for each model $W= 3MX(T-f)$ and therefore it is not given in this table. Moreover from the transition of the second column to the third we have used $T = {e^{ \pm \sqrt {2/3} \psi }} + i\sqrt {2/3} \phi$ without explicitly stating this in the table. In the third column we show the resulting potential density and kinetic energy density which together determine the full Lagrangian density of the inflationary model. The only exception is resulting Lagrangian density for the no-scale inflationary model, because we haven't studied this model in more detail yet.}
\label{Table:Overview no-scale models}
\end{sidewaystable}

\subsection{A toy model in which a heavy field flattens the potential}
\label{Subsection: toy model from axion monodromy}
\begin{figure}[h]
  \begin{center}
    \includegraphics[width=0.38\textwidth,natwidth=360,natheight=317]{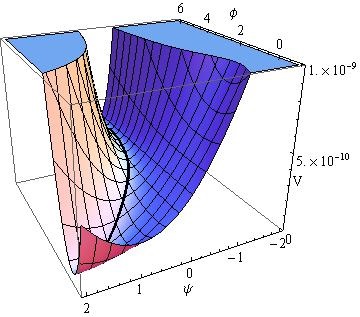}
  \end{center}
  \caption{The potential as given in equation (\ref{Equation:potential axion monodromy}). The potential flattens in the regime $\phi\gg M/g$.}
  \label{Figure:axionpotential}
\end{figure}

In \cite{Dong:2010in} it is explained how interaction of heavy fields with the inflaton can flatten the potential.
The motivation comes from string theory which provides many heavy scalar fields and some light scalar fields which can play the role of the inflaton. An example of a light field is the axion and the flattening effect can be realized in axion monodromy inflation. This is studied in this paper in section 3. As a toy model to explain the flattening effect they use the following potential
\begin{equation}
V=g^2\phi^2\psi^2+M^2(\psi-\psi_0)^2.
\label{Equation:potential axion monodromy}
\end{equation}
The idea is that the heavy fields $\psi$ will try to minimize the potential at any given value of the light field $\phi$. For large values of $\phi$ this will eat the first term and the potential gets flattened. Computing the valley of the potential assuming $\psi$ is massive enough and that it takes the value $\bar{\psi}(\phi)$ you get the following effective potential
$$
\bar{\psi} = \frac{M^2\psi_0}{g^2\phi^2+M^2} \quad \rightarrow \quad V(\phi,\bar{\psi})=\frac{g^2\phi^2}{g^2\phi^2+M^2}M^2\psi_0^2.
$$
There are two different regimes of this effective potential
\begin{align*}
&\phi\gg M/g: \quad V\rightarrow M^2\psi_0^2, \quad H^2\rightarrow \frac{M^2\psi_0^2}{3M_p^2}, \quad \bar{\psi} \rightarrow 0,\\
&\phi\ll M/g: \quad V\rightarrow g^2\psi_0^2 \phi^2, \quad H^2\rightarrow \frac{g^2\psi_0^2}{3M_p^2}\phi^2, \quad \bar{\psi} \rightarrow \psi_0.
\end{align*}
Indeed the potential gets very flat in the regime $\phi\gg M/g$ as is visible in Figure \ref{Figure:axionpotential}.
The bending of the potential from one regime into the other regime takes place around the value $\phi\sim M/g$.\\

For the kinetic terms the following form is assumed
$$
\L_{kin}=\frac{1}{2}G_{\phi\phi}(\psi)(\partial_\mu\phi)^2+\frac{1}{2}(\partial_\mu\psi)^2,
$$
which also captures the kinetic terms studied in subsection \ref{Subsection: no-scale supergravity models}. Two specific examples mentioned in the introduction are canonical kinetic terms $G_{\phi\phi}(\psi)=1$ and one example of non-canonical kinetic terms $G_{\phi\phi}(\psi)=\frac{\psi}{M_P}$ which is actually introduced to give an example in which there is steepening instead of flattening. The non-canonical kinetic term comes without motivation and is furthermore not always positive so one should be careful while studying this example. For now we do not worry about the physical relevance because we use it as a toy model to see whether there are detectable effects coming from the heavy field.

\section{Why do we expect to see something and what are the observational consequences?}
\label{Section: why do we expect to see something and what are the observational consequences?}
\pagestyle{headings}
Before we continue describing our numerical studies we first outline our expectations. As we have seen in the previous chapter any mismatch between the geodesics in field space and the minimum of the potential leads to turns in the trajectory. In the models of inflation described in sections \ref{Subsection: no-scale supergravity models} and \ref{Subsection: toy model from axion monodromy} the kinetic terms are of the following form
\begin{equation}
\L_{kin}=\frac{1}{2}G_{\phi\phi}(\psi)(\partial_\mu\phi)^2+\frac{1}{2}(\partial_\mu\psi)^2.
\label{Equation: Kinetic terms for numerical study}
\end{equation}
The deviation from a geodesic can occur when the kinetic terms are non-canonical, i.e. when $G_{\phi\phi}(\psi)\neq 1$ and really depends on $\psi$. In the models from section \ref{Subsection: no-scale supergravity models} the $\psi$-field gets stabilized at a particular value, in the model from section \ref{Subsection: toy model from axion monodromy} we expect this to happen in the regime $\phi \gg M/g$.  To see intuitively why we expect turns in case of a stabilized $\psi$-field, suppose we have $G_{\phi\phi}(\psi)=\frac{\psi^2}{M_P^2}$. This corresponds to an interpretation of the $\psi$-field as a radial coordinate and the $\phi$-field as an angular coordinate. If the $\psi$-field takes a constant value then it means the radius is constant and inflation happens only along the flat angular direction. We expect that the radial-angular intuition also holds for more general forms of $G_{\phi\phi}(\psi)$ and therefore we expect that the inflaton will have a constant turn rate throughout its trajectory if $\psi$ is stabilized, assuming $\epsilon$ is constant.\\

Moreover in the model discussed in section \ref{Subsection: toy model from axion monodromy} we also expect a turn due to the bend in the potential when the inflaton goes from the flat regime to the second regime. In this case we expect a turn rate which varies in time.

\subsection{Observational consequences}
\label{Subsection: observational consequences}
We will discuss the observational consequences and detectability for both expected types of turns. In case of a constant turn rate and a large mass hierarchy we have seen in the previous chapter that the power spectrum and bispectrum will get renormalized with respect to the single field spectra (denoted with an index 0)
\begin{align*}
\P_\R&= \frac{1}{c_s}\P_{\R 0},\\
B_\R& = \frac{1}{c_s^2}B_{\R 0}.
\end{align*}
Therefore one could compare the relative amplitudes of the power spectrum and the bispectrum to see if there is a deviation from slow-roll. So far only upper bounds on non-Gaussiantity are given and with the current constraints a speed of sound below $c_s=0.04$ is ruled out\footnote{In order to see this we compare equation (97) of \cite{Ade:2013ydc} with equation (1.1) of \cite{Cespedes:2013rda}. This yields $\tilde{c_3}(c_s^{-2}-1)\approx \frac{3}{4}c_s^{-2}$ for a small speed of sound for the type of models we consider, which allows us to read of the graph given in Figure 25 of \cite{Ade:2013ydc}. This means that for a speed of sound of $c_s\lesssim0.04$ we are outside the $2\sigma$ confidence region and we find the constraint $c_s>0.04$.}. If the polarization measurement of BICEP2 is a result of primordial gravitational waves with, then an much stronger bound of $c_s>0.25$ is derived in \cite{Baumann:2014cja}. It is hard to detect non-Gaussianity, but future measurements \cite{Giannantonio:2011ya} might be able to detect non-Gaussianity at a level corresponding to $c_s<0.15$ at first sight\footnote{If one takes equation $(98)$ from \cite{Ade:2013ydc} and put in $A=1/2\cdot(c_s^2-1)$, which corresponds to the value of $c_3$ we computed in the previous footnote, then one can solve $c_s$ for any given value for $f_{NL}^{\text{equil}}$. At the same time, considering Table $7$ from \cite{Giannantonio:2011ya} it might be possible in future to constrain $f_{NL}^{\text{equil}}$ with $\sigma(f_{NL}^{\text{equil}})\sim 10$, such that an detection of $c_s=0.15$ or below should be possible. For $c_s=0.15$ we find $f_{NL}^{\text{equil}}=-13.6$, for $c_s=0.1$ we find $f_{NL}^{\text{equil}}=-31$.}. Moreover, the tensor-to-scalar ratio becomes $r=16\epsilon c_s$ and together with a measurement of tensor spectral index $n_t=-2\epsilon$ this could also reveal a value of $c_s$ different from 1. Although $r$ might have been detected by BICEP2 \cite{Ade:2014xna}, we are probably not so close to a precise estimate of the value of both quantities. In the near future\footnote{In \cite{Baumann:2014cja} it is argued that $c_s=0.47$ is the critical value below which one can still differentiate between canonical single field models and effective theories with a constant speed of sound and that experimental improvements are necessary to be able to detect it.} therefore we probably can only measure this signal when $c_s$ is below $0.2$. Therefore in order to measure the presence of the heavy fields we need a reduction in the speed of sound of about 80 percent. However the effective single field description is only valid under the assumption that there is a large mass hierarchy. In case the mass of the additional field is comparable but larger than the Hubble scale we cannot speak in terms of the speed of sound and we should compute the rescaling of the power spectrum numerically. Therefore more generally in order to measure the presence of the heavy fields we need the amplitude of the power spectrum to increase by a factor of about 5 with respect to the canonical single field prediction.  \\

In case of a localized turn induced by the bend of the potential we expect oscillations in the power spectrum and bispectrum which might be well detectable in a joint analysis of the spectra in the near future if the turn happens within or just before the regime where the observable modes cross the Hubble radius and if the turn is sudden enough \cite{Achucarro:2013cva}, therefore we think that any sudden turn within the observational regime cannot be neglected and should be taken into account. \\

\section{Numerical study of the no-scale supergravity inflationary models from literature}
\label{Section: numerical study of the no-scale supergravity inflationary models from literature}
We first study the no-scale supergravity models from the literature without modifications as listed in Table \ref{Table:Overview no-scale models} and summarize our findings.
\subsection{The modified Cecotti model}
\label{Subsection: the modified Cecotti model}
\begin{figure}
\centering
\begin{subfigure}[b]{.4\textwidth}
  \centering
  \includegraphics[width=1\linewidth,natwidth=360,natheight=327]{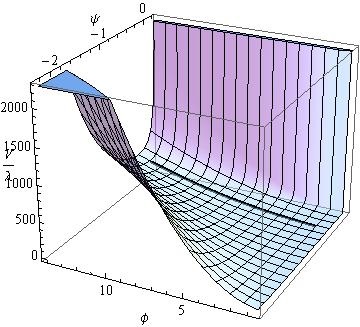}
  \caption{Trajectory}
  \label{Figure:cecotti trajectory}
\end{subfigure}%
\begin{subfigure}[b]{.3\textwidth}
  \centering
  \includegraphics[width=1\linewidth,natwidth=360,natheight=241]{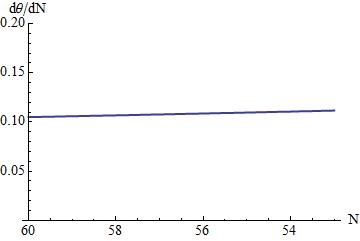}
   \caption{Turn rate}
  \label{Figure:cecotti etaperp}
\end{subfigure}%
\begin{subfigure}[b]{.3\textwidth}
  \centering
  \includegraphics[width=.9\linewidth,natwidth=360,natheight=246]{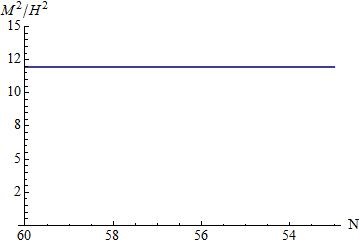}
  \caption{Isocurvature mass}
  \label{Figure:cecotti isocurvature mass}
\end{subfigure}
\caption{A typical (a) trajectory, (b) turn rate and (c) isocurvature mass squared compared to the Hubble parameter for the modified Cecotti model studied in section \ref{Subsection: the modified Cecotti model}. In this particular case we choose $\alpha$ such that $\phi_b=0.2M_p$.}
\label{Figure:cecotti}
\end{figure}

The modified Cecotti model is given in the first row of Table \ref{Table:Overview no-scale models}. We pick $T =  e^{ + \sqrt {2/3} \psi/M_p }M_p + i\sqrt {2/3} \phi$ as in the paper which means that the $\pm$ sign in the table becomes a $+$ sign. Note that by redefining the fields such that $T\rightarrow fT$ we can put $f=1$. This model leads to chaotic inflation because there is a potential barrier at $\psi_b=-\sqrt{3/2}\ln(2\alpha) M_p$ which stabilizes the field $\psi$ at some value close to zero. We need $\alpha>0$ in order to have a potential barrier. Inflation is supposed to happen for $\psi<\psi_b$ because the inflaton will first go in the positive $\psi$ direction until it comes close to the barrier where it gets obstructed and then it inflates along the $\phi$ direction. Therefore we should take $\alpha<0.5$ such that $\psi_b>0$ to enforce that the absolute minimum of the potential stays within reach of the inflaton field to avoid that we end up with some nonzero cosmological constant. We vary $\alpha$ such that $\psi_b \in \{0.1, 0.2, 1, 3\} M_p$ for which the kinetic term for $\phi$ will deviate more strongly from canonical for increasing $\psi_b$. We find however that there is not much difference and for each case the trajectory enforced by the potential is curved with a constant turn rate of order $\etaperp=\frac{d\theta}{dN} \sim 0.1$. The isocurvature mass is about $M^2/H^2\sim 12$ and we find a negligible renormalization of the power spectrum. We show one example in Figure \ref{Figure:cecotti} where we choose $\psi_b=0.2$.

\subsection{The imaginary supersymmetric Starobinsky model}
\label{Subsection: the imaginary supersymmetric Starobinsky model}
\begin{figure}
\centering
\begin{subfigure}[b]{.3\textwidth}
  \centering
  \includegraphics[width=1\linewidth,,natwidth=360,natheight=241]{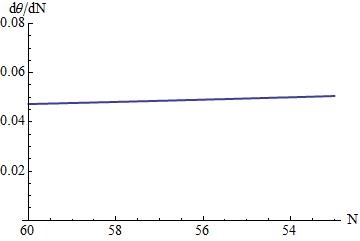}
   \caption{Turn rate}
  \label{Figure:imaginarystarobinsky etaperp}
\end{subfigure}%
\begin{subfigure}[b]{.3\textwidth}
  \centering
  \includegraphics[width=.9\linewidth,natwidth=360,natheight=246]{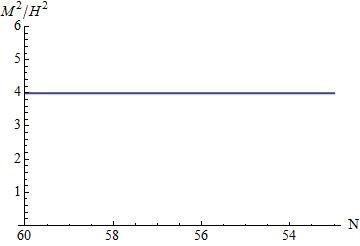}
  \caption{Isocurvature mass}
  \label{Figure:imaginarystarobinsky isocurvature mass}
\end{subfigure}%
\begin{subfigure}[b]{.4\textwidth}
  \centering
  \includegraphics[width=1\linewidth,natwidth=576,natheight=381]{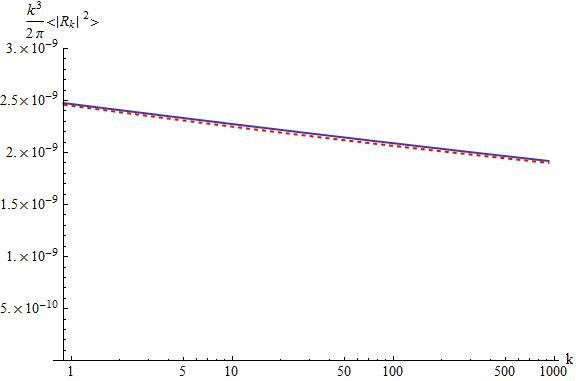}
  \caption{Trajectory}
  \label{Figure:imaginarystarobinsky powerspectrum}
\end{subfigure}
\caption{A typical (a) turn rate and (b) isocurvature mass squared compared to the Hubble parameter for the imaginary supersymmetric Starobinsky model studied in section \ref{Subsection: the imaginary supersymmetric Starobinsky model}. In this particular case we choose $n=1.1$ and $c=1.3$. In (c) the corresponding power spectrum is plotted where the blue line is the full numerical solution and the red dashed line the single field approximation.}
\label{Figure:imaginarystarobinsky}
\end{figure}

The imaginary supersymmetric Starobinsky model is given in the second row of Table \ref{Table:Overview no-scale models}. Again we put $f=1$  and take the $\pm$ sign to be $+$. In order to have a potential barrier $m$ has to be negative. In the corresponding paper \cite{Ferrara:2014ima} they choose $m=-n^{-1}2^{1-n}$ such that the potential is minimized at $\psi = 0$ independent of the value of $\phi$. More generally parameterizing $m=-c^{-1}2^{1-n}$ we have a potential barrier at $\psi_b=\sqrt{3/2}\ln(c)/(n-1)$. We take $n>1$ and $c>1$ such that $\psi_b>0$ in order to have inflation without a cosmological constant when $\psi$ gets stabilized slightly below $\psi_b$. We pick $n\in \{1.1, 1.5, 2\}$ and for each value of $n$ we choose $c$ equal to, less than and higher than $n$. In each case we find a very low turn rate $\etaperp=\frac{d\theta}{dN}<0.1$ and an isocurvature mass of about $M^2/H^2\sim 4$. The results of this model should therefore be reproduced by the SRST approximation in which case we expect and find indeed a negligible correction to the amplitude of the power spectrum compared to the single field approximation. We show one particular example in Figure \ref{Figure:imaginarystarobinsky} where we choose $n=1.1$ and $c=1.3$.

\subsection{The $(R+R^2+R^4)$-extension of the imaginary Starobinsky inflation}
The $(R+R^2+R^4)$-extension of the imaginary supersymmetric Starobinsky model is given in the third row of Table \ref{Table:Overview no-scale models}. Note that $z$ is in units of $M_p^{-4}$. Again we put $f=1$ and take the $\pm$ sign to be $+$. The overall normalization fixes $z$ and therefore the parameter space consists of $M$, $m$ and $n$. The potential barrier stays the same as in the imaginary Starobinsky model and parameterizing $m=-c^{-1}2^{1-n}$ we have $\psi_b=\sqrt{3/2}\ln(c)/(n-1)$. We pick $n\in \{1.1, 1.5\}$ and for each value of $n$ we choose $c$ equal to, less than and higher than $n$. We get the same results as for the imaginary supersymmetric Starobinsky model: in each case we have a very low turn rate $\etaperp=\frac{d\theta}{dN}<0.1$ and an isocurvature mass of about $M^2/H^2\sim 4$. Therefore we do not consider it further.

\subsection{The no-scale inflationary model}
The no-scale inflationary model is given in the last row of Table \ref{Table:Overview no-scale models}. We haven't studied this particular model yet. In this case $\psi$ is not stabilized at a constant value but rather a line in the $(\operatorname{Re} T, \operatorname{Im} T))$-plane is stabilized. This will probably result in a turn rate which increases or decreases in time. It is interesting to see whether the change in the turn rate is visible in the regime when the observable modes cross the Hubble radius. Besides a renormalization of the power spectrum this could for example lead to a change in the spectral index as well. We should investigate this model in the future.

\section{Numerical study of a toy model}
\label{Section: numerical study toy model}
For the first three models given in Table \ref{Table:Overview no-scale models} we haven't found so far interesting deviations from the single field approximation. We are however also restricted by choosing the value where $\psi$ is stabilized, the value of the isocurvature mass and the form of the kinetic terms because we only worked with the particular models offered by the papers. In order to have more grip on these parameters we therefore continue studying a toy model. The potential must allow for tunability of the isocurvature mass and the value where $\psi$ gets stabilized. Furthermore we will also include the potential which has a bend in this toy model. We study non-canonical kinetic terms which only depends on one of the fields $\psi$ i.e. $G_{\phi\phi}=G_{\phi\phi}(\psi)$ but we restrict to the forms found in literature. We can now make the research question more precise.
\begin{framed}
\noindent \textbf{Research question}\\
We study a toy model described by the the Lagrangian density
$$
\L=\frac{1}{2}G_{\phi\phi}(\partial_\mu\phi)^2+\frac{1}{2}(\partial_\mu\psi)^2-V(\phi,\psi),
$$
for which we allow the following kinetic terms
$$
G_{\phi\phi}=1, \quad \quad G_{\phi\phi}=\psi/M_p \quad \text{and} \quad G_{\phi\phi}=e^{\pm \sqrt{2/3}\psi/M_p}.
$$
The potential energy density $V(\phi,\psi)$ is given by
$$
V=\mu^2\phi^2+g^2\phi^2\psi^2+M^2(\psi-\psi_0)^2,
$$
and two interesting limiting cases are such that the field $\psi$ gets stabilized at a constant value ($g=0$) or such that it has a bend ($\mu=0$). In the case $g=0$ and in the flat regime of the potential given by $\mu=0$: can the non-canonical kinetic terms induce a renormalization of the power spectrum with a factor of about 5? In the case $\mu=0$: does the bend in the potential lead to a sudden turn in the trajectory in the regime where the observable modes cross the Hubble radius?
\end{framed}
The full parameter space of our toy model is given by $\{\mu,\ g,\ M,\ \psi_0 \}$ plus the three choices for the kinetic terms. Since the overall normalization of the power spectrum is given by experiment, this means the overall normalization of the potential is fixed and therefore this reduces the number of parameters by one. Moreover, as a starting point in this study we focus on the case $\mu=0$ and the case $g=0$, which means we are left with two continuous parameters and one discrete parameter for each case.

\subsection{The potential with a bend: $\mu=0$}
We get the following potential
$$
V=g^2\phi^2\psi^2+M^2(\psi-\psi_0)^2,
$$
which is given in \cite{Dong:2010in} and illustrates that the heavy fields can flatten the potential. This effect is explained in section \ref{Subsection: toy model from axion monodromy}. In the flat regime we can estimate the mass of the isocurvature mode
$$
V_{NN}\approx 2g^2\phi^2+2M^2, \quad \rightarrow \quad \frac{M_{NN}}{H^2}\sim\frac{V_{NN}}{H^2}=\frac{2g^2\phi^2+2M^2}{M^2\psi_0^2}3M_p^2\gg \frac{12 M_p^2}{\psi_0^2}
$$
where we used that the normal points towards $\psi$ because inflation happens in the $\phi$ direction. During the bend the isocurvature mass will presumably decrease judging the plot of the potential shown in Figure \ref{Figure:axionpotential}. However from this lower bound on the isocurvature mass we are probably safe in saying that the mass of the isocurvature mode depends on the value of $\psi_0/M_p$: the smaller $\psi_0/M_p$ the bigger the isocurvature mass with respect to the Hubble scale. Besides the overall normalization of the potential the two parameters we can vary are $\psi_0$ and the ratio $M/g$. The bigger the ratio $M/g$ the more inflation we have in the regime $\phi\ll M/g$ and the longer it takes for the inflaton to go from $\psi=0$ to $\psi=\psi_0$ because the radius of the bend of the potential increases. The bigger $\psi_0$ the bigger the change in $\bar{\psi}$ so probably there will be more turning in this case. At the same time the isocurvature mass decreases, so here something interesting might happen already with canonical kinetic terms. \\\\
Let us now think for a moment about the allowed ranges for the parameters. The mass scales should be well below the Planck scale $M_{Pl}=\sqrt{8\pi} M_p\sim 5 M_p$. If the interesting scales cross the Hubble radius during the flat regime then the value of the Hubble parameter is given by $\frac{H^2}{M_p^2}=\frac{M^2\psi_0^2}{3M_p^4}$, at $\phi =  \frac{M}{g}$ it becomes half this value. In the flat regime $\epsilon$ is small so in order to compensate for $\epsilon$ to get the right amplitude of the power spectrum we know the Hubble parameter must be small indeed. In the regime of the bend it is a bit unclear what the value for $\epsilon$ will be but say in the most optimistic case we end up with $H = 5\cdot 10^{-5} M_{p}$ (corresponding to $r=0.2$ and such that $M\psi_0 \approx 8 \cdot 10^{-5} M_p^2$). This means if we get $M_{NN}/H \ll 10^6$ we are safe, so we can choose $\psi_0$ very close to 0, but of course every time we need to check that this condition is satisfied. On the other hand we would like $M_{NN}^2/H^2 \gg 1$ in the regime of the bend such that the effective description applies. If this does not hold one also needs to do a full computation of the bispectrum in order to understand the effects on the observables. Of course it is interesting to see what will happen to the power spectrum when you enter the regime $M_{NN}^2/H^2 \sim 1$ but that is a different question. We study this potential for several kinetic terms.

\subsubsection{Canonical kinetic terms: $G_{\phi\phi}=1$}
\label{Subsection: axioncanonical}
\begin{figure}
\centering
\begin{subfigure}[b]{.4\textwidth}
  \centering
  \includegraphics[width=1\linewidth,,natwidth=360,natheight=212]{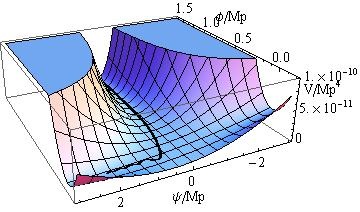}
  \caption{Trajectory}
  \label{Subfigure:axioncanonical1plottrajectory}
\end{subfigure}%
\begin{subfigure}[b]{.3\textwidth}
  \centering
  \includegraphics[width=1\linewidth,natwidth=360,natheight=234]{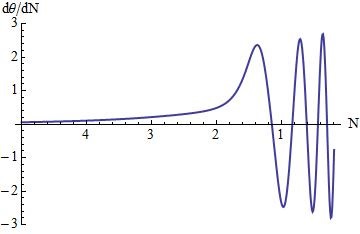}
   \caption{Turn rate}
  \label{Subfigure:axioncanonical1plotetaperp}
\end{subfigure}%
\begin{subfigure}[b]{.3\textwidth}
  \centering
  \includegraphics[width=.9\linewidth,natwidth=360,natheight=253]{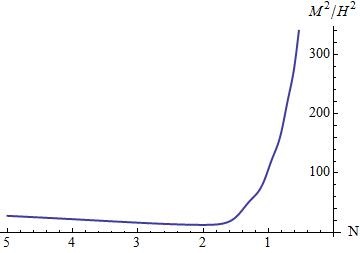}
  \caption{Isocurvature mass}
  \label{Subfigure:axioncanonical1plotisocurvaturemass}
\end{subfigure}
\caption{(a) The trajectory, (b) the turn rate and (c) the isocurvature mass squared compared to the Hubble parameter. All for the potential with a bend ($\mu=1$) with canonical kinetic terms studied in section \ref{Subsection: axioncanonical} and $\psi_0=3 M_p$ and $M/g=0.1 M_p$.   }
\label{Figure:axioncanonical1}
\end{figure}

\begin{figure}
\centering
\begin{subfigure}[b]{.33\textwidth}
  \centering
  \includegraphics[width=.9\linewidth,natwidth=360,natheight=219]{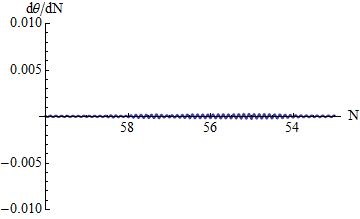}
  \caption{Turn rate}
 \label{Subfigure:axioncanonical2plotetaperp}
\end{subfigure}%
\begin{subfigure}[b]{.33\textwidth}
  \centering
  \includegraphics[width=.9\linewidth,natwidth=360,natheight=242]{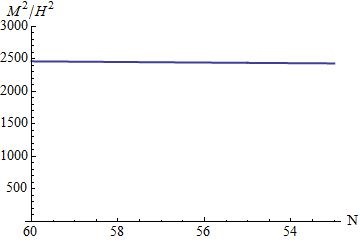}
   \caption{Isocurvature mass}
  \label{Subfigure:axioncanonical2plotisocurvaturemass}
\end{subfigure}%
\begin{subfigure}[b]{.33\textwidth}
  \centering
  \includegraphics[width=.9\linewidth,natwidth=360,natheight=240]{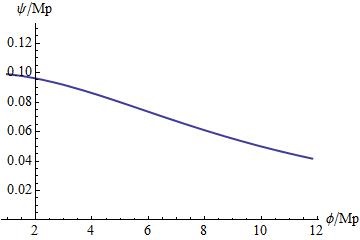}
  \caption{Trajectory in field space}
  \label{Subfigure:axioncanonical2plotphipsi}
\end{subfigure}
\caption{(a) The turn rate, (b) the isocurvature mass squared compared to the Hubble parameter and (c) the trajectory in field space during the 60 to 53 e-folds before the end of inflation. All for the potential with a bend ($\mu=1$) with canonical kinetic terms studied in section \ref{Subsection: axioncanonical} and $\psi_0=0.1 M_p$ and $M/g=10 M_p$.   }
\label{Figure:axioncanonical2}
\end{figure}

We choose $M/g \in \{0.1, 1, 10\} M_p$ and $\psi_0 \in \{0.1, 1, 3\} M_p$. We find that the turn rate is either too low to be visible or too far away from the inflationary regime where the observable modes cross the Hubble radius: 60 to 53 e-folds before the end of inflation. One example where we have a (too sharp) turn at the end of inflation is for $\psi_0=3 M_p$ and $M/g=0.1 M_p$ is illustrated by Figure \ref{Figure:axioncanonical1}. This would have led to oscillations in the power spectrum - although an effective description is not possible here - if it happened much earlier. It makes sense that this happens at the end of inflation because if the turn in the potential is traversed fast enough it means at the same time that inflation ends quickly afterwards, because the bend is positioned close to the end of the trajectory. \\
Studying the regime where the observable modes cross the Hubble radius, in each case the turn rate is too low compared to the isocurvature mass and you can neglect its influence on the observables. We give an example for $\psi_0=0.1 M_p$ and $M/g=10 M_p$ as illustrated by Figure \ref{Figure:axioncanonical2}. Considering the plot of the trajectory in field space this configuration might lead to a decreasing turn rate in the case of non-canonical kinetic terms.

\subsubsection{Non canonical kinetic terms: $G_{\phi\phi}=\psi/M_p$}
\label{Subsection: axionnoncanonical1}
\begin{figure}
  \begin{center}
    \includegraphics[width=0.38\textwidth,natwidth=360,natheight=277]{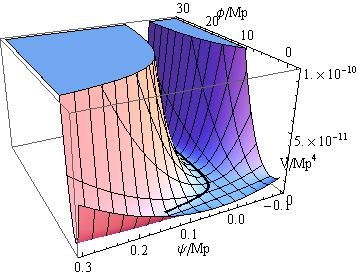}
  \end{center}
  \caption{The trajectory for the potential with a bend ($\mu=1$) with non-canonical kinetic terms studied in section \ref{Subsection: axionnoncanonical1} with $G_{\phi\phi}=\psi/M_p$ and $\psi_0=0.2 M_p$ and $M/g=1 M_p$.}
  \label{Figure:axionnoncanonical1plottrajectory}
\end{figure}

\begin{figure}
\centering
\begin{subfigure}[b]{.33\textwidth}
  \centering
  \includegraphics[width=.9\linewidth,natwidth=360,natheight=254]{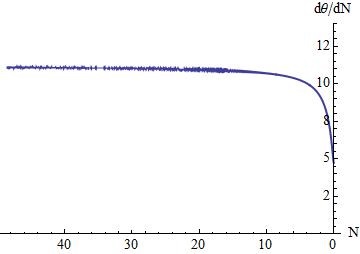}
  \caption{Turn rate}
\label{Figure:axionnoncanonical1plotetaperp}
\end{subfigure}%
\begin{subfigure}[b]{.33\textwidth}
  \centering
  \includegraphics[width=.9\linewidth,natwidth=360,natheight=242]{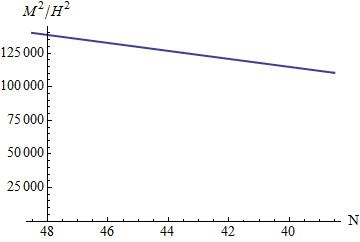}
   \caption{Isocurvature mass}
  \label{Figure:axionnoncanonical1plotisocurvaturemass}
\end{subfigure}%
\begin{subfigure}[b]{.33\textwidth}
  \centering
  \includegraphics[width=.9\linewidth,natwidth=360,natheight=251]{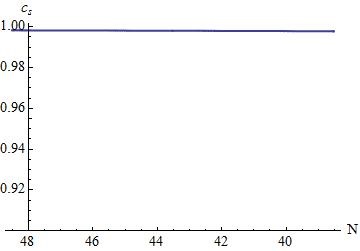}
  \caption{Speed of sound}
  \label{Figure:axionnoncanonical1plotcs}
\end{subfigure}
\caption{(a) The turn rate, (b) the isocurvature mass squared compared to the Hubble parameter and (c) the speed of sound. All for the potential with a bend ($\mu=1$) with non-canonical kinetic terms studied in section \ref{Subsection: axionnoncanonical1} with $G_{\phi\phi}=\psi/M_p$ and $\psi_0=0.2 M_p$ and $M/g=1 M_p$.}
\label{Figure:axionnoncanonical1}
\end{figure}

This kinetic term is not always positive so we need to be careful. If $\psi_0$ gets very close to 0 the quantum fluctuations of the fields can render the kinetic term negative, so this is something we should keep track of during the numerical computations. To omit this difficulty we first study examples in which $\psi_0$ comes not so close to 0 and we choose $M/g \in \{1, 10\} M_p$ and $\psi_0 \in \{0.1, 0.2, 1, 3\} M_p$. We find that if $\psi_0$ gets smaller the $\phi$-field will traverse larger distances in field space. This has to do with the fact that $\dot{\phi}$ will be very large if $\psi$ is small, which is precisely the case if $\psi_0$ is small. During the regime where the observable modes cross the Hubble radius we find a speed of sound very close to 1 for all cases. So the turn rate might be large but this is compensated by a much larger isocurvature mass. We illustrate one example $\psi_0=0.2 M_p$ and $M/g=1 M_p$ in Figures \ref{Figure:axionnoncanonical1plottrajectory} and \ref{Figure:axionnoncanonical1}. Furthermore we do a non-rigorous check of what happens if we decrease $\psi_0$ and we find that we get very large turn rates for which the speed of sound reduces considerably. This seems to be an interesting regime and we should do a more careful analysis in future.

\subsubsection{Non canonical kinetic terms: $G_{\phi\phi}=e^{\pm\sqrt{2/3}\psi/M_p}$}
In case of $G_{\phi\phi}=e^{-\sqrt{2/3}\psi/M_p}$ we have a very low turn rate with heavy masses when $\psi_0$ is small. If you want a bigger turn rate you need to increase $\psi_0$ to a rather high value such that the exponential becomes small, but then you will only get some effect at the end of inflation and moreover the isocurvature mass becomes extremely low, so this is not what we are looking for. In case of $G_{\phi\phi}=e^{+\sqrt{2/3}\psi/M_p}$ we always have a small turn rate because $\psi\geq0$, so here you only get $c_s=1$ to a very good approximation. Conclusion: for the kinetic terms $G_{\phi\phi}=e^{\pm\sqrt{2/3}\psi/M_p}$ the single field description is fine as long as the isocurvature mass is not small.

\subsection{The separable potential which stabilizes $\psi$ at a constant value: $g=0$}
We put $g=0$ to get the following potential
$$
V=\frac{1}{2}\mu^2\phi^2+\frac{1}{2}M^2(\psi-\psi_0)^2,
$$
where one field is stabilized such that inflation is only along one of the fields. The potential is separable and therefore much easier to understand. We take $M$ large compared to $\mu$ such that $\psi$ is the heavy field which gets stabilized at the `naive' value $\psi=\psi_0$. If due to non-canonical terms the turn rate is large we expect that the inflaton gets pushed uphill such that the `naive' computation underestimates the amplitude of the power spectrum for scalar perturbations.\\
Assuming that $\psi$ gets fixed at some value close to $\psi_0$ we know the normal points in the $\psi$-direction and we find
$$
V_{NN}=M^2.
$$
Furthermore for any non-canonical kinetic term of the form $G_{\phi\phi}(\psi)$ we can deduce in the slow-roll regime
$$
\phi^; = -\frac{2 M_p^2}{G_{\phi\phi} \phi},
$$
where we used the fact that in this case we have $\sigma^;=\sqrt{G_{\phi\phi}}\phi^;$ and plugged this into equations (\ref{Equation:tangentvector}) and (\ref{Equation:backgroundprojected}). For constant $\psi$ we can integrate this equation to find
$$
\phi(N)=\sqrt{\frac{2(N-N_e+1)}{G_{\phi\phi}}}M_p,
$$
where $\phi_e=\sqrt{\frac{2}{G_{\phi\phi}}}M_p$ such that $\epsilon=\frac{2 M_p^2}{\phi_e^2 G_{\phi\phi} }=1$. This gives the following estimate for the Hubble parameter
$$
H^2\approx \frac{50 \mu^2}{3 G_{\phi\phi}},
$$
at the time where the observable modes cross the Hubble radius. We are therefore interested in the ranges of the parameters for which
$$
\sqrt{\frac{50}{3 G_{\phi\phi}}}\mu \ll M \ll M_p.
$$
For chaotic inflation we have
$$
H \approx 5\cdot10^{-6} M_p \quad \rightarrow \quad \mu\approx \sqrt{G_{\phi\phi}}\cdot10^{-6} M_p.
$$
For given kinetic terms the normalization of the power spectrum fixes $\mu$ and we can vary the parameters $M/\mu$ and $\psi_0$.

\subsubsection{Non canonical kinetic terms: $G_{\phi\phi}=\psi/M_p$}
\label{Subsection: toymodelnoncanonical1}
\begin{figure}
\centering
\begin{subfigure}[b]{.33\textwidth}
  \centering
  \includegraphics[width=.9\linewidth,natwidth=360,natheight=318]{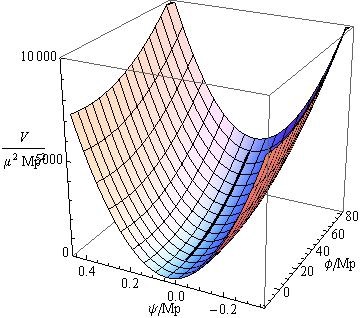}
  \caption{Trajectory}
\label{Figure:toymodelnoncanonical1plotpotential}
\end{subfigure}%
\begin{subfigure}[b]{.33\textwidth}
  \centering
  \includegraphics[width=.9\linewidth,natwidth=360,natheight=241]{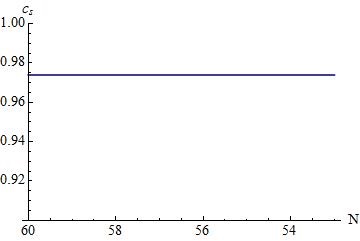}
   \caption{Speed of sound}
  \label{Figure:toymodelnoncanonical1plotcs}
\end{subfigure}%
\begin{subfigure}[b]{.33\textwidth}
  \centering
  \includegraphics[width=.9\linewidth,natwidth=360,natheight=237]{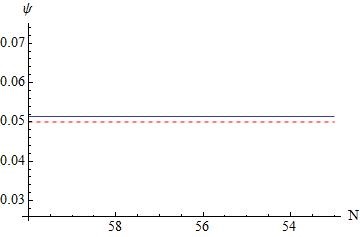}
  \caption{Displacement of $\psi$}
  \label{Figure:toymodelnoncanonical1plotpsi}
\end{subfigure}
\caption{An example of (a) a trajectory, (b) the reduced speed of sound and (c) the displacement of $\psi$ (red dashed line) compared to $\psi_0=0.05 M_p$ (blue line). All for the toy model with the separable potential ($g=1$) with non-canonical kinetic terms studied in section \ref{Subsection: toymodelnoncanonical1} with $G_{\phi\phi}=\psi/M_p$, $\psi_0=0.05 M_p$ and $M/\mu = 300$.}
\label{Figure:toymodelnoncanonical1}
\end{figure}

\begin{figure}
  \begin{center}
    \includegraphics[width=0.5\textwidth,natwidth=576,natheight=381]{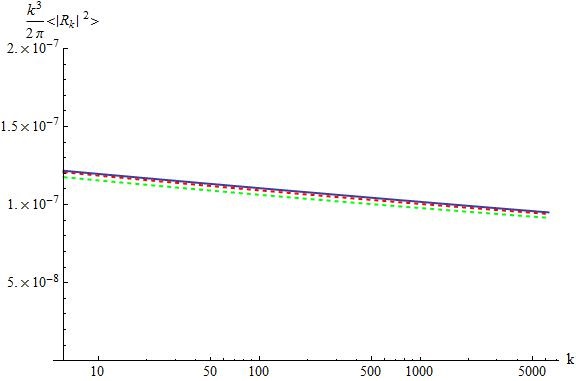}
  \end{center}
  \caption{The power spectrum for the toy model with the separable potential ($g=1$) with non-canonical kinetic terms studied in section \ref{Subsection: toymodelnoncanonical1} with $G_{\phi\phi}=\psi/M_p$, $\psi_0=0.05 M_p$ and $M/\mu = 300$. The blue line represents the full numerical solution. The red dashed line is the theoretical expected power spectrum including the renormalization due to the turn as explained in section \ref{Subsection: observational consequences}. The green dashed line represents the `naive' single field approximation in which the correction due to the turn is not taken into account.}
  \label{Figure:toymodelnoncanonical1plotpowerspectrum}
\end{figure}

We choose $M/\mu \in \{50, 100, 200, 300, 1000\}$ and $\psi_0 \in \{0.01, 0.1, 1\} M_p$. The speed of sound gets only reduced considerably in the case $\psi_0\sim 0.01$. For example for $\psi_0=0.01 M_p$ and $M/\mu = 1000$ we find $M_{NN}^2/H^2 \sim 350$ and $c_s \sim 0.85$. However this kinetic term becomes negative for negative values for $\psi$ including the quantum fluctuations so we should do a more careful analysis in this case to check whether this example is reliable. For $\psi_0=0.1$ the turn rate is about $\etaperp\sim0.6$ and therefore for $M^2/H^2\gg1$ this effect is certainly not visible. The constant turn might cause some trouble when the isocurvature mass gets closer to the Hubble scale. For example taking $\psi_0=0.1 M_p$ and $M/\mu = 50$ we get $M_{NN}^2/H^2 \sim 8$ and $\etaperp\sim 0.6$. The isocurvature mass is expected to be high enough for the isocurvature modes to decay quickly after Hubble radius crossing but the effective theory does not apply. We can compute the power spectrum for this case, but we don't know precisely what will happen to the bispectrum because we cannot speak in terms of a speed of sound. On the other hand if we just compute the speed of sound as if we are working with an effective single field description it is about $c_s \sim 0.9$ which is too close to 1 to be detectable. We show one example where we choose $\psi_0$ in between the values discussed in this section in Figure \ref{Figure:toymodelnoncanonical1} where we take $\psi_0=0.05 M_p$ and $M/\mu = 300$ such that $M_{NN}^2/H^2 \sim 125$ and $c_s \sim 0.975$. We also show the displacement of $\psi$ from the `naive' value $\psi_0$. The resulting power spectrum is shown in Figure \ref{Figure:toymodelnoncanonical1plotpowerspectrum} and compared with the theoretical expected power spectrum given in section \ref{Subsection: observational consequences} and the `naive' single field approximation. We see the power spectrum gets renormalized by a few percent which is not detectable but motivates us to explore what happens for lower values for $\psi_0$ in future.

\subsubsection{Non canonical kinetic terms: $G_{\phi\phi}=e^{\pm\sqrt{2/3}\psi/M_p}$}
In case of $G_{\phi\phi}=e^{-\sqrt{2/3}\psi/M_p}$ we have a very low turn rate even when $\psi_0$ takes the values such that the exponential becomes equal much less than 1. The same applies for $G_{\phi\phi}=e^{\sqrt{2/3}\psi/M_p}$. 
We conclude that for the kinetic terms $G_{\phi\phi}=e^{\pm\sqrt{2/3}\psi/M_p}$ the single field description is fine because the turn rate is extremely low.

\subsection{The full toy-model}
Let us think for a moment about the full potential
$$
V=\mu^2\phi^2+g^2\phi^2\psi^2+M^2(\psi-\psi_0)^2.
$$
Assuming $\psi$ is massive enough and that it takes the value $\bar{\psi}(\phi)$ you get
$$
\bar{\psi} = \frac{M^2\psi_0}{g^2\phi^2+M^2} \quad \rightarrow \quad V(\phi,\bar{\psi})=\frac{g^2\phi^2}{g^2\phi^2+M^2}M^2\psi_0^2+\mu^2\phi^2.
$$
We have again the two regimes
\begin{align*}
&\phi\gg M/g: \quad V\rightarrow M^2\psi_0^2 + \mu^2\phi^2, \quad H^2\rightarrow \frac{M^2\psi_0^2+ \mu^2\phi^2}{3M_p^2}, \quad \bar{\psi} \rightarrow 0,\\
&\phi\ll M/g: \quad V\rightarrow \left(g^2\psi_0^2+ \mu^2\right)\phi^2, \quad H^2\rightarrow \frac{g^2\psi_0^2+ \mu^2}{3M_p^2}\phi^2, \quad \bar{\psi} \rightarrow \psi_0.\\
\end{align*}
The potential still gets flatter in the regime $\phi\gg M/g$ but always has the additional chaotic term $\mu^2\phi^2$. The reasoning applied to the non-canonical kinetic terms $G_{\phi\phi}=e^{\pm\sqrt{2/3}\psi/M_p}$ to the model without $\mu^2\phi^2$ should be still the same so we won't explore that part of parameter space. In case of $G_{\phi\phi}=\psi/M_p$ the turn rate might get cranked up so that is something to check. In case of canonical kinetic terms we again expect for the same reasons that the turn in the potential is either too mild or otherwise it will be close to the end of inflation. But we should check this because the additional chaotic term will speed up the inflaton. This region of parameter space should therefore also be explored in future.

\section{Conclusion and Discussion}
\label{Section:conclusiondiscussionproject2}
In this chapter we have addressed the question\footnote{The research question is more precisely posed in section \ref{Section: numerical study toy model}.} whether certain types of two-field inflationary models, one described by a potential with a bend and the other described by a potential which stabilizes one of the fields and where we allow for non-canonical kinetic terms, will lead to observable signatures in the CMB data, distinguishable from canonical single field inflation. Based on the explored part of parameter space so far we arrive at the tentative conclusion that the single field approximation without taking into account the turns is fine and the heavy degrees of freedom present in these models won't be detectable in the current or near future data because:\\
1) The turn rate is constant but too low to be distinguishable from a single field model \\
2) There is a sudden turn but it happens close to the end of inflation and therefore it does not influence the observable modes. \\
In particular we found no sizable turn rates for the non-canonical kinetic terms $G_{\phi\phi}=e^{\pm\sqrt{2/3}\psi/M_p}$ and the single field description ignoring the influence of turns seems to be very good as long as the isocurvature mass is large enough. However for the kinetic term $G_{\phi\phi}=\psi/M_p$ we found hints of large turn rates which should be investigated in more detail in future. Moreover although the studied models cannot be distinguished from single field models from an observational point of view so far one should actually take into account the rescaling of the tensor-to-scalar ratio $r$ when computing the observables in order to study the validity of the particular model under consideration.  \\

The data analysis performed in this chapter has not finished yet, however. In order to arrive at a more precise conclusion we need to explore more regions in the parameter space as given in section \ref{Section: numerical study toy model} for which we expect that it could describe models which allow for observable effects of the additional fields. For example in case of the non-canonical kinetic term $G_{\phi\phi}=\psi/M_p$ we should explore what happens if we choose $\psi_0$ closer to 0. Another possibly interesting region in parameter space is when we consider the full toy model where both $\mu$ and $g$ are nonzero. \\

Furthermore, going back to the original study of the models in literature outlined in sections \ref{Section: characterization of the models} and \ref{Section: numerical study of the no-scale supergravity inflationary models from literature} we have not studied the `no-scale inflationary model' yet. This model might reveal detectable effects because the stabilization used in the respective paper is different than the stabilization used in the other models from literature and the toy model we studied in section \ref{Section: numerical study toy model}. \\

Finally in section \ref{Subsection: no-scale supergravity models} where we derived the two-field Lagrangians from the K\"ahler potential and superpotential we assumed that the field $X$ can be stabilized at $X=0$ and that we can neglect them in the final theory. For the same reason that turns can reveal the presence of additional fields we should carefully check whether this field is really stabilized there.

\chapter{Conclusion and Outlook}
\label{Chapter:discussionoutlook}
\section{Summary of Conclusions}
In this thesis we have studied two projects. The first project, as discussed in chapter \ref{Chapter:multifieldinflation}, was to compare and translate between studies of multiple field inflation in the literature. The notation and definitions used in these studies differ and therefore we provided dictionaries to translate between the papers. These dictionaries can be found in Table \ref{Dictionary:slowroll}, \ref{Table:dictionarysetup}, \ref{Dictionary:perturbations} and \ref{Dictionary:two-field}. We stressed that the turn parameter is very important to describe the multi-field effects and should be considered separately from the slow-roll parameters. However, historically the slow-roll parameters and the turn parameters are mixed up as can be seen in Table \ref{Dictionary:slowroll}.
Moreover, we studied the approximation schemes used in the papers which are used to solve the perturbation equations. We find that only two different approximation schemes are used and we study their regime of validity and the analytical predictions of the power spectrum of the curvature mode. The overview of the approximation schemes can be found in Table \ref{Table:overviewapproximations}. We concluded that the SRST approximation scheme is very limited in the sense that it can only be used to describe multi-field inflationary models which are effectively simple single field models. The second approximation scheme on the other hand, which is a combination of the slow-roll approximation and a assuming a large mass hierarchy, can be used to study multi-field inflationary models which might have distinct signatures compared to simple single field models.\\

The second project, as discussed in chapter \ref{Chapter:dataanalysis}, was a numerical study of concrete models of multi-field inflation from recent papers in the literature. We investigated whether the current and future experiments might be able to detect the presence of the additional fields in these models or not. We arrived at the tentative conclusion that the single field approximation without taking into account the turns is fine and the heavy degrees of freedom present in these models won't be detectable in the current or near future data for two possible reasons. Either the turn rate is constant but too low to be distinguishable from a single field model, or there is a sudden turn but it happens close to the end of inflation and therefore it does not influence the observable modes.

\section{Outlook}
There are many possibilities of new projects as a continuation of this master's research. First of all the two projects discussed in this thesis leave much room for improvement. Moreover, there are extensions to these projects which place this work in a broader research area and connect it to future developments. \\

The numerical analysis of multi-field models described in chapter \ref{Chapter:dataanalysis} is still in its infancy. The analysis has not saturated all possibly interesting possibilities yet. Some regions of parameter space might reveal interesting physics and should be explored in more detail. In addition, several aspects of the original models on which the data analysis is based, need to be studied more carefully. Moreover, we should continue our quest for multi-field effects in concrete models to understand what kind of turns appear and how they will influence the observables.  \\

The analytical study of multi-field models in chapter \ref{Chapter:multifieldinflation} does not cover the full range of possible multi-field models. In particular, it would be interesting to get more understanding in case that the additional fields are heavy enough for the isocurvature modes to decay, but where the effective theory is expected to break down. By should study these type of models by means of an analytical analysis in parallel with a numerical study. Moreover, we focused mainly on the case of two fields. It would be interesting to generalize the analytical expressions to the case of a multi-field theory containing one light field and the others much heavier.\\

As an extension to these projects, it is highly interesting to learn how to search for signatures of turns in the CMB. Assuming an effective single field description with a reduced speed of sound, we should be able to match different shapes of the speed of sound to the current and future CMB data as in \cite{Achucarro:2013cva}. When more data is released we should be able to falsify models and find support for models consistent with observation in order to learn more about the physics of the early universe.\\

Another important probe of cosmology in the near future will be high precision LSS surveys, e.g. Euclid \cite{Amendola:2012ys}. These surveys will provide a huge amount of cosmological data and are the most promising datasets to establish the presence of non-Gaussianity. The next years it will be worth to specialize on many aspects related to LSS, considering the fact it will become the main probe for cosmology. For example, the interpretation of the measurements comes with some complicating issues of theoretical and astrophysical character and it is important to get more theoretical understanding about these problems. In addition, in order to deal with the complexity of the data it will be extremely useful to consult numerical methods.

\chapter*{Acknowledgements}
Ana, I am very grateful to you for being such a fantastic supervisor. You gave me many learning experiences of how to work as a researcher by coaching me in many aspects of the actual research, in the preparation of important presentations and in finding the right attitude of writing. Moreover, you embedded me in your research group and involved me with all kinds of meetings. I really liked our meetings where you shared your physical insights with me and equally enjoyed the non-physics conversations we had. Thanks for your confidence, patience and your great kindness. I am very happy to continue working with you.\\

Tomislav, thanks for keeping an eye on the project and for your extensive and very useful feedback on the final presentation and this thesis, all provided on a very short notice, even when you were having holidays.  \\

Pablo, it was great to start working with you on a project. You are a very amiable person and a good teacher. I am looking forward to really kick off this project. \\

Furthermore, I would like to thank all my colleagues in Leiden and Utrecht for the wonderful time, the physics conversations and all the useful feedback you gave me. In particular I would like to thank Pablo, Vicente, Jes\'us, Drian, Drazen, Wessel, Bart, Artem, Mark and Bin. \\

Finally, many thanks to my family, friends and Martijn, which always have been most important for me.

\bibliography{mybib}{}
\bibliographystyle{hieeetr}

\begin{appendices}

\chapter{Full Computations of the Analytical Approximations}
\label{Appendix:analytical approximations}
In this appendix we work out some of the underlying computations of the analytical approximations discussed in chapter \ref{Chapter:multifieldinflation}, section \ref{Section:analyticalapproximations}. The names of the sections in this appendix are the same as the corresponding subsections in section \ref{Section:analyticalapproximations}.

\section{Sub-horizon}
We work out the details of the approximation of the low and high frequency in the second part of the sub-horizon regime $ M^2 \gg k^2/a^2\gg H^2$ where we assume a constant turn parameter. The perturbation equations to zeroth order in slow roll become
\begin{align*}
&\ddot{q}^T+H\dot{q}^T+2\etaperp H \dot{q}^N+\left(\frac{k^2}{a^2}-2H^2\right)q^T+4\etaperp H^2q^N=0,\\
&\ddot{q}^N+H\dot{q}^N-2\etaperp H\dot{q}^T+\left(\frac{k^2}{a^2}+H^2\left(m^2-2-(\etaperp)^2\right)\right)q^N+2\etaperp H^2q^T=0.
\end{align*}
Using the following ansatz for the solution of the perturbation equations
\begin{align*}
&q^T = A_- e^{-i\int dt\omega_- }+A_+ e^{-i\int dt\omega_+ },\\
&q^T = B_- e^{-i\int dt\omega_- }+B_+ e^{-i\int dt\omega_+ },
\end{align*}
we get the following equation for $\omega_\pm$ and the amplitudes $A_\pm$ and $B_\pm$
$$
\begin{pmatrix}
 -\omega^2_\pm-i\dot{\omega}_\pm-iH\omega_\pm+\frac{k^2}{a^2}-2H^2 & 2\etaperp\left(-iH\omega_\pm+2H^2\right)  \\
  2\etaperp\left(iH\omega_\pm+H^2\right)  & -\omega^2_\pm-i\dot{\omega}_\pm-iH\omega_\pm+\frac{k^2}{a^2}+H^2\left(m^2-2-(\etaperp)^2\right)
 \end{pmatrix}
\begin{pmatrix}
  A_\pm  \\
  B_\pm
 \end{pmatrix}
 =0.
$$
Assuming we can neglect $\dot{\omega}_\pm$, $H\omega_\pm$, $H^2$ and $\etaperp H^2$ with respect to the other terms we get the following solution
\begin{align*}
&\omega_\pm^2=\frac{k^2}{a^2}+\frac{1}{2}H^2\left(m^2+3(\etaperp)^2\right)\pm \frac{1}{2}\sqrt{H^4\left(m^2+3(\etaperp)^2\right)^2+16\frac{k^2(\etaperp)^2H^2}{a^2}},\\
&A_+\approx\frac{-2i\etaperp H\omega_+}{\omega_+^2-k^2/a^2}B_+,\\
&B_-\approx\frac{-2i\etaperp H\omega_-}{H^2\left(m^2-(\etaperp)^2\right)+k^2/a^2-\omega_-^2}A_-,
\end{align*}
and expanding the square root to first order, using $m^2 +3(\etaperp)^2 \gg k^2/a^2H^2$ and assuming $|\etaperp|\leq m$, we get
$$
\omega_\pm^2\approx \frac{k^2}{a^2}+\frac{1}{2}H^2\left(m^2+3(\etaperp)^2\right)\left(1\pm 1\right) \pm 4\frac{k^2(\etaperp)^2}{a^2\left(m^2+3(\etaperp)^2\right)}.
$$
which is in agreement with the assumption that $\dot{\omega}_\pm$, $H\omega_\pm$ can be neglected with respect to the other terms, since $\dot{\omega}_-=-H\omega_-$ and $\dot{\omega}_+\ll-H\omega_+$ and $k^2/a^2\gg H^2$.

\section{Super-horizon}
We derive the solution of the toy model with constant turn rate $\etaperp$ to zeroth and first order in slow roll. To zeroth order the perturbation equations become
\begin{align*}
q^{T\prime\prime}&-\frac{2}{z^2}q^T = 2\etaperp\left(\frac{1}{z} q^{N\prime} - \frac{2}{z^2}q^N\right),\\
q^{N\prime\prime}&+\frac{1}{z^2}\left(\frac{M^2}{H^2}-2-(\etaperp)^2\right)q^N = -2\etaperp\left(\frac{1}{z}q^{T\prime}+\frac{1}{z^2}q^T\right).
\end{align*}
We use the ansatz $q^I=A^Iz^p$ to solve the equation, this yields
$$
\frac{1}{z^2}
\begin{pmatrix}
  p(p-1)-2 & -2\etaperp(p-2)  \\
  2\etaperp(p+1) & p(p-1)+\left(\frac{M^2}{H^2}-2-(\etaperp)^2\right)
 \end{pmatrix}
\begin{pmatrix}
  A^T  \\
  A^N
 \end{pmatrix}
 =0.
$$
The determinant should be zero
$$
(p-2)(p+1)\left(p^2-p+\frac{M^2}{H^2}-2+3(\etaperp)^2\right)=0,
$$
providing us the solutions
\begin{align*}
q^T=\frac{A}{z}& \quad \text{and} \quad q^N=0, \\
q^T=A z^2& \quad \text{and} \quad q^N=\frac{-6\etaperp}{\frac{M^2}{H^2}-(\etaperp)^2}q^T,\\
q^N=A z^{\frac{1}{2}\left(1\pm\sqrt{9-4(\frac{M^2}{H^2}+3(\etaperp)^2}\right)}& \quad \text{and} \quad q^T=\frac{4\etaperp}{3\pm\sqrt{9-4(\frac{M^2}{H^2}+3(\etaperp)^2)}}q^N.\\
\end{align*}
We have only one growing mode $p=-1$ which corresponds to $\R\sim 1$ and freezes out after horizon crossing. The other solutions are decaying solutions if the isocurvature mode is massive enough, with its effective mass squared given by $M^2+3(\etaperp)^2 H^2$. Note that they are also oscillating rapidly. In order to make more clear what these solutions mean, the full solution is given by
\begin{align*}
q^T_\alpha&=A_\alpha\frac{1}{z}+B_\alpha z^2 +  C_\alpha\frac{4\etaperp z^{\frac{1}{2}\left(1+\sqrt{9-4(\frac{M^2}{H^2}+3(\etaperp)^2}\right)}}{3+\sqrt{9-4(\frac{M^2}{H^2}+3(\etaperp)^2)}} + D_\alpha\frac{4\etaperp z^{\frac{1}{2}\left(1-\sqrt{9-4(\frac{M^2}{H^2}+3(\etaperp)^2}\right)}}{3-\sqrt{9-4(\frac{M^2}{H^2}+3(\etaperp)^2)}}, \\
q^N_\alpha&=B_\alpha \frac{6\etaperp}{(\etaperp)^2-\frac{M^2}{H^2}}z^2+ C_\alpha z^{\frac{1}{2}\left(1+\sqrt{9-4(\frac{M^2}{H^2}+3(\etaperp)^2}\right)} + D_\alpha z^{\frac{1}{2}\left(1-\sqrt{9-4(\frac{M^2}{H^2}+3(\etaperp)^2}\right)},
\end{align*}
Where $A_\alpha, B_\alpha \sim \delta^T_\alpha$ and $C_\alpha, D_\alpha \sim \delta^N_\alpha$ because when the modes are decoupled ($\etaperp=0$) this is true.\\\\
Next we repeat the same computation but then to first order in slow roll. The equations of motion for the perturbation equations become
\begin{align*}
q^{T\prime\prime}&-\frac{2+3\epsilon_H}{z^2}q^T = 2\etaperp\left(\frac{1+\epsilon_H}{z} q^{N\prime} - \frac{2+3\epsilon_H}{z^2}q^N\right),\\
q^{N\prime\prime}&+\frac{1+2\epsilon_H}{z^2}\left(m^2-2+\epsilon_H-(\etaperp)^2\right)q^N = -2\etaperp\left(\frac{1+\epsilon_H}{z}q^{T\prime}+\frac{1+2\epsilon_H}{z^2}q^T\right).
\end{align*}
Using the ansatz $q^I=A^I z^P$ and assuming first order slow roll and expanding to second order in $\epsilon$ the matrix becomes
$$
\left(\begin{smallmatrix}
 p(p-1)-2-3\epsilon-6\epsilon^2 & -2 \etaperp  \left(p-2+(p-3) \epsilon+(p-4) \epsilon ^2\right) \\
 2 \etaperp  \left(p+1+(p+2) \epsilon+(p+3) \epsilon ^2 \right) & p(p-1)+(m^2-2-(\etaperp)^2) +\left(-2 (\etaperp)^2+2
   m^2-3\right)\epsilon  +\left(-3 \eta ^2+3 m^2-4\right)\epsilon ^2
\end{smallmatrix} \right)
\left(\begin{smallmatrix}
  A^T  \\
  A^N
\end{smallmatrix} \right)
 =0,
$$
where we used $m^2\equiv M^2/H^2$. The four solutions are given by
\begin{align*}
&p=-1-\epsilon-\frac{ 11 (\etaperp)^2+m^2}{9 (\etaperp)^2+3 m^2}\epsilon ^2 \quad \text{and} \quad \frac{A^N}{A^T}=-\frac{4 \etaperp  }{9 (\etaperp)^2+3 m^2}\epsilon ^2,\\
&p=2+\epsilon + \frac{ 11 (\etaperp)^2+m^2}{9 (\etaperp)^2+3 m^2}\epsilon ^2 \quad \text{and} \quad \frac{A^N}{A^T}=-\frac{6 \etaperp }{m^2-(\etaperp)^2},\\
&p=\frac{1}{2} \left(1\pm\sqrt{9-4 \left(3 (\etaperp)^2+m^2\right)}\right)\mp\frac{  \left(6 (\etaperp)^2+2 m^2-3\right)}{\sqrt{9-4 \left(3 (\etaperp)^2+m^2\right)}}\epsilon + O(\epsilon^2) \\
&\quad \text{and} \quad \frac{A^T}{A^N}=\frac{4 \etaperp }{3\pm\sqrt{9-4 \left(3 (\etaperp)^2+m^2\right)}}\pm\frac{4 \etaperp }{\sqrt{9-4 \left(3 (\etaperp)^2+m^2\right)} \left(3\pm\sqrt{9-4 \left(3 (\etaperp)^2+m^2\right)}\right)}\epsilon +O\left(\epsilon ^2\right),
\end{align*}
where we expanded the last solutions only to first order in $\epsilon$ because it became too messy and in case of sufficiently high masses these are decaying solutions anyway. The full solution to linear order in $\epsilon$ is therefore given by
\begin{align*}
q^T_\alpha&=\delta^T_\alpha \left(Az^{-1-\epsilon}+ B z^{2+\epsilon}\right) \\&+  \delta^N_\alpha C\frac{4\etaperp}{3+\sqrt{9-4(m^2+3(\etaperp)^2)}}\left(1+\frac{\epsilon}{\sqrt{9-4(m^2+3(\etaperp)^2)}}\right) z^{\frac{1}{2} \left(1+\sqrt{9-4 \left(3 (\etaperp)^2+m^2\right)}\right)-\frac{  \left(6 (\etaperp)^2+2 m^2-3\right)}{\sqrt{9-4 \left(3 (\etaperp)^2+m^2\right)}}\epsilon} \\&+
\delta^N_\alpha D\frac{4\etaperp}{3-\sqrt{9-4(m^2+3(\etaperp)^2)}}\left(1-\frac{\epsilon}{\sqrt{9-4(m^2+3(\etaperp)^2)}}\right) z^{\frac{1}{2} \left(1-\sqrt{9-4 \left(3 (\etaperp)^2+m^2\right)}\right)+\frac{  \left(6 (\etaperp)^2+2 m^2-3\right)}{\sqrt{9-4 \left(3 (\etaperp)^2+m^2\right)}}\epsilon}, \\
q^N_\alpha&=\delta^T_\alpha\left( B \frac{6\etaperp}{(\etaperp)^2-m^2}z^{2+\epsilon}\right)\\&+ \delta^N_\alpha\left(C z^{\frac{1}{2} \left(1+\sqrt{9-4 \left(3 (\etaperp)^2+m^2\right)}\right)-\frac{  \left(6 (\etaperp)^2+2 m^2-3\right)}{\sqrt{9-4 \left(3 (\etaperp)^2+m^2\right)}}\epsilon} + D z^{\frac{1}{2} \left(1-\sqrt{9-4 \left(3 (\etaperp)^2+m^2\right)}\right)+\frac{  \left(6 (\etaperp)^2+2 m^2-3\right)}{\sqrt{9-4 \left(3 (\etaperp)^2+m^2\right)}}\epsilon}\right),
\end{align*}
to first order in slow roll. Note that we find the same growing and decaying solutions as before and in particular that the sourcing of the isocurvature mode by the growing curvature mode is suppressed by $\epsilon^2$.\\

\section{Transition}
In this section we provide more details of the computations made in section \ref{section:transition}.

\subsection{Case 1: effective single field description in case of a large mass hierarchy}
\label{Appendix:transitionmasshierarchy}
We derive the effective perturbation equation for $\R$, the adiabatic conditions under which the effective theory is valid and the computation of the predicted power spectrum.
\subsubsection{Effective perturbation equation for $\R$}
As explained in section \ref{Section:superhorizon} we integrate out the heavy modes when $z<m$. The low frequency solution allows us to neglect the time derivatives of $\S$ which gives
$$
\left(\frac{k^2}{a^2H^2}+m^2-(\etaperp)^2+\eta_H(\eta_H-3+\epsilon)-\eta_H^;\right)\S=2\etaperp \R^;,
$$
here the semi-colon represents a derivative with respect to number of e-folds $N$, resulting into the following equations of motion for $\R$
$$
\left(\frac{d}{dN}+3-\epsilon-2\eta_H \right)\left(\R^;+\frac{4(\etaperp)^2}{\frac{k^2}{a^2H^2}+m^2-(\etaperp)^2+\eta_H(\eta_H-3+\epsilon)-\eta_H^;}\R^;\right)+ \frac{k^2}{a^2H^2}\R=0,
$$
Defining the speed of sound $c_s$ as
$$
\frac{1}{c_s^2}\equiv 1+\frac{4(\etaperp)^2}{\frac{k^2}{a^2H^2}+m^2-(\etaperp)^2+\eta_H(\eta_H-3+\epsilon)-\eta_H^;},
$$
we get the following effective equation
$$
\frac{1}{c_s^2}\left[\R^{;;}+\left(3-\epsilon-2\eta_H-2\frac{c_s^;}{c_s}\right)\R^;+\frac{c_s^2k^2}{a^2H^2}\R\right]=0.
$$
Rewriting this as
$$
\frac{c_s^2}{a^3\epsilon H}\left(\frac{\R^;a^3H\epsilon}{c_s^2}\right)^;-\frac{c_s^2}{a^2H^2}\partial_i^2\R=0,
$$
allows us to derive the following effective quadratic action up to a constant
$$
S_{\text{eff}} \sim M_P^2\int dNd^3\v{x}\ \frac{a^3\epsilon H}{c_s^2}\left[(\R^;)^2-\frac{c_s^2}{a^2H^2}(\partial_i \R)^2\right],
$$
where the square of the Planck mass is included because of dimensional analysis. Using our knowledge of canonical single field inflation the constant of proportionality is equal to unity.

\subsubsection{Adiabatic conditions}
We assumed that we can neglect the derivatives of $\S$ in the equations of motion, which means the effective theory is only valid if
\begin{align*}
&3\left|\S^;\right| \ll \left| \left(\frac{k^2}{a^2H^2}+m^2-(\etaperp)^2\right)\S \right|,\\
&\left|\S^{;;}\right| \ll \left| \left(\frac{k^2}{a^2H^2}+m^2-(\etaperp)^2\right)\S \right|.
\end{align*}
Using $\R^;\sim\frac{k}{aH}\R$ and $\S=\frac{1}{2}(c_s^{-2}-1)\R^;$ and $\left(\frac{k}{aH}\right)^;\approx-\frac{k}{aH}$ these conditions translate into
\begin{align*}
&3\left|\frac{\left(c_s^{-2}-1\right)^;}{c_s^{-2}-1}+\frac{k}{aH}-1\right| \ll \left| \frac{k^2}{a^2H^2}+m^2-(\etaperp)^2 \right|,\\
&\left|\frac{\left(c_s^{-2}-1\right)^{;;}}{c_s^{-2}-1}+\frac{k^2}{a^2H^2}+2\left(\frac{k}{aH}-1\right)\frac{\left(c_s^{-2}-1\right)^;}{c_s^{-2}-1}-3\frac{k}{aH}+1\right| \ll \left| \frac{k^2}{a^2H^2}+m^2-(\etaperp)^2 \right|,
\end{align*}
where we neglected the slow roll parameters. In addition we can use $m^2\gg\frac{k^2}{a^2H^2}\gg1$ and denote $u\equiv 1- c_s^{-2}$ to simplify the condition to
\begin{align*}
&\left|\frac{u^;}{u}\right| \ll \frac{1}{3}\left|m^2-(\etaperp)^2 \right|,\\
&\left|\frac{u^{;;}}{u}+2\left(\frac{k}{aH}-1\right)\frac{u^;}{u}\right| \ll \left| m^2-(\etaperp)^2 \right|,
\end{align*}
the latter condition implies that if the following \textit{adiabatic conditions} are satisfied the effective theory is valid
\begin{align*}
&\left|\frac{u^;}{u}\right| \ll \min\left(\frac{|m^2-(\etaperp)^2|}{3}, \frac{\left|m^2-(\etaperp)^2\right|}{\left|2k/aH-2\right|} \right) ,\\
&\left|\frac{u^{;;}}{u}\right| \ll \left| m^2-(\etaperp)^2 \right|,
\end{align*}
independent on the fact whether $u$ is small or not. It is also possible to express everything in term of the turn parameter and its derivatives. We have
\begin{align*}
&\frac{u^;}{u}\approx-2\xiperp(1+u),\\
&\frac{u^{;;}}{u}\approx-2{\xiperp}^;(1+u)+4(\xiperp)^2(1+3u+2u^2),
\end{align*}
where we neglected derivatives with respect to $m$, $\frac{k}{aH}$ and the slow-roll parameters. Now when $u\ll 1$ we get therefore approximately
\begin{align*}
&\left|\frac{{\etaperp}^;}{\etaperp}\right| \ll \sqrt{|m^2-(\etaperp)^2|} ,\\
&\left|{\xiperp}^;\right| \ll |m^2-(\etaperp)^2|.
\end{align*}
These conditions ensure that the timescale of the duration of the turn is much bigger than the period of the oscillations about the flat direction of the potential.

\subsubsection{Power spectrum}
If the adiabatic conditions are satisfied it is possible to solve the effective single field equation (\ref{Equation:effectiveequationR}) for $\R$. We consider the two cases of a constant turn rate $\etaperp$ for all times and a transient turn for which $c_s$ is assumed to go to 1 for the limits $\tau\rightarrow -\infty$ and $\tau\rightarrow 0$. Assuming a constant turn rate, the initial conditions for $\R$ change as explained in section \ref{Section:subhorizon}. The low frequency $\omega_-$ as defined in equation (\ref{Equation:frequenciessubhorizon}) becomes
$$
\omega_-^2=\frac{k^2}{a^2}\left(1-\frac{4(\etaperp)^2}{m^2+3(\etaperp)^2}\right)\approx \frac{k^2c_s^2}{a^2}
$$
where the latter estimate relies on $k^2/a^2H^2 \ll m^2$. To see this invert the expression for the speed of sound
$$
c_s^2=1-\frac{4(\etaperp)^2}{\frac{k^2}{a^2H^2}+m^2+3(\etaperp)^2}.
$$
The solution is written as
\begin{align*}
&q^T_\alpha = (A_-)_\alpha e^{i\int dt\omega_- }+(A_+)_\alpha e^{i\int dt\omega_+ },\\
&q^T_\alpha = (B_-)_\alpha e^{i\int dt\omega_- }+(B_+)_\alpha e^{i\int dt\omega_+ }.
\end{align*}
The quantization conditions determine the normalization and can be used to approximate $A_-$ and $B_+$
$$
(A_-)_\alpha=\frac{1}{\sqrt{2a\omega_-}}\delta^T_\alpha \quad \text{and} \quad (B_+)_\alpha=\frac{1}{\sqrt{2a\omega_+}}\delta^N_\alpha,
$$
where we assumed that $A_+$ and $B_-$ do not contribute much to the amplitude of $q^I$ (and therefore assuming that at the beginning of the second part of the sub-horizon regime $(\etaperp)^2\ll m^2$.) If we compute an effective theory we neglect the rapidly oscillating $\alpha=2$ modes, this leads to the following effective equation as function of conformal time
$$
q^{\prime\prime}+\left(k^2c_s^2-\frac{c_s}{a\sqrt{\epsilon}}\left(\frac{a\sqrt{\epsilon}}{c_s}\right)^{\prime\prime}\right)q=0,
$$
with $q=q^T_1/c_s$. The initial conditions are given by
$$
q_i=\frac{1}{c_s \sqrt{2k c_s}} \quad \text{and} \quad q_i^\prime = \frac{i \sqrt{k c_s}}{c_s \sqrt{2}}.
$$
In case of a more or less constant turn rate we can neglect the time variation of $c_s$ (also the $k^2/a^2H^2$ part) and defining the time variable $d\tau_c=c_s d\tau$ the equation becomes
$$
\frac{d^2 q}{d\tau_c^2}+\left(k^2-\frac{1}{a\sqrt{\epsilon}}\frac{d^2 \left(a\sqrt{\epsilon}\right)}{d\tau_c^2}\right)q=0,
$$
which is the Mukhanov-Sasaki equation. To zeroth order in slow-roll the solution is given by
$$
q_k = \frac{e^{ik c_s\tau}}{c_s \sqrt{2kc_s}}\left(1-\frac{i}{k\tau c_s}\right),
$$
which yields
$$
\R_k = \frac{1}{\sqrt{2\epsilon}aM_P}\frac{e^{ik c_s\tau}}{ \sqrt{2kc_s}}\left(1-\frac{i}{k\tau c_s}\right).
$$
On super-horizon scales $k\tau c_s  \rightarrow 0$ we therefore get
$$
\R_k = \frac{1}{\sqrt{2\epsilon}aM_P} \frac{1}{i\sqrt{2k^3c_s^3}\tau}\approx -\frac{H}{i{\sqrt{2\epsilon k^3 c_s}aM_P}},
$$
where we used
$$
\H=-\frac{1}{\tau_c}(1+\epsilon_\ast+\epsilon_\ast^2), \quad k\tau_c^\ast=-(1+\epsilon_\ast+\epsilon_\ast^2).
$$
This yields the following power spectrum
$$
\P_\R=\frac{H^2_\ast}{8\epsilon_\ast \pi^2 c_s M_P^2}\approx \frac{1}{c_s}\P_{\R 0},
$$
which is about $\frac{1}{c_s}$ times the power spectrum $\P_{\R 0}$ one would achieve when naively computing the truncated single field version.
Actually one should also take into account the change of $H_\ast$ and $\epsilon_\ast$ in order to compare them. When $c_s<1$ the modes will cross the horizon earlier which means that $H_\ast$ becomes bigger and $\epsilon_\ast$ comparable or smaller than the single field version. This means the power spectrum will be enhanced even more. We can make an estimate how much $H_\ast$ changes
$$
H_{\ast 1}\approx H_{\ast 0}-H^;_{\ast 0}\Delta N \approx H_{\ast 0}\left(1+\epsilon\frac{1-c_s}{c_s}\right),
$$
which is indeed negligible when $c_s$ is close to 1. Furthermore we can compute the spectral indices and $r$
$$
n_s-1=\frac{d\ln \P_\R}{d \ln k}=\frac{d\ln \P_\R}{d N} \frac{dN}{d \ln k}=\frac{-2\epsilon+\eta_H}{1-\epsilon}\approx-2\epsilon+\eta_H,
$$
$$
n_t=\frac{d\ln \P_T}{d \ln k}=\frac{-2\epsilon}{1-\epsilon}\approx-2\epsilon,
$$
$$
r=16\epsilon_\ast c_s = c_s r_0,
$$
where we used that
$$
\P_T=2\left(\frac{H_\ast}{\pi M_P}\right)^2.
$$

\subsection{Case 2: full solution to the perturbation equations using the SRST approximation}
We derive the diagonalization of the perturbation equation for $q$, the constraints on the validity of the approximation scheme and the prediction of the power spectrum.
\subsubsection{Diagonalization of the perturbation equation q}
We start from the equation
$$
D_\tau^2q^a+k^2q^a+\Omega^a_b q^b=0, \quad \Omega^a_b=a^2H^2\begin{pmatrix}
 -2+\epsilon-\eta_H(-3+\epsilon+\eta_H-\xi_H)+(\etaperp)^2  & \etaperp\left(3-\epsilon-2\eta_H-\xiperp\right) \\
\etaperp\left(3-\epsilon-2\eta_H-\xiperp\right)  &  -2+\epsilon +\frac{M^2_{NN}}{H^2}
 \end{pmatrix}.
$$
During transition which is only a short period of time we can take the first order slow roll approximation ($\epsilon=\epsilon_\H$, $\eta_H, \xi_H =0$). Assuming in addition the first order slow turn approximation ($\etaperp\approx const$, $\xiperp=0$) we have
$$
\frac{\Omega^a_b}{a^2H^2}=\begin{pmatrix}
 -2+\epsilon+(\etaperp)^2  & \etaperp(3-\epsilon) \\
 \etaperp(3-\epsilon)  &  -2+\epsilon +\frac{M^2_{NN}}{H^2}
 \end{pmatrix},
$$
which is approximately constant and therefore it can be diagonalized to
\begin{align*}
&\frac{(U^{-1}\Omega U)^a_b}{a^2H^2}=\begin{pmatrix}
 -2+\epsilon+m^2_-  & 0 \\
 0  &   -2+\epsilon+m^2_+  \end{pmatrix}, \\
&m^2_\pm=\frac{1}{2}\left((\etaperp)^2+\frac{M^2_{NN}}{H^2}\pm\sqrt{4(3-\epsilon)^2(\etaperp)^2+\left(\frac{M^2_{NN}}{H^2}-(\etaperp)^2\right)^2}\right),\\
&U=\begin{pmatrix}
 \cos\theta  &  -\sin\theta \\
 \sin\theta  &  \cos\theta \end{pmatrix}, \quad \tan2\theta=\frac{2\etaperp(3-\epsilon)}{(\etaperp)^2-\frac{M^2_{NN}}{H^2}}.
\end{align*}
Defining
$$
u^I(\tau)\equiv (U^{-1})^I_J q^J(\tau),
$$
we get
\begin{equation}
\begin{pmatrix}
 D_\tau^2 + k^2+a^2H^2\left(-2+\epsilon +m^2_-\right)  & 0 \\
 0  & D_\tau^2 + k^2 +a^2H^2\left(-2+\epsilon +m^2_+\right)  \end{pmatrix} \begin{pmatrix} u^T   \\ u^N  \end{pmatrix}  =0,
\label{Equation:transitionqnoturns2appendix}
\end{equation}

\subsubsection{Constraints on the validity of the approximation scheme}
In the papers \cite{Peterson:2010np, GrootNibbelink:2001qt} equation (\ref{Equation:transitionqnoturns2appendix}) is solved by Hankel functions. However, this equation is only equivalent to the Bessel equation if $D_\tau^2=\partial_\tau^2$, which is true if there are no turns during the transition period. Therefore in general we actually should start from equation (\ref{Equation:eomqtilde}) and diagonalize it such that we can solve it with Hankel functions. For two fields $R$ is defined by the following equation
$$
R'=aH\etaperp\left(\begin{smallmatrix} 0 & -1 \\ 1 & 0 \end{smallmatrix}\right) R
$$
Define
$$
\alpha(\tau)=\int_{\tau_-}^\tau d\tilde{\tau}aH\etaperp=\int_{N_-}^N d\tilde{N}\etaperp,
$$
then
$$
R(\tau)=R(\tau_-)e^{\alpha(\tau)\left(\begin{smallmatrix} 0 & -1 \\ 1 & 0 \end{smallmatrix}\right)}=R(\tau_-)\begin{pmatrix} \cos\alpha(\tau) & -\sin\alpha(\tau) \\ \sin\alpha(\tau) & \cos\alpha(\tau) \end{pmatrix}.
$$
We introduce a new variable
$$
\bar{q}^I_{\v{k}}(\tau)\equiv R(\tau_\H)^I_J\tilde{q}^J_{\v{k}}(\tau).
$$
The equations of motion for $\bar{q}^I_{\v{k}}$ become
$$
\bar{q}^{I\prime\prime}_{\v{k}}+k^2\bar{q}^I_{\v{k}}+\bar{\Omega}^I_J \bar{q}^J_{\v{k}}=0,
$$
with
\begin{equation}
\label{Equation:omegabar}
\bar{\Omega}^I_J(\tau)=\left[\begin{pmatrix} \cos\Delta\alpha & \sin\Delta\alpha \\ -\sin\Delta\alpha & \cos\Delta\alpha \end{pmatrix}\Omega(\tau) \begin{pmatrix} \cos\Delta\alpha & -\sin\Delta\alpha \\ \sin\Delta\alpha & \cos\Delta\alpha \end{pmatrix}\right ]^I_J, \quad \Delta\alpha=\alpha(\tau)-\alpha(\tau_\H).
\end{equation}
Using the variable substitution $z=-k\tau$ we get
$$
\frac{d^2\bar{q}^I_{\v{k}}}{dz^2}+\bar{q}^I_{\v{k}}+\frac{\bar{\Omega}^I_J}{k^2}\bar{q}^J_{\v{k}}=0.
$$
Comparing this to the Bessel equation
$$
z^2\frac{d^2y}{dz^2}+z\frac{dy}{dz}+(z^2-\beta^2)y=0,
$$
which becomes for the variable $x=\sqrt{z}y$
$$
\frac{d^2x}{dz^2}+x-\frac{\beta^2-\frac{1}{4}}{z^2}x=0.
$$
We can identify
$$
x=\bar{q}^I_{\v{k}}, \quad \quad (\beta^2)^I_J=\frac{1}{4}\id^I_J-\frac{1}{(1-\epsilon_\H)^2}\frac{\bar{\Omega}^I_J}{a^2H^2},
$$
where we used $a^2H^2\approx (1+2\epsilon_\H)k^2/z^2$. Now diagonalizing $\beta$ yields the two Hankel functions as solutions to this equation. But this identification only works when $\beta$ is a constant matrix, which means that $\Delta\alpha$ should be constant and therefore approximately zero during transition. This corresponds to neglecting the turns which allows us to use equation (\ref{Equation:transitionqnoturns2appendix}) with $D^2_\tau$ replaced by $\partial^2_\tau$. We can make an estimate of how small the turn must be such that this still a reasonable approximation. Using the first order slow roll approximation we can derive
$$
\H\approx k e^{(1-\epsilon)(N-N_\H)},
$$
therefore $N_-$ and $N_+$ are bounded by
\begin{align*}
&\H^2_-\ll k^2 \longrightarrow N_-  \leq N_\H-2.3\\
&\H^2_+\ll k^2 \longrightarrow N_+ \geq N_\H+2.3.
\end{align*}
We have the following upper bound of the absolute value of $\Delta\alpha$ at any time during transition
$$
\left|\Delta\alpha(N)\right|=\left|\int_{N_\H}^N d\tilde{N}\etaperp\right|\leq \left|(N-N_\H)\right||\etaperp|_{\max},
$$
or if $\etaperp$ has approximately a Gaussian shape with its dispersion $\sigma$ about two e-folds or less we have the estimate
$$
\left|\Delta\alpha(N)\right|\lesssim\sigma\sqrt{2\pi}|\etaperp|_{\max}.
$$
If we take the beginning and the end of transition $2.3$ e-folds from horizon crossing and demand that $\left|\Delta\alpha(N)\right|<0.1$ at any time during transition we find the following bound on the turn rate
\begin{align*}
|\etaperp|&\lesssim 0.04, \quad \text{if }\etaperp\text{ is more or less constant during transition},\\
|\etaperp|_{\max}&\lesssim \frac{0.04}{\sigma}, \quad \text{if }\etaperp\text{ has approximately a Gaussian shape with $\sigma\leq 2$}.
\end{align*}
This should agree with the assumption of slow turn. Moreover we still need to assume that $\beta$ is constant which in addition demands that $\etaperp$ is constant because for example
$$M^2_{TT}=V_{TT}=\sigma^;(V_T^;+\etaperp V_N)=H^2(\etaperp)^2+H^2O(\epsilon,\eta_H, \xi_H).$$
Therefore, the validity of the usual solution in terms of Hankel function depends strongly on whether the following conditions are satisfied or not
\begin{align}
& \epsilon, \etaperp\approx const \quad \text{and} \quad |\etaperp|\lesssim 0.04.
\end{align}

\subsubsection{Predictions}
If these conditions are fulfilled the solutions of the equations of motion for $u^I$ are given by a linear combination of the Hankel functions of the first and second kind
\begin{equation}
u^I_\alpha=\sqrt{z}H^{(1)}_{\nu^I}(z) A^I_\alpha+\sqrt{z}H^{(2)}_{\nu^I}(z) B^I_\alpha,
\label{Equation:Hankel_u2appendix}
\end{equation}
with
\begin{align*}
(\nu^T)^2&=\frac{1}{4}+\frac{2-\epsilon_\H-m^2_-}{(1-\epsilon_\H)^2}\approx\frac{9}{4}+3\epsilon_H+9\frac{(\etaperp)^2}{m^2},\\
(\nu^N)^2&=\frac{1}{4}+\frac{2-\epsilon_\H-m^2_+}{(1-\epsilon_\H)^2}\approx \frac{9}{4}+3\epsilon_H-(1+2\epsilon_H)m^2-9\frac{(\etaperp)^2}{m^2},
\end{align*}
where we assumed first order slow roll and $m^2=\frac{M^2_{NN}}{H^2}=O(10)\etaperp$ or $m^2\gg\etaperp$.
To first order in slow roll $\nu^N$ is not imaginary as long as
$$
m^2\leq\frac{9}{4}-\frac{3}{2}\epsilon-4(\etaperp)^2.
$$
The constants $A^I_\alpha$ and $B^I_\alpha$ in equation (\ref{Equation:Hankel_u2appendix}) have to be matched to the initial conditions for $q^I_\alpha$ when $z=z_-\gg1$. The Hankel functions can be estimated in this limit by
\begin{align*}
\sqrt{z}H^{(1)}_\nu(z)&=\sqrt{\frac{2}{\pi}}e^{i(z-\nu\pi/2-\pi/4)},\\
\sqrt{z}H^{(2)}_\nu(z)&=\sqrt{\frac{2}{\pi}}e^{-i(z-\nu\pi/2-\pi/4)},
\end{align*}
This yields
\begin{align*}
&A^T_\alpha=0, \quad B^T_\alpha=\sqrt{\frac{\pi}{4k}}\left(\cos\theta\delta^T_\alpha+\sin\theta\delta^N_\alpha\right)e^{-i\nu^T\pi/2-i\tilde{\lambda}^T},\\
&A^N_\alpha=0, \quad B^N_\alpha=\sqrt{\frac{\pi}{4k}}\left(\cos\theta\delta^N_\alpha-\sin\theta\delta^T_\alpha\right)e^{-i\nu^N\pi/2-i\tilde{\lambda}^N},
\end{align*}
with $\tilde{\lambda}=\lambda-\pi/4$.
At the end of transition we can use the limit $z\ll 1$ and the second Hankel function can be estimated as
$$
\sqrt{z}H^{(2)}_\nu(z)=\frac{i\sqrt{2}\Gamma(\nu)}{\pi}\left(\frac{2}{z}\right)^{\nu-1/2}+\frac{\sqrt{2}(1-i\cot(\pi\nu))}{\Gamma(1+\nu)}\left(\frac{2}{z}\right)^{-\nu-1/2}.
$$
Note that for $\nu=3/2$ we get the familiar solutions $z^{-1}$ and $z^2$. Making use of the fact that $\nu^T\approx 3/2$ we can neglect the second term for $u^T$ and we find
\begin{align*}
&u^T_\alpha(z_+)=\sqrt{\frac{\pi}{4k}}\left(\cos\theta\delta^T_\alpha+\sin\theta\delta^N_\alpha\right)e^{-i\nu^T\pi/2-i\tilde{\lambda}}\frac{i\sqrt{2}\Gamma(\nu^T)}{\pi}\left(\frac{2}{z_+}\right)^{\nu^T-1/2},\\
&u^N_\alpha(z_+)=\sqrt{\frac{\pi}{4k}}\left(\cos\theta\delta^N_\alpha-\sin\theta\delta^T_\alpha\right)e^{-i\nu^N\pi/2-i\tilde{\lambda}} \left(\frac{i\sqrt{2}\Gamma(\nu^N)}{\pi}\left(\frac{2}{z_+}\right)^{\nu^N-1/2}+\frac{\sqrt{2}(1-i\cot(\pi\nu^N))}{\Gamma(1+\nu^N)}\left(\frac{2}{z_+}\right)^{-\nu^N-1/2} \right).
\end{align*}
Transforming back to $q^I_\alpha$ we get
\begin{align*}
q^T_\alpha(z_+) = &\sqrt{\frac{\pi}{4k}}\left(\cos^2\theta\delta^T_\alpha+\cos\theta\sin\theta\delta^N_\alpha\right)f(\nu^T)\left(\frac{2}{z_+}\right)^{\nu^T-1/2}\\ &+\sqrt{\frac{\pi}{4k}}\left(\cos\theta\sin\theta\delta^N_\alpha-\sin^2\theta\delta^T_\alpha\right)\left(f(\nu^N)\left(\frac{2}{z_+}\right)^{\nu^N-1/2}+ g(\nu^N)\left(\frac{2}{z_+}\right)^{-\nu^N-1/2}\right),\\
q^N_\alpha(z_+)= &\sqrt{\frac{\pi}{4k}}\left(\cos^2\theta\delta^N_\alpha-\cos\theta\sin\theta\delta^T_\alpha\right)\left(f(\nu^N)\left(\frac{2}{z_+}\right)^{\nu^N-1/2}+g(\nu^N)\left(\frac{2}{z_+}\right)^{-\nu^N-1/2}\right)\\ &+\sqrt{\frac{\pi}{4k}}\left(\cos\theta\sin\theta\delta^T_\alpha+\sin^2\theta\delta^N_\alpha\right)f(\nu^T)\left(\frac{2}{z_+}\right)^{\nu^T-1/2} .
\end{align*}
Now there are multiple possibilities. If $m^2\ll1$ then $\nu^N\sim 3/2$ and we can neglect the decaying solution $z^{\nu^N+1/2}$ as well, but the first term is obviously a growing solution and this might produce a large spectrum for the isocurvature modes. If $m^2\gtrsim \frac{9}{4}$ then the real part of $\nu^N$ becomes zero and both terms $z^{\pm\nu^N+1/2}$ go like $z^{1/2}$ and these modes decay like massive modes. We consider both cases separately.\\

If $m^2\gtrsim \frac{9}{4}\gg \etaperp$ we can approximate $\cos\theta =1$, $\sin\theta=\frac{-\etaperp (3-\epsilon)}{m^2}$ and $\nu^T=\frac{3}{2}+\epsilon_\ast$ such that we can derive the following expression for the curvature and isocurvature perturbations
\begin{align*}
&\R_\alpha(z_+)\approx \frac{H(1-\epsilon_\ast)(1+C\epsilon_\ast)}{k\sigma^;\sqrt{2k} }\left(\delta^T_\alpha-\frac{\etaperp_\ast (3-\epsilon_\ast)}{m^2}\delta^N_\alpha\right)\left(\frac{2}{z_+}\right)^{\epsilon_\ast}e^{-i5\pi/4-i\tilde{\lambda}^T},\\
&\S_\alpha(z_+)\approx \frac{H(1-\epsilon_\ast)(1+C\epsilon_\ast)}{k\sigma^;\sqrt{2k} }\left(-\frac{\etaperp_\ast (3-\epsilon_\ast)}{m^2}\delta^T_\alpha\right) \left(\frac{2}{z_+}\right)^{\epsilon_\ast}e^{-i5\pi/4-i\tilde{\lambda}^T}
\end{align*}
with $C\equiv \frac{\Gamma^\prime(3/2)}{\Gamma(3/2)} \approx 0.03649$. This yields the following power spectra
\begin{align*}
&\P_\R\approx\frac{H^2_\ast}{8\epsilon_\ast \pi^2  M_P^2}\left(1+2(C-1)\epsilon_\ast\right)\left(1+\frac{9(\etaperp_\ast)^2}{m^4}\right),\\
&\P_\S\approx\frac{H^2_\ast}{8\epsilon_\ast \pi^2  M_P^2}\left(1+2(C-1)\epsilon_\ast\right)\left(\frac{9(\etaperp_\ast)^2}{m^4}\right).\\
\end{align*}

If $m^2\ll1$ we can approximate 
$\nu^T=\nu^N=\frac{3}{2}+\epsilon_\ast$ such that we get
\begin{align*}
&q^N_\alpha(z_+)=\sqrt{\frac{\pi}{4k}}f\left(3/2+\epsilon_\ast\right)\delta^N_\alpha\left(\frac{2}{z_+}\right)^{1+\epsilon_\ast},\\
&q^T_\alpha(z_+)=\sqrt{\frac{\pi}{4k}}f\left(3/2+\epsilon_\ast\right)\delta^T_\alpha\left(\frac{2}{z_+}\right)^{1+\epsilon_\ast}.
\end{align*}
The curvature and isocurvature perturbations become therefore
\begin{align*}
&\R_\alpha(z_+)\approx \frac{H(1-\epsilon_\ast)(1+C\epsilon_\ast)}{k\sigma^;\sqrt{2k} }\delta^T_\alpha\left(\frac{2}{z_+}\right)^{\epsilon_\ast},\\
&\S_\alpha(z_+)\approx \frac{H(1-\epsilon_\ast)(1+C\epsilon_\ast)}{k\sigma^;\sqrt{2k} }\delta^N_\alpha\left(\frac{2}{z_+}\right)^{\epsilon_\ast},
\end{align*}
meaning they evolve independently and we get both a spectrum of curvature modes and isocurvature modes at the end of the transition regime
\begin{align*}
&\P_\R\approx\frac{H^2_\ast}{8\epsilon_\ast \pi^2  M_P^2}\left(1+2(C-1)\epsilon_\ast\right),\\
&\P_\S\approx\frac{H^2_\ast}{8\epsilon_\ast \pi^2  M_P^2}\left(1+2(C-1)\epsilon_\ast\right).
\end{align*}

\end{appendices}

\end{document}